\documentclass[ejs,preprint]{imsart}

\RequirePackage[numbers]{natbib}
\RequirePackage[colorlinks,citecolor=blue,urlcolor=blue]{hyperref}

\usepackage{adjustbox}
\usepackage[ruled,vlined,noend,linesnumbered]{algorithm2e}
\usepackage{amsfonts,amsmath,amssymb,amsthm}
\usepackage{bbm}
\usepackage{booktabs}
\usepackage{float}
\usepackage[edges]{forest}
\usepackage[margin=3cm]{geometry}
\usepackage{listings}
\usepackage{mathtools}
\mathtoolsset{showonlyrefs=false}
\usepackage{pgfplots}
\usepackage{physics}
\usepackage{silence}
\usepackage{standalone}
\usepackage{subcaption}
\usepackage{tikz}
\usepackage{xcolor}

\usepgfplotslibrary{groupplots}
\pgfplotsset{compat=1.14}
\pgfplotsset{filter discard warning=false}
\doi{10.1214/154957804100000000}
\pubyear{0000}
\volume{0}
\firstpage{1}
\lastpage{1}

\startlocaldefs

\newcommand*{\bbR}{\mathbb{R}}  %

\newtheorem{example}{Example}[section]

\newtheorem{proposition}{Proposition}[section]
\newtheorem{remark}{Remark}[section]

\SetKwFor{ParFor}{for}{in parallel do}{end}
\SetKwFor{ParForSet}{for}{in parallel set}{end}
\SetKwFor{ForSample}{for}{sample}{end}
\SetKwFor{ForSet}{for}{set}{end}
\SetKwFor{ForCompute}{for}{compute}{end}
\SetKwBlock{Locals}{Local Variables:}{end}
\SetKwProg{Fn}{Function}{}{}
\SetKwFunction{LGSSMStep}{LGSSM-MH-Step}

\SetKwInOut{Ret}{Return}
\DontPrintSemicolon

\newcommand{\ft}[3]{#1_{#2:#3}}  %

\newcommand{\bigO}{\mathcal{O}}
\DeclarePairedDelimiterX{\infdivx}[2]{(}{)}{%
#1\;\delimsize\|\;#2%
}
\newcommand{\kl}{\mathrm{KL}\infdivx}

\renewcommand{\citep}{\cite}
\renewcommand{\citet}{\cite}

\endlocaldefs

\begin{document}
\begin{frontmatter}
  \title{Auxiliary MCMC samplers for parallelisable inference in high-dimensional latent dynamical systems}
  \runtitle{Auxiliary MCMC for state-space models}
  \begin{aug}
    \author{\fnms{Adrien}~\snm{Corenflos}\ead[label=e1]{adrien.corenflos@warwick.ac.uk}}
    \address{Department of Statistics, University of Warwick\\
    Department of Electrical Engineering and Automation, Aalto University\\
    \printead[presep={}]{e1}}
    \and
    \author{\fnms{Simo}~\snm{S\"arkk\"a}\ead[label=e2]{simo.sarkka@aalto.fi}}
    \address{Department of Electrical Engineering and Automation, Aalto University\\
      \printead[presep={}]{e2}}
    \runauthor{A. Corenflos and S. S\"arkk\"a}
  \end{aug}

  \begin{abstract}
    Sampling from the full posterior distribution of high-dimensional non-linear, non-Gaussian latent dynamical models presents significant computational challenges. While Particle Gibbs (also known as conditional sequential Monte Carlo) is considered the gold standard for this task, it quickly degrades in performance as the latent space dimensionality increases. Conversely, globally Gaussian-approximated methods like extended Kalman filtering, though more robust, are seldom used for posterior sampling due to their inherent bias.
    We introduce novel auxiliary sampling approaches that address these limitations. By incorporating artificial observations of the system as auxiliary variables in our MCMC kernels, we develop both efficient exact Kalman-based samplers and enhanced Particle Gibbs algorithms that maintain performance in high-dimensional latent spaces. Some of our methods support parallelization along the time dimension, achieving logarithmic scaling when implemented on GPUs. Empirical evaluations demonstrate superior statistical and computational performance compared to existing approaches for high-dimensional latent dynamical systems.
  \end{abstract}


  \tableofcontents

\end{frontmatter}

\section{Introduction}
\label{sec:introduction}
State-space models \citep[SSMs, see, e.g.,][]{DelMoral2004Book,sarkka2023bayesian,Chopin2020book}, otherwise known as hidden Markov models, are a class of dynamic statistical models routinely employed to model phenomena in bio-medicine, epidemiology, chemistry, or economics. 
For a given finite horizon $T>0$, they are fully described by the joint distribution over their latent states and the observations, which, when it exists, can be identified with its density
\begin{equation}\label{eq:ssm}
    p(\ft{x}{0}{T}, \ft{y}{0}{T})
        \coloneqq p_0(x_0) \left\{\prod_{t=0}^T h_t(y_t \mid x_t)\right\} \left\{\prod_{t=1}^T p_t(x_t \mid x_{t-1}) \right\}.
\end{equation}
In this formulation, $p_0$ represents the initial distribution of the state $x_0$, while $p_t$ and $h_t$ represent the (conditional) transition and emission distributions for the states $x_t \in \bbR^{d_x}$ and observations $y_t \in \bbR^{d_y}$, respectively. 

Inference in SSMs typically recovers different meanings depending on the context: filtering is concerned with sampling, or computing expectations with respect to the conditional distribution $p(x_t \mid \ft{y}{0}{t})$, where $\ft{y}{0}{t} = \{y_i; i=0, 1, \dots,t\}$; marginal smoothing is concerned with the same problems for the quantity $p(x_t \mid \ft{y}{0}{T})$, $t \leq T$; and pathwise smoothing is concerned with sampling or computing expectations with respect to the quantity $p(\ft{x}{0}{T} \mid \ft{y}{0}{T})$. 

In many cases, the ``true'' generative model, consisting of the initial distribution $p_0$, the transition distributions $p_t$, and emission distributions $h_t$, is unknown, and one needs to estimate it from the observed data. A typical way is to assume parametric forms for $p_0(x_0 \mid \theta)$, $p_t(x_t \mid x_{t-1}, \theta)$, and $h_t(y_t \mid x_t, \theta)$, as well as a prior distribution $p(\theta)$ for the parameters, resulting in a joint distribution
\begin{equation}
    \label{eq:parametric-ssm}
    p(\ft{x}{0}{T}, \ft{y}{0}{T}, \theta)
    \coloneqq p_0(x_0\mid \theta) \left\{\prod_{t=0}^T h_t(y_t \mid x_t, \theta)\right\} \left\{\prod_{t=1}^T p_t(x_t \mid x_{t-1}, \theta) \right\} p(\theta).
\end{equation}
Under these notations, the parameter estimation problem then consists of computing either deterministic or probabilistic estimates of the posterior distribution over the parameters $p(\theta \mid y_{0:T})$. In this work, we will focus on computing probabilistic estimates for the pathwise smoothing distribution $p(\ft{x}{0}{T} \mid \ft{y}{0}{T})$ and the joint state-parameter posterior distribution $p(\theta, x_{0:T} \mid y_{0:T})$ (which marginally recovers $p(\theta \mid y_{0:T})$).

Throughout the rest of the article, for notational simplicity and when this is not harmful, the dependency on the parameters $\theta$ will be implicit, and the methods will be presented for models with fixed parameters, i.e., we will write $p(\ft{x}{0}{T}, \ft{y}{0}{T})$ and similar for the related conditional distributions. 

In this article, we consider a slight generalisation of~\eqref{eq:ssm}, as given by the larger class of models
\begin{equation}\label{eq:gen-ssm}
    \pi(\ft{x}{0}{T})
        \propto g(x_0, x_1, \ldots, x_T) \, p_0(x_0)\, \left\{\prod_{t=1}^T p_t(x_t \mid x_{t-1}) \right\}.
\end{equation}
It is easy to see that this class comprises, as a special case, the pathwise smoothing distribution $p(x_{0:T} \mid y_{0:T})$ of~\eqref{eq:ssm} by setting $g(x_0, x_1, \ldots, x_T) = \prod_{t=0}^T h_t(y_t \mid x_t)$.
It also recovers the class of Feynman-Kac models~\citep[see, e.g.,][]{DelMoral2004Book}
\begin{equation}\label{eq:feynman-kac}
    \pi(\ft{x}{0}{T})
        \propto g_0(x_0) \, p_0(x_0) \, \left\{\prod_{t=1}^T g_t(x_t, x_{t-1}) \, p_t(x_t \mid x_{t-1}) \right\},
\end{equation}
for a Markovian potential function $g(x_0, x_1, \ldots, x_T) = g_0(x_0) \,\prod_{t=1}^T g_t(x_t, x_{t-1})$, which is typically the setting in which the so-called particle filtering methods apply~\citep[Ch. 5]{Chopin2020book}.

The most popular two classes of methods for inference in SSMs are the Gaussian approximation-based methods (i.e., Kalman filters and smoothers), and the sequential Monte Carlo (SMC) based methods (i.e., particle filters and smoothers). These methods, their benefits, and their drawbacks are briefly reviewed next in Sections~\ref{subsec:kalman-intro} and~\ref{subsec:smc-intro}.

\subsection{Gaussian approximated state-space models}
    \label{subsec:kalman-intro}
    Gaussian approximations rely on the fact that when the SSM at hand is linear Gaussian (LGSSM), then the filtering and marginal smoothing distributions are Gaussian as well, and their means and covariances can be computed sequentially and in closed form~\citep[see, e.g.,][]{sarkka2023bayesian,barfoot2017state}. This is leveraged in Gaussian approximations to the filtering and marginal smoothing solutions of general SSMs. Typically, such approximations rely on Taylor linearisation, leading to the classical extended Kalman filtering \citep[see, e.g.,][]{Jazwinski:1970}, or on sigma-point linearisations, first introduced in \citet{julier2004unscented,wan2000unscented}. 
        
The state of the art for these methods consists in iteratively reusing the approximated marginal smoothing distributions to refine the Gaussian approximation of the SSM at hand~\citep{bell1994iterated,Garcia:2017,Tronarp2018iterative}. Doing so makes it possible to handle SSMs for which the reverse Markov chain representing the smoothing distribution is a slow-mixing process, that is, SSMs which have ``sticky'' transitions kernels and for which the filtering transition largely differs from the smoothing one. These recursive methods have been shown to be equivalent to certain minimisation programs (such as Gauss--Newton) for some given loss functions and to be (locally) convergent. For a review, we refer the reader to~\citet{tronarp2020iterative} and~\citet[Ch. 10, 13, and 14]{sarkka2023bayesian}.

Finally, it has been recently shown \citep{Sarkka2021temporal,yaghoobi2021parallel,Yaghoobi2022sqrt} that (extended/sigma-point) Kalman filtering and smoothing can be parallelised in time (PIT), resulting in a computational complexity of $\bigO(\log(T))$ on parallel hardware such as graphics processing units (GPUs), comparing to their classical $\bigO(T)$ complexity on sequential hardware. This is particularly fruitful in the iterated context, as in \citet{yaghoobi2021parallel,Yaghoobi2022sqrt}, where the operation needs to be repeated until eventual convergence of the smoothing solution. 
Markov chain Monte Carlo algorithms, which the present article is concerned with, are one such class of iterated methods.

An important drawback of all the Gaussian approximation-based methods is that they (in all but the LGSSM case) result in biased estimates of the true non-Gaussian filtering as well as marginal and pathwise smoothing distributions. A bias is also present in the normalisation constant estimate (marginal likelihood of the observations) of the model, which makes parameter estimation procedures biased as well. This bias was the motivation for introducing Monte Carlo filtering methods~\citep{gordon1993novel} which we review next.

\subsection{Sequential Monte Carlo}
    \label{subsec:smc-intro}
    Sequential Monte Carlo (SMC) methods~\citep[see, e.g.,][]{Chopin2020book} are alternatives to Gaussian-approximated posteriors which represent the filtering and smoothing distributions using Monte Carlo samples. They proceed by propagating the trajectory sequentially via an importance sampling-resampling routine. Notably, SMC methods usually provide a representation of the full pathwise smoothing distribution as a byproduct of its representation of the filtering one. This representation converges when the number of samples tends to infinity~\citep{kitagawa1996smoothing}. However, in practice, the resulting paths degenerate for time steps $t \ll T$. This has justified the introduction of backward methods to rejuvenate the trajectories far from the endpoint~\citep{godsill2004monte}, and their resulting convergence improvements have been studied, for example in \citet{douc2011sequential}, and under a more general framework, in \citet{Dau2022complexity}. %

Importantly, because particle filtering provides an unbiased likelihood estimate, it can be used to perform asymptotically exact parameter and state estimation in state-space models. A particularly useful class of methods leveraging this property are the particle Markov chain Monte Carlo (pMCMC) methods \citep{andrieu2009pseudomarginal,Andrieu2010particle}, which are based on constructing MCMC schemes either as a Metropolis--Rosenbluth--Teller--Hastings (MRTH) algorithm~\citep{metropolis1953equation,Hastings:1970}, or a Gibbs-like sampler~\citep{geman1984stochastic}. We refer to these as pseudo-marginal and particle Gibbs (pGibbs), respectively. %

The aforementioned two methods sample consistently from the (joint) pathwise smoothing and parameter posterior distributions in general SSMs, but fail when the latent space dimension is large (or equivalently, when the observations are too informative compared to the prior dynamics). Backward sampling methods~\citep{whiteley2010discussion,lindsten2014particle} can be, to some extent, used to mitigate this problem. However, the failure is due to the inherent property that the set of particles available to describe the smoothing distribution comes from the forward filtering pass in the first place~\citep{daum2003curse}. This problem can, to some extent, be mitigated by using observation-informed proposals, sometimes inherited from the approximations of Section~\ref{subsec:kalman-intro} applied \emph{locally}~\citep[see, e.g.][]{van2000unscented}. Doing so, however, still fails as the dimension becomes larger.

Recently, \citet{finke2021csmc} and \citet[Ch. 4]{malory2021bayesian} independently proposed two related particle Gibbs algorithms that alleviate this issue by a generic localisation trick rather than approximation methods. \citet{finke2021csmc} in particular showed that under a proper scaling of their algorithms, the methods bypass the curse of dimensionality present in classical particle MCMC methods.

Finally, it was recently shown in \citet{corenflos2022sequentialized} that divide-and-conquer methods can provide consistent PIT solutions for particle smoothing and pGibbs algorithms at the cost of additional variance in the resulting estimates, providing an SMC counterpart to the algorithms of \citet{Sarkka2021temporal,yaghoobi2021parallel,Yaghoobi2022sqrt}.

\subsection{Motivation and contributions}
    \label{subsec:contribs-intro}
    As a summary of the sections above, the Gaussian approximated smoothing solutions, whilst being more robust than SMC methods (and extensions thereof), provide coarse approximations of the full posterior and lack the unbiasedness and convergence properties of SMC. They therefore cannot be used for exact Bayesian inference in general SSMs. Furthermore, while Gaussian approximations are regularly used \emph{locally} within particle filtering, and therefore particle MCMC~\citep[see, e.g.][]{van2000unscented}, they are seldom used to design global MCMC kernels~\citep[see, e.g., the introduction of][for a discussion on the difficulty of designing MCMC kernels for state-space models]{Andrieu2010particle}. 
On the other hand, SMC methods allow for asymptotically exact sampling of posterior SSM distributions but suffer from a curse of dimensionality that restricts their use to low-dimensional state spaces. This is true even when \emph{locally} informative proposal distributions are used and is a feature of pGibbs~\citep[Proposition 2.2]{finke2021csmc} that is inherited from particle filtering in general. 

In view of this, we develop general methods to perform statistically and computationally efficient inference in large-dimensional latent dynamical systems. To do so, we will consider two routes, which, at first, may seem unrelated but happen to be two specific instances of the same algorithm. The first one consists in designing an MCMC kernel based on SSM-specific Gaussian approximations and linearisations, while the second one relies on using localisation and linearisation techniques in a modified particle Gibbs algorithm. In both cases, we will pay particular attention to opportunities for parallelising the method on GPUs, specifically along the time dimension using techniques inherited from~\citet{Sarkka2021temporal,yaghoobi2021parallel,Yaghoobi2022sqrt} in Gaussian-approximated case and from~\citet{corenflos2022sequentialized} in the particle Gibbs case.

These two approaches are respectively based on (i) \citet{titsias2018} who design auxiliary MCMC gradient-based inference in high-dimensional latent Gaussian models, which we review in Section~\ref{subsec:auxiliary-general}; (ii) \citet{finke2021csmc} who reduce the curse of dimensionality in pGibbs methods by using localisation and exchangeable proposals within the underlying conditional SMC algorithm. 
At heart, both methods --- the former explicitly, the latter implicitly, as is explained in Section~\ref{subsec:smc-samplers} --- consist in augmenting the target distribution $\pi$ with auxiliary variables: using our SSM notation, $\pi(x_{0:T}, u_{0:T}) = \pi(x_{0:T}) \prod_{t=0}^T \mathcal{N}(u_t; x_t, \frac{\delta_t}{2} \Sigma_t)$ which marginally recovers the original distribution $\pi(x_{0:T})$. 
The inference is then performed in two steps summarised in Algorithm~\ref{alg:aux-sampler-gen} in which the choice of the kernel used in step~\ref{step:aux-2} is, in our specific context, either a custom MRTH kernel~\citep{titsias2018} or a pGibbs kernel for a modified model~\citep{finke2021csmc}. 
\begin{algorithm}[!htb]
\SetAlgoLined
\DontPrintSemicolon
\caption{Auxiliary MCMC}\label{alg:aux-sampler-gen}
\KwResult{An updated trajectory $x^{k+1}_{0:T}$}
\Fn{\textsc{aux-MCMC}$\big(x^k_{0:T}\big)$}{
    Sample $u^{k}_{0:T}  \sim \prod_{t=0}^T \mathcal{N}(u_{t}; x^k_t, \frac{\delta_t}{2} \Sigma_t)$\;\label{step:aux-1}
    Sample $x^{k+1}_{0:T} \sim K(\cdot \mid x^k_{0:T})$ \tcp{from a $\pi(\ft{x}{0}{T} \mid \ft{u^{k}}{0}{T})$-invariant kernel}\label{step:aux-2}
    \Return{$x^{k+1}_{0:T}$}
}
\end{algorithm}

This perspective motivates our contributions outlined below.
\begin{enumerate}
    \item In Section~\ref{sec:auxiliary_samplers}, we show that, in the case of generalised Feynman--Kac models~\eqref{eq:gen-ssm} with Gaussian dynamics, the auxiliary proposals of \citet{titsias2018} recover the posterior distribution of an auxiliary LGSSM. We leverage this to reduce their time and space complexity to $\bigO(T)$ rather than $\bigO(T^2)$. We then extend this to non-Gaussian prior dynamics using local Gaussian approximants. Furthermore, in Section~\ref{subsec:lgssm-sampling}, we introduce parallel-in-time samplers for the pathwise smoothing distribution of LGSSMs based on a prefix-sum implementation akin to \citet{Sarkka2021temporal}, resulting in an overall $\bigO(\log T)$ MCMC algorithm on parallel hardware.
    \item In Section~\ref{sec:pgibbs_samplers}, we describe how \citet{finke2021csmc} is an instance of the auxiliary sampler. 
    This novel perspective allows us to introduce novel, guided, auxiliary particle Gibbs methods by explicitly incorporating prior and gradient information in the form of locally optimal proposals for the auxiliary target model.
    Doing so improves on~\citet{finke2021csmc} in the highly-informative observation regime, but also reduces some of its drawbacks in the weakly-informative regime, essentially providing a more robust version of the method.
    Additionally, we discuss how this new perspective on~\citet{finke2021csmc} allows for the development of statistically efficient, gradient-informed parallel-in-time particle Gibbs samplers that can be efficiently implemented on GPUs.
    \item In Section~\ref{sec:experiments}, we apply the proposed methods to perform inference on a multidimensional stochastic volatility model~\citep{finke2021csmc}, a high-dimensional spatio-temporal model with fat-tailed observations taken from~\citet{cruscino2022highdim}, and on a joint state-parameter inference problem for a non-linear stochastic differential equation~\citep{mider2021continuous}. 
    Special attention is paid to understanding the statistical as well as computational trade-offs of our methods, in particular in terms of how the sequential and parallel counterparts of the methods (when they exist) compare.
    Finally, in Section~\ref{subsec:failure-modes}, we highlight the respective failure modes, potential pitfalls, and limitations of the proposed methods, also providing a heuristic for explaining their performance in the other experiments.
\end{enumerate}

\section{Auxiliary Kalman samplers}
\label{sec:auxiliary_samplers}
In this section, we first review the auxiliary samplers of \citet{titsias2018} for latent Gaussian models $\pi(x) \propto \exp(f(x))\,\mathcal{N}(x; 0, C)$. We then show how, in the case of latent Gaussian dynamics models, they can be specialised to reduce the time and memory complexity to linear in the number of time steps rather than quadratic. Finally, we discuss how linearisation methods can be used to extend the method to non-linear dynamics.

\subsection{Auxiliary gradient-based samplers}\label{subsec:auxiliary-general} 
    Auxiliary gradient-based methods were introduced in \citet{titsias2018} as a way to construct prior-informed proposals in MCMC samplers for Gaussian latent models with a density $\pi(x) \propto \exp(f(x))\,\mathcal{N}(x; 0, C)$\footnote{As well as, under a trivial change of variables, for models with non-zero prior mean.}, where $x \in \mathbb{R}^{d_x}$. 
They were shown to outperform classical pre-conditioned (prior-informed) and gradient-based (likelihood-informed) samplers, such as pre-conditioned Crank--Nicholson~\citep{cotter2013mcmc} or manifold MCMC methods~\citep{Girolami2011mMCMC} for latent Gaussian models. 
This impressive performance is both due to their better representation of the covariance of the posterior distribution~\citep[Section 3.4]{titsias2018}, and their computational advantage compared to classical methods, resulting in an improved effective sample size~\citep[ESS, see, e.g.,][Ch. 11]{gelman2013bayesian} per unit of time even when the effective sample size itself was lesser~\citep[Table 2]{titsias2018}.

Auxiliary gradient-based samplers rely on augmenting the target $\pi$ with an auxiliary variable $u$:
\begin{equation}\label{eq:augmented-proposal-general}
    \pi(x, u) \propto \exp(f(x))\,\mathcal{N}(x; 0, C)\, \mathcal{N}\left(u; x, \frac{\delta}{2} I\right),    
\end{equation}
where $\delta > 0$ is a step size, so that the marginal of $\pi(x, u)$ is $\pi(x)$. Auxiliary samplers then proceed by linearising $f$ around the current state $x$ of the Markov chain to obtain a Gaussian proposal distribution
\begin{equation}\label{eq:aux-general}
    \begin{split}
    q(y \mid x, u)
        &\propto \exp(\nabla f(x)^{\top}y)\,\mathcal{N}(y; 0, C)\, \mathcal{N}\left(u; y, \frac{\delta}{2} I\right)\\
        &= \mathcal{N}\left(y; \frac{2}{\delta} A \left(u + \frac{\delta}{2} \nabla f(x)\right), A\right),
    \end{split}
\end{equation} 
where $A = \frac{\delta}{2} (C + \frac{\delta}{2} I)^{-1} C = (C^{-1} + \frac{2}{\delta} I)^{-1}$. Sampling from $\pi(x, u)$ (and therefore from $\pi(x)$ by discarding the intermediate auxiliary steps) is then done via Hastings-within-Gibbs~\citep{muller1993}:
\begin{enumerate}
    \item Sample $u \mid x \sim \mathcal{N}(u; x, \frac{\delta}{2} I)$.
    \item Propose $y \sim q(\cdot \mid x, u)$ targeting $\pi(\cdot \mid u) \propto \pi(\cdot, u)$, and accept the move with the corresponding acceptance probability.
\end{enumerate}
A more efficient counterpart of this, targeting $\pi(x)$ directly, can be given by integrating the proposal distribution~\eqref{eq:aux-general} with respect to $\mathcal{N}(u; x, \frac{\delta}{2} I)$:
\begin{align}\label{eq:marginal-general}
    q(y \mid x) = \mathcal{N}\left(y; \frac{2}{\delta} A \left(x + \frac{\delta}{2} \nabla f(x)\right), \frac{2}{\delta} A^2 + A\right).
\end{align} 

This marginalised version skips the intermediate sampling step of the auxiliary variable, and is provably better -- both empirically and in terms of Peskun ordering~\citep{peskun1973optimum,tierney1998note,leisen2008extension} -- than its auxiliary version.
As a result, the marginalised version can use step sizes $\delta$ roughly twice as large~\citep[see Tables 1, 2, and 3 in][]{titsias2018} for the same acceptance rate, at virtually no additional computational complexity.

A crucial property of both these instances of the auxiliary sampler is that, for all $\delta > 0$, %
the matrices $A$ and $C$ share the same eigenspace~\citep[Section 3.3]{titsias2018}. This ensures that, after an initial spectral decomposition of $C$, calibrating the value of $\delta$ (to achieve a target acceptance rate) can be done at a negligible cost compared to the actual sampling process itself.
However, when $C$ depends on a parameter $\theta$, changing $\theta$ will not keep the eigenspace invariant. This means that when using either of these samplers within a Hastings-within-Gibbs routine targeting a joint model $\pi(x, \theta) \propto \exp(f(x))\,\mathcal{N}(x; 0, C_{\theta}) \, p(\theta)$, the spectral decomposition of $C_{\theta}$ has to be recomputed every time the value of $\theta$ changes. 
This is computationally prohibitive for large dimensional $x$, costing $\bigO(d_x^3)$ operations in general.
However, this can be mitigated thanks to the following observation \citep{titsias2018}: under a reparametrisation of $u$, which corresponds to considering the augmented target
\begin{equation}\label{eq:augmented-proposal-z-form}
    \pi(x, u) \propto \exp(f(x))\,\mathcal{N}(x; 0, C)\, \mathcal{N}\left(u; x + \frac{\delta}{2} \nabla f(x), \frac{\delta}{2} I\right),    
\end{equation}
rather than~\eqref{eq:augmented-proposal-general},
the proposal distribution $q(y \mid x, u)$ can be made independent of the current state of the chain $x$. This makes joint updates of $x$ and $\theta$ in parametric models possible, rather than using Gibbs steps to sample $x$ conditionally on $\theta$, and $\theta$ conditionally on $x$, thereby improving the mixing rate of the sampled Markov chain. This improvement, however, does not change the need for updating the spectral decomposition of $C_{\theta}$ and comes at the price of lower statistical efficiency than the non-reparametrised version for non-parametric models.

In the remainder of this article, despite its statistical efficiency, we do not consider the marginalised proposal~\eqref{eq:marginal-general}, and we consider the auxiliary sampler as defined in~\eqref{eq:augmented-proposal-general}.
This is because, as explained in the following section,~\eqref{eq:marginal-general} does not preserve the Markovian structure of our target models, making it computationally less efficient than the auxiliary samplers, which do. 
Similarly, we do not consider the empirically inferior reparametrised version~\eqref{eq:augmented-proposal-z-form} because its main advantage, namely that the resulting proposal, conditionally on the auxiliary variable, is independent of the current state of the chain~\citep[see][Section 3.3]{titsias2018} does not extend directly to non-Gaussian priors, which we consider in the rest of this article.
Nonetheless, for Gaussian prior dynamics, our methodology is directly compatible with~\eqref{eq:augmented-proposal-z-form} and can be used almost \emph{mutatis mutandis} within our framework.\footnote{This parametrisation was also leveraged extensively in the follow-up work to the present article~\citet{corenflos2024particlemala} treating of auxiliary pGibbs methods.}

\subsection{Auxiliary Kalman samplers}\label{subsec:auxiliary-lgssm} 
    The distribution $\pi(x) \propto \exp(f(x))\,\mathcal{N}(x; 0, C)$ covers latent Gaussian models in general, and in particular covers models with latent Gaussian dynamics\footnote{This was in fact explicitly used in \citet[][Ch. 15]{Chopin2020book}, where the authors successfully apply \citet{titsias2018} to a one-dimensional stochastic volatility model with latent Gaussian dynamics. The fact that the sampler corresponded to a linear Gaussian state-space model was, however, not noted by the authors.}:
\begin{equation}\label{eq:gaussian-FK}
\begin{split}
\pi(\ft{x}{0}{T})
    &\propto g(x_0, \ldots, x_T)\, \mathcal{N}(x_0; m_0, P_0)\, \prod_{t=1}^T \mathcal{N}(x_t; F_{t-1}\, x_{t-1} + b_{t-1}, Q_{t-1}).
\end{split}
\end{equation}
However, \emph{directly} treating these as latent Gaussian models with the methods of \citet{titsias2018} would incur a computational complexity of $\bigO(T^2 d_x^2)$, with an initial pre-processing step that scales as $\bigO(T^3 d_x^3)$, and a memory cost of $\bigO(T^2 d_x^2)$ corresponding to the size of the underlying covariance matrix $C$. 
Nonetheless, as what pointed out by a reviewer, it is possible to reduce the computational complexity to linear in $T$ by leveraging direct sparse Cholesky decompositions~\citep[see also][for such an approach]{dellaportas2023scalable}, which are largely available on sequential hardware~\citep[see, e.g.,][]{Yanqing2008cholmod}.
Such sparse methods are however not as readily available on parallel hardware such as GPUs or TPUs~\citep[see, nonetheless][for a mixed CPU-GPU implementation achieving some speed-up, typically $\times 3$]{RENNICH2016140}, and more plausible alternatives such as conjugate gradient methods~\citep{hestenes1952methods} are not direct and require additional tuning. 
Instead of leveraging general sparse linear algebra techniques, it is possible, in the case of a model like~\eqref{eq:gaussian-FK}, to directly formulate the auxiliary sampler as an LGSSM.
\begin{remark}\label{rem:future}
    Whithin the context of our work, this approach presents several advantages: (i) classical sequential~\citep{sarkka2023bayesian} and parallel filtering and smoothing algorithms~\citep{Sarkka2021temporal} can be applied almost \emph{mutatis mutandis}, see Section~\ref{subsec:lgssm-sampling}, (ii) the formulation makes it easy to then extend the method to non-linear dynamics using linearisation techniques developed in the signal processing literature as discussed in Section~\ref{subsec:auxiliary-ssm-sampler}, and (iii) the links with sequential Monte Carlo methods are more apparent, as we will see in Section~\ref{sec:pgibbs_samplers}.
\end{remark}

In order to formulate~\eqref{eq:aux-general} as a LGSSM, we emulate \citet{titsias2018} and consider the augmented target distribution 
\begin{equation}\label{eq:augmented-gaussian-fk}
    \pi(\ft{x}{0}{T}, u_{0:T}) \propto \pi(\ft{x}{0}{T}) \prod_{t=0}^T \mathcal{N}\left(u_t; x_{t}, \frac{\delta}{2} \Sigma_t\right),
\end{equation}
where $\delta > 0$ and, for all $t=0, \ldots, T$, $\Sigma_t$ is some positive definite matrix in $\mathbb{R}^{d_x \times d_x}$. Note that when $\Sigma_t = I$ is the identity matrix for all $t$, this recovers the proposal~\eqref{eq:augmented-proposal-general}.

Let us define $\gamma$ via $\exp(\gamma(x_0, x_1, \ldots, x_T)) := g(x_0, x_1, \ldots, x_T)$, and linearise it around the previously sampled trajectory $x_{0:T}$, $\gamma(z_{0:T}) \approx \gamma(x_{0:T}) + \langle v_{0:T}, z_{0:T} - x_{0:T}\rangle$, where $v_t = {\pdv{\gamma}{x_t}}(x_{0:T})$ for all $t$, and $\langle a_{0:T}, b_{0:T} \rangle$ denotes the sum of inner products $\sum_{t=0}^T \langle a_{t}, b_{t} \rangle$. 
Under these notations, we can define the auxiliary proposal
\begin{equation}\label{eq:gaussian-FK-proposal}
    \begin{split}
        q(z_{0:T} \mid u_{0:T}, x_{0:T}) \propto \mathcal{N}(z_0; m_0, P_0) 
            &\left\{\prod_{t=1}^T \mathcal{N}(z_t; F_{t-1}z_{t-1} + b_{t-1}, Q_{t-1})\right\} \\ 
            &\left\{\prod_{t=0}^T\mathcal{N}\left(u_{t} + \frac{\delta}{2} \Sigma_t v_{t}; z_{t}, \frac{\delta}{2} \Sigma_t\right)\right\},
    \end{split}
\end{equation}
which corresponds to the pathwise smoothing distribution of an LGSSM with unchanged dynamics compared to~\eqref{eq:gaussian-FK}, and observations given by $u_{t} + \frac{\delta}{2} \Sigma_t v_{t}$ for an observation model $\mathcal{N}\left(\cdot; z_{t}, \frac{\delta}{2} \Sigma_t\right)$, $t=0, 1, \ldots, T$. 
Sampling from this distribution, and evaluating its likelihood can be done using Kalman filtering and smoothing techniques in $\bigO(T)$ steps~\citep[see, e.g.][Ch. 6 and Ch. 12]{sarkka2023bayesian}, \citet[Section 3.2]{Doucet:2010}, \citet[Section 3.2]{Chopin2020book}, and Appendix~\ref{app:kalman} for more details.
In fact, this representation is key to reducing the memory requirements to linear in $T$ as well as the computational complexity from cubic to linear or even logarithmic in $T$ for parallel hardware. 
We come back to this last point in Section~\ref{subsec:lgssm-sampling}. 

To summarise, sampling from $\pi(x_{0:T}, u_{0:T})$ is then done via Hastings-within-Gibbs~\citep{muller1993}: (i) sample $u^k_{0:T} \mid x^k_{0:T} \sim \prod_{t=0}^T\mathcal{N}(u_t; x^k_t, \frac{\delta}{2} \Sigma_t)$, (ii) propose $x^*_{0:T} \sim q(\cdot \mid x^k_{0:T}, u^k_{0:T})$ targeting $\pi(\cdot \mid u^k_{0:T}) \propto \pi(\cdot, u^k_{0:T})$, and (iii) accept the move with the corresponding acceptance probability.
We insist that this proposal is statistically equivalent to the auxiliary method of \citet{titsias2018} for a choice of constant $\Sigma_t = I$, but exhibits better computational complexity than their implementation due to the LGSSM structure. 
Marginalising it over $u_{0:T}$, recovering~\eqref{eq:marginal-general}, however, would destroy the proposal Markovian structure, removing this advantage completely. 
Similarly, in general, a second order approximations of $\gamma$ would result in fully dependent observations, so that the proposal distribution would not correspond to a LGSSM anymore.

Nonetheless, when the potentials are separable, as is the case for state-space models, we can easily use second-order approximations. 
Indeed, when $g(x_{0:T}) = \prod_{t=0}^T g_t(x_t)$, or equivalently, when $\gamma(x_{0:T}) = \sum_{t=0}^T \gamma_t(x_t)$, we can write 
\begin{equation}\label{eq:linearised-log-potential-second}
    \gamma(z_{0:T}) \approx \gamma(x_{0:T}) + \langle v_{0:T}, z_{0:T} - x_{0:T}\rangle + \frac{1}{2} \sum_{t=0}^T (z_t - x_t)^{\top} \Lambda_t (z_t - x_t),
\end{equation} 
where $\Lambda_t$ is the Hessian matrix of $\gamma_t$ evaluated at $x_t$. By rearranging the terms, we can derive the resulting proposal distribution as
\begin{equation}\label{eq:second-order}
    \begin{split}
        q(z_{0:T} \mid u_{0:T}, x_{0:T}) &\propto \mathcal{N}(z_0; m_0, P_0) \left\{\prod_{t=1}^T \mathcal{N}(z_t ; F_{t-1}z_{t-1} + b_{t-1}, Q_{t-1})\right\}\left\{\prod_{t=0}^T\mathcal{N}\left(\omega_{t}; z_{t}, \Omega_t\right)\right\},
    \end{split}
\end{equation}
with $\Omega_t = \left(\frac{2}{\delta} \Sigma_t^{-1} - \Lambda_t\right)^{-1}$ and $\omega_t = \Omega_t\left(\frac{2}{\delta} \Sigma_t^{-1} u_t + v_t - \Lambda_t x_t\right)$.
This proposal is well defined as an LGSSM as soon as $\delta$ is small enough and will recover the exact auxiliary target when the original model $\pi(x_{0:T})$ is Gaussian.

Finally, when the dynamics are not Gaussian, it is often possible to transform the model at hand into an equivalent representation of $\pi(x_{0:T})$ with Gaussian dynamics by setting
\begin{equation}
    \begin{split}
        p_0(x_0)                &\leftarrow \mathcal{N}(x_0; m_0, P_0), \quad
        p_t(x_t \mid x_{t-1})   \leftarrow \mathcal{N}(x_t ; F_{t-1}x_{t-1} + b_{t-1}, Q_{t-1}), \\
        g(x_{0:T})              &\leftarrow g(x_{0:T}) \frac{p_0(x_0)}{\mathcal{N}(x_0; m_0, P_0)} \prod_{t=1}^T \frac{p_t(x_t \mid x_{t-1})}{\mathcal{N}(x_t ; F_{t-1}x_{t-1} + b_{t-1}, Q_{t-1})},
    \end{split}
\end{equation}
for a choice of $m_0$, $P_0$, $F_{t-1}$, $b_{t-1}$, and $Q_{t-1}$, enabling the use of the auxiliary sampler~\eqref{eq:gaussian-FK-proposal}.
While this is sometimes a natural thing to do~\citep[see, e.g.,][for an application swapping a reflected Brownian motion prior for a standard Brownian motion one]{Karppinen2022bridge}, it can also happen that there is no natural way to make such a Gaussian appear in the model. 
This justifies the need for introducing a new class of auxiliary samplers.

\begin{remark}
    As pointed out by a reviewer, when using a second-order approximation of the potential, the resulting proposal distribution is agnostic to the choice of the Gaussian prior. This is a direct consequence of the fact that the second-order approximation of the potential is a quadratic function, and that the resulting proposal distribution is a Gaussian distribution. As a consequence, while there is sometimes no natural choice for introducing a Gaussian prior in the model when using first-order linearisation, all such choices are equivalent when using a second-order approximation and will only affect the computational aspects of the algorithm. However, it is not plausible that the second-order approximation would result in a Markovian structure for the latent variables, making this approach highly inefficient in practice.
    Nonetheless, it may be possible to derive practical Hessian approximations that preserve Markovianity by construction, hopefully offering another approach to designing efficient auxiliary samplers for non-linear non-Gaussian models.
    We leave this as an open question for future research.
\end{remark}

\subsection{New auxiliary samplers for models with non-Gaussian dynamics}\label{subsec:auxiliary-ssm-sampler}
    In Section~\ref{subsec:auxiliary-lgssm}, we have made an explicit link between the auxiliary samplers of \citet{titsias2018} and Kalman filtering when the latent model has Gaussian dynamics. 
This linearity of the latent model corresponds to the assumption of linear Gaussian dynamics in the case of state-space models. 
This is a rather strong modelling assumption that is not easily verified, or enforced, in practice. 
In this section, we present an approach which uses local approximations of the dynamics model by conditional Gaussian transitions, akin to extended Kalman linearisation~\citep[see, e.g.,][Ch. 7]{sarkka2023bayesian}.

Let us assume that our target distribution is given by
\begin{equation}
    \label{eq:non-gaussian-target}
    \pi(\ft{x}{0}{T}) \propto p_0(x_0) \left\{\prod_{t=1}^T p_t(x_t \mid x_{t-1}) \right\} g(x_{0:T}),
\end{equation}
where the latent dynamics model $p_0(x_0) \left\{\prod_{t=1}^T p_t(x_t \mid x_{t-1}) \right\}$ is not necessarily Gaussian anymore.
Similarly as in Section~\ref{subsec:auxiliary-lgssm}, we can form the augmented target distribution
\begin{equation}
    \label{eq:augmented-ssm}
    \begin{split}
        \pi(\ft{x}{0}{T}, \ft{u}{0}{T})
        \coloneqq &p_0(x_0) \left\{\prod_{t=1}^T p_t(x_t \mid x_{t-1}) \right\} g(x_{0:T}) \,\left\{\prod_{t=0}^T\,\mathcal{N}(u_t; x_t, \frac{\delta}{2}\Sigma_t)\right\}\\
    \end{split}
\end{equation}
where $\delta > 0$, and for all $t$, $\Sigma_t$ is a positive definite matrix.

In order to form a proposal distribution $q(z_{0:T} \mid u_{0:T}, x_{0:T})$ for $\pi(\ft{x}{0}{T} \mid \ft{u}{0}{T})$, we can first linearise the potential function 
\begin{equation}
    \label{eq:linearisation}
    g(x_{0:T}) \approx \exp\left\{\gamma(x_{0:T}) + \sum_{t=0}^T \left(v_t^\top (z_t - x_t)\right)\right\}
\end{equation}
around $x_{0:T}$, for $v_t = \nabla_{x_t} \gamma(x_{0:T})$ as in Section~\ref{subsec:auxiliary-lgssm}, forming the intermediary (intractable in general) proposal distribution
\begin{equation}
    \label{eq:augmented-proposal}
    \begin{split}
        \tilde{q}(\ft{z}{0}{T}\mid \ft{u}{0}{T}, v_{0:T}, x_{0:T})
        \coloneqq & p_0(z_0) \left\{\prod_{t=1}^T p_t(z_t \mid z_{t-1}) \right\} g(x_{0:T}) \\
        &\left\{\prod_{t=0}^T\mathcal{N}\left(u_{t} + \frac{\delta}{2} \Sigma_t v_{t}; z_{t}, \frac{\delta}{2} \Sigma_t\right)\right\},
    \end{split}
\end{equation}
where, contrary to~\eqref{eq:gaussian-FK-proposal} we make the dependency on $v_{0:T}$ explicit despite the redundancy with $x_{0:T}$ at this specific stage.

This can then be further approximated by forming a linear Gaussian approximation to the dynamics model $p_0(z_0) \left\{\prod_{t=1}^T p_t(z_t \mid z_{t-1}) \right\}$, whereby we can approximate $p_0(z_0) \approx \mathcal{N}(z_0; m_0, P_0)$, via its first two moments, and $p_t(x_t \mid x_{t-1}) \approx \mathcal{N}(z_t; F_{t-1} z_{t-1} + b_{t-1}, Q_{t-1})$ for $t=1, \ldots, T$.

In principle, the latter approximation can, for example, be obtained by minimising the Kullback--Leibler~\citep[KL,][]{kullback1951kl} divergence between the true and the approximated transition model
\begin{equation}
    \label{eq:KL-divergence}
    \begin{split}
        \kl{\mathcal{N}(z_t; F_{t-1} z_{t-1} + b_{t-1}, Q_{t-1})}{p_t(z_t \mid z_{t-1})}
    \end{split}
\end{equation}
as a function of $F_{t-1}$, $b_{t-1}$, and $Q_{t-1}$. 
However, the optimal solution to this problem will in general depend on the value of $z_{t-1}$ and is therefore not a well-defined problem.
Instead, we can minimise the \emph{expected} KL divergence with respect to a reference random variable distributed as $\mathcal{N}(x_{t-1}, \Gamma_{t-1})$, centred on the current state $x_{t-1}$ and with a user-chosen covariance matrix $\Gamma_{t-1}$. This leads to the generalised statistical linear regression (GSLR) framework of~\citet{Tronarp2018iterative} which we review in Appendix~\ref{app:gslr}.
In practice, the solution to the KL minimisation problem~\eqref{eq:KL-divergence} recovers classical state-space model linearisation techniques~\citep[for a review of these, we refer to][]{sarkka2023bayesian} such as the extended Kalman filter, which we detail in Example~\ref{eq:gslr-example}, but also allows for more sophisticated approximations.

\begin{example}\label{eq:gslr-example}
    Suppose that the latent dynamics model has additive noise, that is, it is given by $X_t = f(X_{t-1}) + \epsilon_{t-1}$, where $f$ is a smooth function and $\epsilon_{t-1}$ is a centred Gaussian noise term with covariance $Q_{t-1}$. 
    Clearly, $p_t(z_t \mid z_{t-1})$ is then \emph{conditionally} Gaussian, with mean $f(z_{t-1})$ and covariance $Q_{t-1}$.
    For a given $z_{t-1}$, we then compute the KL divergence~\eqref{eq:KL-divergence} as
    \begin{equation}
        \label{eq:KL-divergence-example}
        \begin{split}
            \quad 
            &\kl{\mathcal{N}(z_t; F_{t-1} z_{t-1} + b_{t-1}, Q_{t-1})}{\mathcal{N}(z_t; f(z_{t-1}), Q_{t-1})}\\
            &= \mathrm{Cte} + \frac{1}{2}\left(F_{t-1} z_{t-1} + b_{t-1} - f(z_{t-1})\right)^{\top} Q_{t-1} \left(F_{t-1} z_{t-1} + b_{t-1} - f(z_{t-1})\right)
        \end{split}
    \end{equation}
    where $\mathrm{Cte}$ is a constant that does not depend on $F_{t-1}$, $b_{t-1}$, or $Q_{t-1}$.
    A first order linearisation of $f$ around $x_{t-1}$ of the right-hand side of~\eqref{eq:KL-divergence-example} then gives the approximation $F_{t-1} \approx \nabla f(x_{t-1})$ and $b_{t-1} \approx f(x_{t-1}) - F_{t-1} x_{t-1}$, independent of the choice of $\Gamma_{t-1}$.
\end{example}

These linear approximations, together with the known (or approximated) first two moments $m_0$ and $P_0$ of $p_0(x_0)$, can then be used to form a proposal distribution defined as an auxiliary LGSSM smoothing distribution with density
\begin{equation}\label{eq:auxiliary-LGSSM}
    \begin{split}
        q(z_{0:T} \mid u_{0:T}, v_{0:T}, x_{0:T}) 
            \propto \mathcal{N}(z_0; m_0, P_0) 
                &\left\{\prod_{t=0}^T\mathcal{N}\left(u_{t} + \frac{\delta}{2} \Sigma_t v_{t}; z_{t}, \frac{\delta}{2} \Sigma_t\right)\right\}\\ 
                &\left\{\prod_{t=1}^T \mathcal{N}(z_t ; F_{t-1}z_{t-1} + b_{t-1}, Q_{t-1})\right\}.
    \end{split}
\end{equation}
This proposal distribution is then included as part of a Metropolis--Rosenbluth--Teller--Hastings (MRTH) acceptance-rejection step.
The resulting sampler corresponds to Algorithm~\ref{alg:gslr-sampler}.
\begin{algorithm}[!htb]
    \SetAlgoLined
    \DontPrintSemicolon
    \caption{General Auxiliary Kalman sampler}\label{alg:gslr-sampler}
    \KwResult{An updated trajectory $z_{0:T}$}
    \Fn{\textsc{AuxKalmanSampler}$\big(x_{0:T})$}{
        \tcp{Generate the auxiliary observations}
        \lForSample{$t=0, 1, \ldots, T$}{
            $u_t \mid x_t \sim \mathcal{N}(\cdot; x_t, \frac{\delta}{2} \Sigma_t)$
        }
        \tcp{Form the proposal $q(\cdot \mid u_{0:T}, v_{0:T}, x_{0:T})$ in~\eqref{eq:auxiliary-LGSSM}}
        \For{$t=0\ldots, T$}{
            \uIf{$t>0$}{
                Form an approximation $\mathcal{N}(z_t; F_{t-1} z_{t-1} + b_{t-1}, Q_{t-1}) \approx p_t(z_t \mid z_{t-1})$ around $x_{t-1}$\;\label{step:forw-1}
            }
            Set $v_t = \nabla_{x_t} \gamma(x_{0:T})$\label{step:forw-2}
        }
        Sample $x^*_{0:T} \sim q(\cdot \mid u_{0:T}, v_{0:T}, x_{0:T})$ and compute $L^* = \frac{q(x^*_{0:T}, u_{0:T}, v_{0:T} \mid x_{0:T})}{q(u_{0:T}, v_{0:T} \mid x_{0:T})}$\;
        \tcp{MRTH step}
        Form the reversed proposal $q^*(x_{0:T} \mid u_{0:T}, v^{*}_{0:T}, x^*_{0:T})$ following steps~\ref{step:forw-1} and~\ref{step:forw-2} around $x^*_{0:T}$ and compute $L = \frac{q^*(x_{0:T}, v^{*}_{0:T}, u_{0:T} \mid x^*_{0:T})}{q^*(v^{*}_{0:T}, u_{0:T} \mid x^*_{0:T})}$\;
        With probability $\min\left(1, \frac{p(x^*_{0:T}, u_{0:T}) L}{p(x_{0:T}, u_{0:T}) L^*}\right)$, set $z_{0:T} = x^*_{0:T}$\label{step:mrth}\;
        Otherwise, set $z_{0:T} = x_{0:T}$\;
        \Return{$z_{0:T}$}
    }
\end{algorithm}
Evaluating the augmented density~\eqref{eq:augmented-gaussian-fk} appearing in the acceptance ratio of the MRTH algorithm, line~\ref{step:mrth}, is easily done. 
Therefore, to effectively implement the steps above we only need to understand how to sample from the smoothing distribution $x^*_{0:T} \sim q(\cdot \mid u_{0:T}, v_{0:T}, x_{0:T})$ of the LGSSM at hand, and compute the corresponding smoothing density $q(x^*_{0:T}, v_{0:T}, u_{0:T} \mid x_{0:T}) / q(v_{0:T}, u_{0:T} \mid x_{0:T})$. We come back to this point in Section~\ref{subsec:lgssm-sampling}.

We end this section by noting that, while we assumed that we had linearised the potential $g$ prior to finding an approximation to the dynamics, the two tasks can be tackled simultaneously. This is particularly useful when the potential is obtained as a product of observation models $h(y_t \mid x_t)$, as in the case of state-space models, for which we are able to compute approximations
\begin{equation}\label{eq:observation-approx}
    h(y_t \mid z_t) \approx \mathcal{N}(y_t; H_t z_t + c_t, R_t),
\end{equation}
around $x_t$ for all $t$. 
In this case, we can apply exactly the same linearisation procedure to the observation model as we did to the dynamics model, form the proposal distribution~\eqref{eq:auxiliary-LGSSM} by combining the two linear approximations into a proposal model
\begin{equation}\label{eq:two-lin-prop}
    \begin{split}
        q(z_{0:T} \mid u_{0:T}, y_{0:T}, x_{0:T}) 
            &\propto \mathcal{N}(z_0; m_0, P_0) 
                \left\{\prod_{t=1}^T \mathcal{N}(z_t ; F_{t-1}z_{t-1} + b_{t-1}, Q_{t-1})\right\}\\ 
                &\left\{\prod_{t=0}^T\mathcal{N}\left(u_{t}; z_{t}, \frac{\delta}{2} \Sigma_t\right)\right\}\left\{\prod_{t=0}^T \mathcal{N}(y_t ; H_t z_t + c_t, R_t)\right\},
    \end{split}        
\end{equation}
and then proceed almost identically to Algorithm~\ref{alg:gslr-sampler}.
Forming such approximations is described in more detail in Appendix~\ref{app:gslr}.

\subsection{Sampling and evaluating the posterior of LGSSMs}\label{subsec:lgssm-sampling}
    In the previous sections, we have described a new auxiliary-variable-based MCMC algorithm for Markovian models, which, after a choice of linearisation, amounts to sampling from a linear Gaussian state-space model $q(z_{0:T} \mid u_{0:T}, x_{0:T})$ depending on the current state of the chain $x_{0:T}$ and the auxiliary variables $u_{0:T}$ and then accepting the move with the corresponding acceptance probability within a MRTH step.
To use it within Algorithm~\ref{alg:aux-sampler-gen}, we therefore only need to understand how to sample from the proposal distribution $q(z_{0:T} \mid u_{0:T}, x_{0:T})$ and evaluate it.
Thankfully, the resulting distribution is the posterior distribution of an LGSSM, for which efficient sampling and evaluation methods exist~\citep[see, e.g.,][for a comprehensive treatment of the topic]{barfoot2017state,sarkka2023bayesian}.
In this section, we quickly review the classical forward-filtering backward-sampling algorithm for LGSSMs, which, when implemented on sequential hardware, has a time and complexity of $\bigO(T)$. 
We then explain how this can be improved to $\bigO(\log T)$ using either prefix-sum algorithms~\citep[see, e.g.,][]{blelloch1989scans} or divide-and-conquer strategies.
More details on the different methods, including implementation details, are provided in Appendix~\ref{app:kalman}.

\subsubsection{Forward-filtering backward-sampling for LGSSMs}\label{subsec:ffbs-lgssm}
The forward-filtering backward-sampling~\citep[FFBS,]{Carter1994OnGS,Fruhwirth1994data} algorithm is a classical method to sample from the posterior distribution of an LGSSM.
Given a state-space model
\begin{equation}\label{eq:ssm-again}
    p(x_{0:T}, y_{0:T}) = p_0(x_0) \prod_{t=1}^T p_t(x_t \mid x_{t-1}) h_t(y_t \mid x_t),    
\end{equation}
as in~\eqref{eq:ssm}, we can compute the filtering densities $p(x_t \mid y_{0:t})$ recursively as
\begin{equation}\label{eq:filtering-explicit}
    \begin{split}
        p(x_t \mid y_{0:t}) &\propto h_t(y_t \mid x_t) \int p_t(x_t \mid x_{t-1}) p(x_{t-1} \mid y_{0:t-1}) \dd x_{t-1}, \quad t=1, \ldots, T,\\
        p(x_0 \mid y_0) &\propto h_0(y_0 \mid x_0) p_0(x_0).
    \end{split}
\end{equation}
When the initial distribution, transition and observation models are linear Gaussian, the quantities in~\eqref{eq:filtering-explicit} are Gaussian too and can be computed in closed form.
Moreover, this recursion, as a by-product, also computes the marginal likelihood of the observations $p(y_{0:T})$ via 
\begin{equation}\label{eq:marginal-likelihood}
    \begin{split}
        p(y_{0:T}) &= p(y_0) \prod_{t=1}^T p(y_t \mid y_{0:t-1}) = p(y_0) \prod_{t=1}^T \int h_t(y_t \mid x_t) p(x_t \mid y_{0:t-1}) \dd x_t
    \end{split}
\end{equation}
where each term $p(y_t \mid y_{0:t-1})$ (resp. $p(y_0)$) is the normalisation constant of the filtering density $p(x_t \mid y_{0:t})$ with respect to $p(x_t \mid y_{0:t-1})$ (resp. $p(x_0 \mid y_0)$ with respect to $p_0(x_0)$).

Once all the filtering densities have been computed, the backward sampling step consists in sampling from the conditional distribution $p(x_{0:T} \mid y_{0:T})$ recursively as
\begin{equation}\label{eq:backward-sampling}
    \begin{split}
        x_T &\sim p(x_T \mid y_{0:T}),\\
        x_t &\sim p(x_t \mid x_{t+1}, y_{0:T}), \quad t=T-1, \ldots, 0,
    \end{split}
\end{equation}
noting that
\begin{equation}\label{eq:backward-recursion}
    \begin{split}
        p(x_t \mid x_{t+1}, y_{0:T}) 
            &= p(x_t \mid x_{t+1}, y_{0:t}) \propto p(x_{t+1} \mid x_t) p(x_t \mid y_{0:t}),
    \end{split}
\end{equation}
which, under the same hypothesis as above, is Gaussian and can be computed in closed form.
Given that 
\begin{equation}\label{eq:auxiliary-LGSSM-repeat}
    \begin{split}
        q(z_{0:T} \mid u_{0:T}, v_{0:T}, x_{0:T}) 
            \propto \mathcal{N}(z_0; m_0, P_0) 
                &\left\{\prod_{t=0}^T\mathcal{N}\left(u_{t} + \frac{\delta}{2} \Sigma_t v_{t}; z_{t}, \frac{\delta}{2} \Sigma_t\right)\right\}\\ 
                &\left\{\prod_{t=1}^T \mathcal{N}(z_t ; F_{t-1}z_{t-1} + b_{t-1}, Q_{t-1})\right\},
    \end{split}
\end{equation}
introduced in~\eqref{eq:auxiliary-LGSSM}, is the posterior distribution of an LGSSM with observations $y_t = u_{t} + \frac{\delta}{2} \Sigma_t v_{t}$, we can therefore sample from it using the decomposition above, and evaluate the marginal likelihood $q(u_{0:T}, v_{0:T} \mid x_{0:T})$ using~\eqref{eq:marginal-likelihood}.
This offers a solution to also compute $q(x^*_{0:T} \mid u_{0:T}, v_{0:T}, x_{0:T}) = q(x^{*}_{0:T},u_{0:T}, v_{0:T} \mid x_{0:T})/q(u_{0:T}, v_{0:T} \mid x_{0:T})$ appearing in Algorithm~\ref{alg:gslr-sampler}.
The same applies to all other instances of the method we presented above.

\subsubsection{Parallel-in-time sampling of LGSSMs}\label{subsec:parallel-lgssm}
Due to its recursive structure, the method described in Section~\ref{subsec:ffbs-lgssm} has a time complexity of $\bigO(T)$, which can be prohibitive for large values of $T$.
While this complexity is optimal on sequential hardware, where the computation of the filtering densities has to be done sequentially, it can be improved to $\bigO(\log T)$ on parallel hardware, such as GPUs or TPUs.
In this section, we describe two methods to achieve this: a prefix-sum approach and a divide-and-conquer approach, which can be used to sample from the proposal distribution $q(z_{0:T} \mid u_{0:T}, x_{0:T})$ in Algorithm~\ref{alg:aux-sampler-gen} in $\bigO(\log T)$ time.
Both algorithms rely on first computing the filtering densities $p(x_t \mid y_{0:t})$ in $\bigO(\log T)$, which, when the model is linear Gaussian, can be done using~\citet{Sarkka2021temporal}.

In order to simplify the description of these methods, we assume that the backward distribution $p(x_t \mid x_{t+1}, y_{0:T}) = \mathcal{N}(x_t; E_t x_{t+1} + f_t, L_t)$ has a linear Gaussian form, which is the case for the LGSSM model~\eqref{eq:ssm-again}, and that the coefficients $E_t$, $f_t$, and $L_t$ have been precomputed and are available for sampling in $\bigO(1)$ time. 
Further details on the prefix-sum and divide-and-conquer approaches, including a review of the filtering densities computation of~\citet{Sarkka2021temporal}, and the formulation of the backward distribution parameters $E_t$, $f_t$, and $L_t$, are provided in Appendix~\ref{app:kalman}.

\paragraph{Prefix-sum approach to parallel FFBS.}
Prefix-sum algorithms are a class of parallel algorithms that compute the cumulative ``sum'' of an array of elements in logarithmic time under sufficient parallelism.
Formally, given a sequence of elements $x_0, x_1, \ldots, x_{T-1}$ and an operator $\oplus$, the prefix sum of the sequence is the sequence $y_0, y_1, \ldots, y_{T-1}$ such that $y_t = \bigoplus_{i=0}^t x_i$.
Provided that the operator $\oplus$ is associative, the prefix sum can be computed in $\bigO(\log T)$ time using $\bigO(T)$ processors.

Suppose now that we have access to $X_{t+1}$. 
Then, to sample from $X_t \mid \{X_{t+1}, y_{0:T}\}$, we can sample from the Gaussian noise $X_t \sim \mathcal{N}(0, L_t)$ and compute $X_t \gets E_t X_{t+1} + f_t + X_t$.
However, this operation, as seen as an operation on $X_t, X_{t+1}$ only, is not associative, and we cannot apply the prefix sum directly.
On the other hand, the same method can be seen as an operation on the quadruplet $(E_t, f_t, L_t, X_{t+1})$ via the operator $\circ$ defined as
\begin{equation}\label{eq:sampling-op}
    \begin{split}
        &(E_{t-1}, f_{t-1}, L_{t-1}, X_{t}) \circ (E_t, f_t, L_t, X_{t+1})\\
        &= (E_{t-1} E_{t}, \,E_{t-1} f_{t-1} + f_t, \,E_{t-1} L_{t} E_{t-1}^{\top} + L_{t-1}, \,X_t + E_{t} X_{t+1} + f_t).
    \end{split}
\end{equation}
Because it collects the composition of the transition matrices $E_t$, the offset vectors $f_t$, and the transition covariance $L_t$ in a single operation, the operator $\circ$ is associative, and we can apply the prefix sum to the sequence $(E_t, f_t, L_t, X_{t+1})$ to sample from $X_{0:T} \mid y_{0:T}$ in $\bigO(\log T)$ time.
A formal statement of this result, as well as a more efficient implementation of the algorithm, relying on propagating only $E_t$ and $X_t$, are provided in Appendix~\ref{subsubsec:prefix-sum-sampling}.

\paragraph{Divide-and-conquer approach to parallel FFBS.}
Another approach to parallelise the FFBS algorithm is to use a divide-and-conquer strategy, which consists in computing bridging distributions between time steps in a hierarchical manner.
Formally, we can recursively sample the distribution $p(x_{0:T} \mid y_{0:T})$ by 
\begin{enumerate}
    \item first sampling from $p(x_T \mid y_{0:T})$ and $p(x_0 \mid x_T, y_{0:T})$, 
    \item then $p(x_{\lfloor T/2 \rfloor} \mid x_0, x_T, y_{0:T})$, 
    \item then $p(x_{\lfloor T/4 \rfloor} \mid x_0, x_{\lfloor T/2 \rfloor}, y_{0:T})$ and $p(x_{\lfloor T/4 \rfloor} \mid x_{\lfloor T/2 \rfloor}, x_{T}, y_{0:T})$,
    \item and so on,
\end{enumerate}
until we exhaust all the time steps.

Because, at each level $k$ in the recursion (apart from the first one), we sample from $2^k$ distributions in parallel, the total number of non-parallel steps is $\bigO(\log T)$, and the divide-and-conquer approach therefore has a time complexity of $\bigO(\log T)$ on parallel hardware.
Of course, if implemented naively, this approach would require computing the bridging distributions $p(x_s \mid x_l, x_u, y_{0:T})$ for all $l < s < u$ at each level of the recursion, which would be computationally prohibitive.
Instead, it is possible to compute these via an initial reversed recursion, 
whereby $p(x_t \mid x_{t+1}, y_{0:T})$ is initialised as $\mathcal{N}(x_t; E_t x_{t+1} + f_t, L_t)$, which then allows to compute $p(x_t \mid x_{t+2^0}, x_{t-2^0}, y_{0:T})$, then $p(x_t \mid x_{t+2^1}, x_{t-2^1}, y_{0:T})$, and so on, for all times $t$ that appear at level $\lfloor \log T\rfloor - k$ of the recursion. 
This method, as well as a description of how to efficiently compute the bridging distributions $p(x_s \mid x_l, x_u, y_{0:T})$ arising in the recursion above, are provided in Appendix~\ref{subsubsec:divide-and-conquer}.

\section{Auxiliary particle Gibbs samplers}
\label{sec:pgibbs_samplers}
We have so far been concerned with designing global MCMC proposals that leveraged local LGSSM approximations of the target distribution. These proposals, while expected to work particularly well when the prior is almost Gaussian and the potential relatively non-informative, nonetheless constitute ``global acceptance methods'', and, as such, present at least two limitations:
\begin{enumerate}
    \item Because they accept or reject a full trajectory at once, a single unfortunate proposed time-step can lead to a rejection of the whole trajectory, even if the rest of the trajectory is correct. 
    In other terms, the method is not robust to heterogenously informative observations.
    \item Even though the method is efficient when the prior is informative, it will still collapse when the number of time steps goes to infinity.
\end{enumerate}
\begin{remark}\label{rem:separable}
    While intuitive, the second limitation can be formalised in the case when the model is fully separable, that is, when $\pi(x_{0:T}) = \prod_{t=0}^T p(x_t) g(x_t)$ for some Gaussian prior $p$ and likelihood function $g$. In this case, the acceptance probability of the Kalman-based MCMC kernel will be the product of the acceptance probabilities of the individual time steps and therefore, the acceptance probability of the full trajectory will go to zero as $T$ goes to infinity unless $\delta$ decreases to zero as well. 
    Understanding the exact rate at which $\delta$ should in general decrease to zero is a difficult problem, one that is not addressed in this article.
    Nonetheless, when $p \equiv 1$ is an improper prior, it can be seen that~\citet{titsias2018} recovers the MALA algorithm~\citet{besag1994comments} and therefore, one can expect that the results of~\citet{roberts1996mala} correspond to a worst-case scenario for the method, i.e., $\delta$ should then decrease to zero at a rate of $\bigO(1/T^{1/3})$ at most.
\end{remark}

For the above reasons, in this section, we turn ourselves to the successful class of particle MCMC algorithms, in particular particle Gibbs algorithms~\citep{Andrieu2010particle}, which have been shown to be very robust to increasingly many time steps $T$~\citep{lee2020coupled,karjalainen2024mixingtime}, and show how the same auxiliary observation trick can be leveraged to design efficient particle MCMC samplers for Feynman--Kac models.
Intuitively, these will be more robust to the highly informative observations as they essentially form \emph{local} MCMC moves~\citep[Section 2.2]{finke2021csmc}.
For instance, for the degenerate case of fully separable models, factorising in time as in Remark~\ref{rem:separable}, ``trajectories'' will be accepted independently of each other, and the algorithm will be less affected by the presence of a single bad time-step.

In the remainder of this section, we first quickly recall the basic particle Gibbs algorithm, and a recent high-dimensional extension due to~\citet{finke2021csmc}.
We then show how~\citet{finke2021csmc} can be understood as an instance of a more general method, relying on a similar auxiliary observation trick as the one used in Section~\ref{sec:auxiliary_samplers}.
This novel perspective then allows us to introduce novel auxiliary particle Gibbs methods, extending~\citet{finke2021csmc} to incorporate prior and gradient information in the form of ``locally optimal proposals''~\citep[also called guided proposals in][Ch. 16]{Chopin2020book}, and discuss when these can be parallelised efficiently on GPUs along the time dimension, similarly to the methods of Section~\ref{subsec:lgssm-sampling}.

\subsection{SMC and particle Gibbs algorithms}
\label{subsec:pgibbs-refresh}
    Particle Gibbs algorithms are Gibbs-like MCMC samplers that target the posterior distribution of Feynman--Kac models~\citep{Andrieu2010particle,lindsten2012ancestor,lindsten2014particle}. 
In their simplest form, they consist in running a particle filter algorithm conditioned on the current state of the MCMC chain ``surviving'' the resampling step. This kernel, called conditional SMC (cSMC), can be proven to be ergodic for the pathwise smoothing distribution under the weak hypothesis that the potential functions are bounded above~\citep[see][and references within]{lee2020coupled}. 
In Algorithm~\ref{alg:csmc}, we reproduce the original version~\citep{Andrieu2010particle} of a cSMC kernel with $N \geq 2$ particles, targeting the posterior distribution of a generic Feynman--Kac model $\pi(x_{0:T}) \propto g_0(x_0) \, p_0(x_0) \left\{\prod_{t=1}^T g_t(x_t, x_{t-1}) \, p_t(x_t \mid x_{t-1}) \right\}$.
\begin{algorithm}[!htb]
\SetAlgoLined
\DontPrintSemicolon
\caption{Conditional SMC}\label{alg:csmc}
\KwResult{An updated trajectory $z_{0:T}$}
\Fn{\textsc{cSMC}$\big(x_{0:T}$, $N\big)$}{
    \tcp{Forward propagation}
    \For{$n=1, 2, \ldots, N-1$}{
        Sample $X^n_0 \sim p_0$ and set $w^n_0 = g_0(X^n_0)$
    }
    Set $X_0^N = x_0$, $w^N_0 = g_0(x_0)$\;
    \For{$t=1,\ldots, T$}{
        \For{$n=1, \ldots, N-1$}{
            Sample $A^n_t$ with $\mathbb{P}(A^n_t = k) \propto w^k_{t-1}$\;
            Sample $X^n_t \sim p_t(\cdot \mid X^{A^n_t}_{t-1})$ and set $w_t^n = g_t(X^n_t, X^{A^n_t}_{t-1})$\label{line:smc-sample}
        }
        Set $X_t^N = x_t$, $w^N_t = g_t(x_t \mid x_{t-1})$\;
    }
    \tcp{Genealogy selection}\label{line:genealogy}
    Sample $B_T$ with $\mathbb{P}(B^n_T = k) \propto w^k_{T}$ and set $z_T = X^{B_T}_T$\;
    \For{$t=T-1,\ldots, 0$}{
        Set $B_t = A_{t+1}^{B_{t+1}}$, $z_t = X^{B_t}_t$
    }
    \Return{$z_{0:T}$}
}
\end{algorithm}

Other versions of this algorithm exist, in particular, when it is possible to evaluate the density $p_t(x_t \mid x_{t-1})$ as a function of $x_t$ and $x_{t-1}$, we can modify the representation of the Feynman--Kac model as
\begin{equation}
    \pi(x_{0:T}) \propto \tilde{g}_0(x_0) \, \tilde{p}_0(x_0) \left\{\prod_{t=1}^T \tilde{g}_t(x_t, x_{t-1}) \, \tilde{p}_t(x_t \mid x_{t-1}) \right\}
\end{equation}
provided that the identity $\tilde{g}_t = \frac{g_t \cdot p_t}{\tilde{p}_t}$ holds for all $t=0, 1, \ldots, T$, in which case $p_t$ and $g_t$ can be replaced by $\tilde{p}_t$ and $\tilde{g}_t$ in Algorithm~\ref{alg:csmc} while keeping the same posterior target $\pi(x_{0:T})$ invariant.
This key property will be used extensively in the remainder of this section.

Additionally, when the density $p_t$ can be evaluated, we can also rejuvenate the selection of the genealogy (step~\ref{line:genealogy} in Algorithm~\ref{alg:csmc}), allowing for lower degeneracy in the early time steps. 
The most notable two such methods are the backward and ancestor sampling methods~\citep[respectively]{whiteley2010discussion,lindsten2014particle}. 
The former~\citep{whiteley2010discussion}, in particular, has been the subject of much interest in the literature, and has recently been shown to improve the mixing of the algorithm from $\bigO(T)$ to $\bigO(\log T)$ in the number of time steps~\citep{karjalainen2024mixingtime}.
In other terms, when implemented with the backward sampling method, the resulting Markov chain will require $\bigO(\log T)$ iterations to achieve stationarity, while the `naive' version of Algorithm~\ref{alg:csmc} will require $\bigO(T)$ iterations.
The full implementation is given in Algorithm~\ref{alg:local-csmc-backward} in Appendix~\ref{subsec:backward-sampling}.
Another method, useful in our context, is that of \citet[Section 3]{corenflos2022sequentialized}, which implements a parallel-in-time conditional SMC, particularly amenable to when the proposal/dynamics model is separable, that is, when $p_t(x_t \mid x_{t-1}) = p_t(x_t)$ does not depend on $x_{t-1}$ as is the case for some samplers in this article: for example the sampler presented next in Algorithm~\ref{alg:local-csmc}, see Section~\ref{subsec:smc-samplers} for more details.

While widely used in practice, particle Gibbs algorithms suffer from degeneracy inherent to importance sampling methods when the dimension of the latent space increases.
In order to counteract this issue, several methods have been proposed, such as using spatial blocking~\citep{Singh2017blocking} or divide-and-conquer strategies~\citep{cruscino2022highdim}, but these are not always applicable as they require a specific structure in the model and can be complicated to implement and tune.
Recently, \cite{finke2021csmc} proposed a localised cSMC algorithm, recognising that the degeneracy of the particle filter came from the fact that the proposals used therein (line~\ref{line:smc-sample} of Algorithm~\ref{alg:csmc}) did not depend on the current state of the Markov chain $x_{0:T}$; a property that can be understood as it generalising the Metropolis--Hastings algorithm for independent proposals.
From this observation, they proposed to modify the cSMC algorithm to emulate the random walk Metropolis--Hastings algorithm, by using a proposal that depends on the current state of the chain.
We summarise this approach in Algorithm~\ref{alg:local-csmc} (RW-cSMC).

\begin{algorithm}[!htb]
    \SetAlgoLined
    \DontPrintSemicolon
    \caption{Random-walk Conditional SMC}\label{alg:local-csmc}
    \KwResult{An updated trajectory $x_{0:T}$}
    \Fn{\textsc{RW-cSMC}$\big(x_{0:T}$, $N$, $\lambda_{0:T} \in \bbR^{T+1}\big)$}{
    \tcp{Forward propagation}
    \For{$t=0, \ldots, T$}{
    \For{$n=1, \ldots, N-1$}{
        Sample $U_t \sim \mathcal{N}(x_t, \frac{\delta_t}{2}I)$ then $X^n_t \sim \mathcal{N}(U_t, \frac{\delta_t}{2}I)$\; \label{line:aux-step}
    }
    Set  $X_t^{N+1} = x_t$
    }
    \For{$n=1, \ldots, N$}{Set $w_{0}^n = g_0(X_0^n) p_0(X^n_0)$}
    \For{$t=1,\ldots, T$}{
    \For{$n=1, \ldots, N-1$}{
    Sample $A^n_t$ with $\mathbb{P}(A^n_t = k) \propto w^k_{t-1}$\;
    Set $w_t^n = g_t(X^n_t, X^{A^n_t}_{t-1}) p_t(X^n_t\mid  X^{A^n_t}_{t-1})$\;
    }
    Set $X_t^N = x_t$, $w^N_t =  g_t(x_t, x_{t-1}) p_t(x_{t} \mid x_{t-1})$\;
    }
    \tcp{The genealogy selection is left unchanged compared to Algorithm~\ref{alg:csmc}}
    }
\end{algorithm}

This algorithm takes its name from the fact that it generalises the Gaussian random-walk Metropolis--Hastings (RWMH) algorithm to more than a single time step.
Indeed, given the current state of the chain $x_{t}$ at time $t$, the proposed particles $X^{1:N}_t$ are all marginally distributed as $\mathcal{N}(x_t, \delta_t I)$, and the acceptance probability of the proposal is given by the ratio of the  $g_t$ of the current and proposed states, which is symmetric in the current and proposed states.
Importantly, RW-cSMC exhibits a similar asymptotic scaling as RWMH in terms of the dimension of the state-space, and a similar scaling as cSMC in terms of the time dimension, making it a good candidate for high-dimensional state-space models.
See~\citet{finke2021csmc} for quantitative details and different instances of the algorithm.
In the following section, we offer another interpretation of Algorithm~\ref{alg:local-csmc} in terms of an auxiliary variable sampler, better suited to extensions.

\subsection{Particle Gibbs for Feynman--Kac models with auxiliary observations}
    \label{subsec:smc-samplers}
    For the class of Feynman--Kac models~\eqref{eq:feynman-kac}, we can emulate the construction of Section~\ref{sec:auxiliary_samplers} to form the following auxiliary target
\begin{equation}\label{eq:auxiliary-feynman-kac}
    \pi(\ft{x}{0}{T}, \ft{u}{0}{T})
        \propto g_0(x_0) \, p_0(x_0) \left\{\prod_{t=1}^T g_t(x_t, x_{t-1}) \, p_t(x_t \mid x_{t-1}) \right\}\left\{\prod_{t=0}^T \mathcal{N}\left(u_t; x_t, \frac{\delta_t}{2} \Sigma_t\right)\right\},
\end{equation}
corresponding to a model with an augmented potential function $g_t(x_t, x_{t-1}) \, \mathcal{N}\left(u_t; x_t, \frac{\delta_t}{2} \Sigma_t\right)$ at each time step $t$.
In order to sample from $\pi(\ft{x}{0}{T}, \ft{u}{0}{T})$, it is, therefore, enough to implement an abstract algorithm given by Algorithm~\ref{alg:aux-pgibbs}.

\begin{algorithm}[!htb]
\SetAlgoLined
\DontPrintSemicolon
\caption{Auxiliary cSMC}\label{alg:aux-pgibbs}
\KwResult{An updated trajectory $x^{k+1}_{0:T}$}
\Fn{\textsc{aux-cSMC}$\big(x^k_{0:T}, u^k_{0:T}\big)$}{
    Sample $u^{k+1}_{0:T}  \sim \prod_{t=0}^T\mathcal{N}(u_{t}; x^k_t, \frac{\delta_t}{2} \Sigma_t)$\;\label{step:pgibbs-1}
    Sample $x^{k+1}_{0:T} \sim K(\cdot \mid x^k_{0:T})$ \tcp{from a $\pi(\ft{x}{0}{T} \mid \ft{u^{k+1}}{0}{T})$-invariant \textsc{cSMC} kernel}\label{step:pgibbs-2}
    \Return{$x^{k+1}_{0:T}, u^{k+1}_{0:T}$}
}
\end{algorithm}

Clearly, in Algorithm~\ref{alg:aux-pgibbs}, if $(x^k_{0:T}, u^k_{0:T})$ are distributed according to $\pi$, then $(u^{k+1}_{0:T}, x^k_{0:T})$ are too after line~\ref{step:pgibbs-1}, so that $x^k_{0:T}$ is distributed according to $\pi(\cdot \mid u^{k+1}_{0:T})$, and therefore $(x^{k+1}_{0:T}, u^{k+1}_{0:T})$ are still distributed according to $\pi$ after line~\ref{step:pgibbs-2}. Otherwise said, this algorithm can be seen as a ``true'' particle Gibbs algorithm~\citep{Andrieu2010particle} for the choice of an improper prior $\pi(u_{0:T}) \equiv 1$ for the auxiliary variables.

At first sight, this may seem like a very bad idea, and it appears like we have made the problem more difficult than it was originally, and this is probably the reason why (to the best of our knowledge) this has not been \emph{explicitly} proposed before. 
Indeed, instead of considering the potential function $g_t(x_t, x_{t-1})$, we are now considering the potential function $g_t(x_t, x_{t-1}) \, \mathcal{N}\left(u_t; x_t, \frac{\delta_t}{2} \Sigma_t\right)$ at each time step $t$. This new potential function becomes very informative as $\delta_t$ gets smaller, which is known to induce high variance weights in particle filtering and smoothing algorithms~\citep[see, e.g.][Section 10.3.1]{Chopin2020book} akin to increasing the dimension.
However, rather than seeing $\mathcal{N}\left(u_t; x_t, \frac{\delta_t}{2} \Sigma_t\right)$ as describing an auxiliary observation, we can leverage the symmetry of Gaussian distributions to look at it as the generative model $\mathcal{N}\left(x_t; u_t, \frac{\delta_t}{2} \Sigma_t\right)$ instead. 

Namely, we can swap the roles of $p_t(x_t \mid x_{t-1})$ and $\mathcal{N}\left(u_t; x_t, \frac{\delta_t}{2} \Sigma_t\right)$ in the auxiliary Feynman--Kac model~\eqref{eq:auxiliary-feynman-kac} to obtain the following modified model
\begin{equation}\label{eq:modified-auxiliary-feynman-kac}
    \pi(\ft{x}{0}{T} \mid \ft{u}{0}{T})
        \propto \tilde{p}_0(x_0 \mid u_0)\, \left\{\prod_{t=1}^T \tilde{p}_t(x_t \mid x_{t-1}, u_t)\right\}\tilde{g}_0(x_0) \left\{\prod_{t=1}^T \tilde{g}_t(x_t, x_{t-1}) \right\},
\end{equation}
for the modified dynamics $\tilde{p}_t(x_t \mid x_{t-1}, u_t) = \mathcal{N}\left(x_t; u_t, \frac{\delta_t}{2} \Sigma_t\right)$ and potential functions $\tilde{g}_t = g_t \cdot p_t$, $t=0, 1, \ldots, T$.
\begin{remark}\label{rem:modified-auxiliary-feynman-kac}
    This procedure amounts to moving the auxiliary likelihood $\mathcal{N}(u_t; x_t, \frac{\delta_t}{2} \Sigma_t)$ from the potential function to the dynamics, and the ``true'' dynamics $p_t(x_t \mid x_{t-1})$ to the potential function, which is a common principle we leverage in all the methods we propose in this section.
\end{remark}

This change of perspective immediately makes the problem much simpler, as we are now given a model with an informative and separable prior for which we can implement Step~\ref{step:pgibbs-2} of Algorithm~\ref{alg:aux-pgibbs} via Algorithm~\ref{alg:csmc}. 
Moreover, because the auxiliary prior model is separable across time, the method of \citet{corenflos2022sequentialized} applies directly\footnote{While in \citet{corenflos2022sequentialized} it was derived for likelihood terms $g_t(x_t)$ rather than $g_t(x_t, x_{t-1})$ this was a notational simplification, and all the results derived within in fact hold for bivariate potentials.}, and a parallel-in-time particle Gibbs can be implemented to reduce the computational complexity to $\bigO(\log T)$ on parallel hardware, the construction of which we describe in Appendix~\ref{app:pGibbs}.
We also note that, contrarily to \citet[][Section 5.2]{corenflos2022sequentialized}, in this specific case, doing so would not necessarily come at a loss of statistical efficiency compared to sequential conditional SMC counterparts. This is due to the fact that the sequential algorithms would also rely on sampling from the same independent proposals. 

In hindsight, it is easy to see that, when $\Sigma_t = I$, $t=0, \ldots, T$, this method is \emph{exactly} the same one as the one proposed in \citet[Algorithm 3 and extensions]{finke2021csmc} who instead phrase it as a form of conditional SMC with exchangeable proposals. 
Informally, rather than independently proposing the particles $x^{1:N}_t$ from $p_t(x_t \mid x_{t-1})$, they use a correlated proposal $\tilde{p}_t(x^1_t, \ldots, x^N_t)$ which induces an exchangeable dependency across particles, that is, $\tilde{p}_t(x^1_t, \ldots, x^N_t) = \tilde{p}_t(x^{\sigma(1)}_t, \ldots, x^{\sigma(N)}_t)$ for any permutation $\sigma$. 
As done in \citet{finke2021csmc}, and first introduced in the context of classical MCMC in \citet{tjelmeland2004using}, in the case of Gaussian variables, taking a conditional sample $p(\ldots, x^{k-1}_t, x^{k+1}_t, \ldots\mid x^k_t)$ can for instance be achieved by first sampling a ``centering'' variable $u_t \sim \mathcal{N}(x^k_t, \frac{\delta_t}{2}I)$ and then the remainder of the variables from $\prod_{i\neq k} \mathcal{N}(x_t^i; u_t, \frac{\delta_t}{2}I)$. 
This directly corresponds to the proposal and weighting mechanism of Algorithm~\ref{alg:aux-pgibbs} for the modified Feynman--Kac model~\eqref{eq:modified-auxiliary-feynman-kac} and justifies the following proposition.
\begin{proposition}\label{prop:finke-equivalence}
    The method of \citet{finke2021csmc}, given in Algorithm~\ref{alg:local-csmc}, implements Algorithm~\ref{alg:aux-pgibbs} with $u^{k+1}_t \sim  \mathcal{N}\left(u_t; x_t^{k}, \frac{\delta_t}{2} I\right)$ and proposal distributions $\tilde{p}(x_{t} \mid x_{t-1}) \sim \mathcal{N}\left(\cdot; u^{k+1}_t, \frac{\delta_t}{2} I\right)$ for different choices of kernels $K$: embedded HMM~\citep{neal2003embedded}, conditional SMC~\citep{Andrieu2010particle}, conditional SMC with forced move~\citep{Chopin2015particle}, and conditional SMC with backward sampling~\citep[see also Algorithm~\ref{alg:csmc}]{whiteley2010discussion}.
\end{proposition}

In other terms, the results of~\citet{finke2021csmc} apply too, and for a given choice of a standard conditional SMC -- with and without backward sampling -- Algorithm~\ref{alg:aux-pgibbs} for the proposals $\tilde{p}(x_{t} \mid x_{t-1}) = \mathcal{N}\left(x_t; u^{k+1}_t, \frac{\delta_t}{2} I\right)$ avoids the curse of dimensionality.
Formally, under a scaling $\delta = \bigO(1/d_x)$, it is stable for increasingly large $d_x$ (as well as $T$)~\citep[for details and assumptions, see Proposition 3.4 in][]{finke2021csmc}.
This new perspective on Algorithm~\ref{alg:local-csmc} is rich in consequences: the entirety of the literature on particle Gibbs can be applied to step~\ref{step:pgibbs-2} of Algorithm~\ref{alg:aux-pgibbs}, and we can expect that the curse of dimensionality can be controlled in this case too, provided that the auxiliary variables are used to design the proposal.

\subsection{Adapted proposals in particle Gibbs with auxiliary observations}
    \label{subsec:guided-smc-sampler}
    In the previous section, we have described an algorithm that recovers \citet[Algorithm 3 and extensions]{finke2021csmc}. However, explicitly introducing the auxiliary variable allows us to decouple the state of the Markov chain and the generative model so that we can incorporate additional statistical information in the auxiliary particle Gibbs sampler beside simple locality. Formally, we can implement ``locally-adapted'' particle filters for $\pi(x_{0:T} \mid u_{0:T})$ that improve the statistical properties of \citet{finke2021csmc}. While this can be applied to many models, we demonstrate how this can be done for differentiable models and for those that have (approximately) conditional Gaussian transitions and arbitrary potential functions.

\subsubsection{Differentiable models}\label{subsubsec:diff}
When the potential functions $g_t$ are differentiable, it is possible to incorporate first or second-order information from the potential. Indeed, we have
\begin{equation}
    \begin{split}
    \exp(\gamma(x_{0:T})) \approx \exp\left(\gamma(u_{0:T}) + \sum_{t=0}^T\pdv{\gamma(u_{0:T})}{u_t}\cdot (x_t - u_t)\right) \eqqcolon \prod_{t=0}^T \hat{g}_t(x_t \mid u_{0:T}).
    \end{split}
\end{equation}
Now, as in Section~\ref{sec:auxiliary_samplers}, we can form the proposal distributions
\begin{equation}\label{eq:differentiable-model}
    \begin{split}
        \tilde{p}_t(x_t \mid x_{t-1}, u_{0:T}) 
            &\propto \mathcal{N}\left(x_t; u_t, \frac{\delta_t}{2} \Sigma_t\right) \, \hat{g}_t(x_t \mid u_{0:T}) \propto \mathcal{N}\left(x_t; u_t + \frac{\delta_t}{2} \Sigma_t \pdv{\gamma(u_{0:T})}{u_t}, \frac{\delta_t}{2} \Sigma_t\right)
    \end{split}
\end{equation}
and similarly for $\tilde{p}_0$.
Omitting the dependency on $u_{0:T}$ on the right handside for simplicity, we can therefore reformulate the auxiliary Feynman--Kac model as
\begin{equation}\label{eq:differentiable-Feynman-Kac}
    \begin{split}
        \pi(\ft{x}{0}{T} \mid \ft{u}{0}{T})
        &\propto p_0(x_0)\left\{\prod_{t=1}^T p_t(x_t \mid x_{t-1})\right\}\\
        & \times g_0(x_0)\left\{\prod_{t=1}^T g_t(x_t, x_{t-1})\right\} \left\{\prod_{t=0}^T \mathcal{N}\left(u_t; x_t, \frac{\delta_t}{2} \Sigma_t\right)\right\} \\
        &\propto \tilde{p}_0(x_0)\left\{\prod_{t=1}^T \tilde{p}_t(x_t \mid x_{t-1})\right\} \tilde{g}_0(x_0) \left\{\prod_{t=1}^T \tilde{g}_t(x_t, x_{t-1}) \right\},
    \end{split}
\end{equation}
for
\begin{equation}\label{eq:differentiable-Feynman-Kac-2}
    \tilde{g}_t(x_t, x_{t-1}) = g_t(x_t, x_{t-1}) \, p_t(x_t \mid x_{t-1})\frac{\mathcal{N}\left(x_t; u_t, \frac{\delta_t}{2} \Sigma_t\right)}{\mathcal{N}\left(x_t; u_t + \frac{\delta_t}{2} \Sigma_t \pdv{\gamma(u_{0:T})}{u_t}, \frac{\delta_t}{2} \Sigma_t\right)}
\end{equation}
and similarly for $\tilde{g}_0$.

Similarly to Section~\ref{subsec:auxiliary-lgssm}, when the potential function is separable, i.e., when we have $\gamma(x_{0:T}) = \sum_{t=0}^T \gamma_t(x_t)$, it is also possible to use second-order linearisation whilst not relinquishing the Feynman--Kac structure required to implement Algorithm~\ref{alg:csmc}. And, finally, when $p_t$ is also differentiable, we can also include information from it in the sampler by considering $\exp(\gamma) = \prod_{t=0} p_t \, g_t$ rather than simply using $g_t$.
We can then plug these choices for $\tilde{p}$ and $\tilde{g}$ inside~\eqref{eq:modified-auxiliary-feynman-kac} to then recover a gradient-informed equivalent representation of $\pi(x_{0:T} \mid u_{0:T})$ that will still be local, as \citet{finke2021csmc}, but will have proposal distributions that are approximately ``locally-optimal''~\citep[in the sense of, e.g.,][Ch. 10]{Chopin2020book} for the auxiliary target.
Interestingly, these new proposal distributions are also fully separable in time, so that they can immediately be used in the parallel-in-time particle Gibbs algorithm of \citet{corenflos2022sequentialized}.

\subsubsection{Approximately Gaussian transitions}\label{subsubsec:guided}
Consider now the case when the prior process is conditionally Gaussian (this extends, as in Section~\ref{subsec:auxiliary-lgssm}, to the more general case when the prior is not conditionally Gaussian but its conditional means and covariances are tractable). 
We can easily design a model~\citep[this is called a guided proposal in][Section 10.3.2]{Chopin2020book} locally adapted to the auxiliary observation $u_t$ as
\begin{equation}\label{eq:guided-prop}
    \begin{split}
        \tilde{p}_t(x_t \mid x_{t-1}, u_t) 
            \propto \mathcal{N}\left(u_t; x_t, \frac{\delta_t}{2} \Sigma_t \right) \,\mathcal{N}\left(x_t; m^X_{t-1}(x_{t-1}), C^X_{t-1}(x_{t-1})\right) \propto \mathcal{N}\left(x_t; \mu_t, \Lambda_t\right),
    \end{split}
\end{equation}
for $\mu_t = m^X_{t-1}(x_{t-1}) + K_{t-1}[u_t - m^X_{t-1}(x_{t-1})]$ and $\Lambda_t = C^X_{t-1}(x_{t-1}) - K_{t-1} C^X_{t-1}(x_{t-1})$, where $K_{t-1} = C^X_{t-1}(x_{t-1}) \left[C^X_{t-1}(x_{t-1}) + \frac{\delta_t}{2}\Sigma_t\right]^{-1}$. 
A similar form is available for $\tilde{p}_0$.
Using this new proposal, and making the dependency on $u_t$ implicit for notational simplicity, an equivalent Feynman--Kac model will then take the form
\begin{equation}\label{eq:re-re-modified-auxiliary-feynman-kac}
    \pi(\ft{x}{0}{T} \mid \ft{u}{0}{T})
        \propto \tilde{p}_0(x_0)\left\{\prod_{t=1}^T \tilde{p}_t(x_t \mid x_{t-1})\right\}\tilde{g_0}(x_0) \left\{\prod_{t=1}^T \tilde{g_t}(x_t, x_{t-1}) \right\},
\end{equation}
where $\tilde{p}_t$ is given by~\eqref{eq:guided-prop}, and 
\begin{equation}\label{eq:re-re-modified-auxiliary-feynman-kac-2}
   \tilde{g}_t(x_t, x_{t-1}) = \frac{g_t(x_t, x_{t-1}) \, p_t(x_t \mid x_{t-1})}{\tilde{p}_t(x_t \mid x_{t-1})}\mathcal{N}\left(u_t; x_t, \frac{\delta_t}{2} \Sigma_t \right).
\end{equation} 
The resulting auxiliary Feynman--Kac model~\eqref{eq:re-re-modified-auxiliary-feynman-kac} can then be sampled from using Algorithm~\ref{alg:csmc} where the particles are sampled from the proposal $\tilde{p}_t$ of~\eqref{eq:guided-prop} and the weights are computed using the potential functions $\tilde{g}_t$ of~\eqref{eq:re-re-modified-auxiliary-feynman-kac-2}.

Using such a proposal model, contrary to the independent auxiliary proposal cases, is not parallelisable in time, and will scale as $\bigO(T)$, even on parallel hardware. On the other hand, when the potential is weakly informative compared to the dynamics, we can expect them to have better statistical properties, as they explicitly incorporate these inside the proposal model. 
We also note that the construction proposed in~\eqref{eq:guided-prop} and~\eqref{eq:re-re-modified-auxiliary-feynman-kac} extends to other methods developed to leverage approximate Gaussian conjugacy relationships in state-space models, for instance, they are directly compatible with Laplace approximations of the potential~\citep[see, e.g.][Section 10.5.3]{Chopin2020book} or Rao--Blackwellisation~\citep{Murphy2001}.

\subsubsection{Hybrid proposal models}\label{subsubsec:guided-diff}
It is worth highlighting that the two approaches presented above are not mutually exclusive. 
Indeed, we can combine an approximately Gaussian transition model together with a first or second-order linearisation of the potential function, thereby obtaining hybrid adapted proposals that may work better than their individual components taken in isolation. 

With the notations above, this would, for example, correspond to 
\begin{equation}\label{eq:guided-prop-grad}
    \begin{split}
        \tilde{p}_t(x_t \mid x_{t-1}, u_{0:T}) 
            &\propto \mathcal{N}\left(u_t; x_t, \frac{\delta_t}{2} \Sigma_t \right) \,\mathcal{N}\left(x_t; m^X_{t-1}(x_{t-1}), C^X_{t-1}(x_{t-1})\right) \hat{g}_t(x_t \mid u_{0:T})\\
            &\propto \mathcal{N}\left(u_t + \frac{\delta_t}{2}\Sigma_t \pdv{\gamma(u_{0:T})}{x_t}; x_t, \frac{\delta_t}{2} \Sigma_t \right) \,\mathcal{N}\left(x_t; m^X_{t-1}(x_{t-1}), C^X_{t-1}(x_{t-1})\right),
    \end{split}
\end{equation}
if the linearisation point of $\gamma$ was taken to be $u_{0:T}$. This can then be simplified explicitly as in~\eqref{eq:guided-prop} to obtain gradient-informed, guided proposals. 
Similarly as in Sections~\ref{subsubsec:diff} and~\ref{subsubsec:guided}, we can then formulate the modified potential functions $\tilde{g}_t = g_t \cdot p_t / \tilde{p}_t$ and $\tilde{g}_0 = g_0 \cdot p_0 / \tilde{p}_0$ to obtain a new representation of the auxiliary Feynman--Kac model which can be sampled from using Algorithm~\ref{alg:csmc}.

Finally, other linearisation/combination choices are also possible, and the willing statistician is free to fully leverage the flexibility brought by introducing the auxiliary observations $u_{0:T}$. Understanding which is the best choice will typically be application specific, although we expect the methods presented in this section to provide a competitive test-bed for more advanced methods. 

\subsection{Extension to pseudo-marginal methods}
    \label{subsec:pseudo-marginal}
    While the particle Gibbs approach to sampling from~\eqref{eq:auxiliary-feynman-kac} is perhaps the most natural, it is also possible to instead consider a pseudo-marginal approach~\citep{andrieu2009pseudomarginal} as given by the particle marginal Metropolis-–Hastings (PMMH) sampler of \citet{Andrieu2010particle}. Consider a proposal distribution $q(u'_{0:T} \mid u_{0:T})$, for example, $\prod_{0}^T \mathcal{N}(u'_t; u_t, \frac{\delta_t}{2}\Sigma_t)$. Similarly to PMMH, because sequential Monte Carlo provides an unbiased estimate $\hat{\mathcal{Z}}_N(u_{0:T)}$ of the normalising constant for $\pi(x_{0:T} \mid u_{0:T})$, we can marginally target $\pi(x_{0:T})$ using a PMMH methodology. We succinctly summarise this extension in Algorithm~\ref{alg:auxiliary-pmcmc}.
\begin{algorithm}[!htb]
\SetAlgoLined
\DontPrintSemicolon
\caption{Auxiliary pseudo-marginal sampler}\label{alg:auxiliary-pmcmc}
\KwResult{An updated trajectory $x^{k+1}_{0:T}$}
\Fn{\textsc{aux-pm}$\big(x^k_{0:T}$, $u^k_{0:T},$ $\hat{\mathcal{Z}}^k_N\big)$}{
    Sample $u'_{0:T} \sim q(\cdot \mid u^k_{0:T})$\label{step:pmcmc-1}\;
    Sample $x'_{0:T}$ and $\hat{\mathcal{Z}}'_N$ using a particle filter targeting $\pi(\ft{x}{0}{T} \mid \ft{u'}{0}{T})$\label{step:pmcmc-2}\;
    Set $x^{k+1}_{0:T}$ to $x'_{0:T}$ with probability $\frac{\hat{\mathcal{Z}}'_N q(u^k_{0:T} \mid u'_{0:T})}{\hat{\mathcal{Z}}^k_N q(u'_{0:T} \mid u_{0:T})}$, otherwise, set it to $x^k_{0:T}$\;
    \Return{$x^{k+1}_{0:T}$}
}
\end{algorithm}

This method is related to the method of \citet{Deligiannidis2018correlated}. They show that, by correlating the noise introduced by the particle filter, the pseudo-marginal algorithm can be made to scale better with time series of increasing lengths $T$. This is because it results in correlated likelihood ratios $\frac{\hat{\mathcal{Z}}'_N}{\hat{\mathcal{Z}}^k_N}$ which exhibit lower variance than they would have otherwise. 

By using a proposal distribution adapted to the auxiliary target at hand, in a similar spirit as for the auxiliary particle Gibbs sampler of Algorithm~\ref{alg:aux-pgibbs}, we can hope to also benefit from a reduced variance of the likelihood estimates ratio in Algorithm~\ref{alg:auxiliary-pmcmc}. This, however, is not because the two estimates are correlated, but rather because they will both exhibit lower variance individually than their non-augmented counterparts. 
Contrary to \citet{Deligiannidis2018correlated}, this method necessitates the evaluation of the full (unnormalised) density of the Feynman--Kac model at hand, and will likely not perform well for a very large $T$. On the other hand, and in contrast to the correlated pseudo-marginal method~\citep[see the comments in Theorem 3 and Section 5.3]{Deligiannidis2018correlated}, Algorithm~\ref{alg:auxiliary-pmcmc} is likely to perform well in higher dimensions, due to the localisation of the proposals. Both approaches are furthermore not incompatible and could be used together. The benefit of doing so compared to simply using a particle Gibbs sampler, which (under backwards sampling) is stable for an increasing number of observations too~\citep{lee2020coupled}, is however not clear, and we leave the study of this question open for future work.

\section{Experimental evaluation}\label{sec:experiments}
In this section, we aim to empirically evaluate the statistical and computational behaviours of our proposed methods. To this end, we consider four sets of examples. In all cases we compare to state-of-the-art methods, that is, either the original method of~\citet{finke2021csmc} or \citet{mider2021continuous}.
\begin{itemize}
    \item The first model is a multivariate stochastic volatility model known to be challenging for Gaussian approximations and used as a benchmark in, for example, \citet{guarniero2017iterated,finke2021csmc}. This model has latent Gaussian dynamics, and an observation model which are both differentiable with respect to the latent state, so that all the methods of Section~\ref{subsec:auxiliary-lgssm} and Section~\ref{sec:pgibbs_samplers} apply. We consider the same parametrisation as in \citet{finke2021csmc}, which makes the system lack ergodicity and the standard particle Gibbs samplers not converge.
    \item The second one is a spatio-temporal model with independent latent Gaussian dynamics and is used in \citet{cruscino2022highdim} as a benchmark for high dimensional filtering. This model is akin to a type of dynamic random effect model in the sense that the latent states only interact at the level of the observations. This model is used to illustrate how latent structure can be used to design computationally efficient Kalman samplers that beat cSMC ones when the runtime is taken into account.
    \item The third model performs joint parameter and state estimation for a discretely observed stochastic differential equation. This model was used in \citet{mider2021continuous} to assess the performance of their forward-guiding backwards-filtering method. We demonstrate here how to use auxiliary samplers for the same purpose and show the competitiveness of our approach.
    \item While the three first examples highlight the benefits of our approach, the final example is a very simple, but illustrative toy-example, aimed at isolating their respective failure modes which we already alluded to in Sections~\ref{sec:auxiliary_samplers} and~\ref{sec:pgibbs_samplers}.
\end{itemize}

Throughout this section, when using an auxiliary cSMC sampler, be it the sequential or the parallel-in-time formulation, we use $N = 25$ particles and a target acceptance rate of $50\%$ across all time steps. This is more conservative than the recommendation of \citet{finke2021csmc}, corresponding to $1 - (1 + N)^{-1/3}\approx 66\%$. The difference stems from the fact that it may happen that the methods do not reach the relatively high acceptance rate implied by the more optimistic target for all time steps, even with very small $\delta$ values. As a consequence, the sampler is ``stuck'' by only proposing very correlated trajectories in some places. We believe that this is mostly due to the largely longer time series considered here as well as to the use of multinomial resampling which prevents achieving the optimal acceptance rate of $N / (N+1)$ when $\delta \ll 1$. Softening this constraint resulted in empirically better mixing. Furthermore, for all the samplers, and following \citet[][]{titsias2018,finke2021csmc}, we consider $\delta_t \Sigma_t = \delta_t I$, with a single $\delta_t = \delta$ being constant across time steps for the Kalman samplers. We then calibrate $\delta_t$ to achieve the desired acceptance rate (globally for Kalman samplers or per time step for the cSMC samplers) and the actual acceptance rate is reported below. Finally, we note that all the posterior distributions recovered from all the proposed methods were coherent, so we only report mixing statistics throughout.
Finally, we note that the choice of $N=25$ particles for the cSMC samplers is somewhat arbitrary, and driven mostly by computational (memory) resources. 
The cSMC algorithm in Algorithm~\ref{alg:csmc} together with the modification in Algorithm~\ref{alg:local-csmc-backward} is known to be robust to increasing $T$, even for a fixed $N\geq 2$~\citep{lee2020coupled}; as such, the choice of $N$ is not critical for the convergence of the algorithm.
Nonetheless, its convergence rate improves polynomially with $N$~\citep[Theorem 1]{karjalainen2024mixingtime}, and larger $N$ values would improve the mixing properties of the algorithm.
In simple terms, while it is beneficial to take $N$ as large as possible, there is no minimum $N$ required for the algorithm to converge.

The implementation details for all the experiments are as follows: whenever we say that a method was run on a CPU, we have used an AMD\textsuperscript{\textregistered} Ryzen Threadripper 3960X with 24 cores, and whenever the method has been run on a GPU, we used an Nvidia\textsuperscript{\textregistered} GeForce RTX 3090 GPU with 24~GB memory. All experiments were implemented in Python~\citep{Rossum2009Python} using the JAX library~\citep{jax2018github} which natively supports CPU and GPU backends as well as automatic differentiation that we use to compute the gradients required. The code to reproduce the experiments listed below can be found at the following address:~\url{https://github.com/AdrienCorenflos/aux-ssm-samplers}.%

\subsection{Multivariate stochastic volatility model}
    \label{subsec:stoch-vol}
    We consider the same multivariate stochastic volatility example as in \citet[Section E.3.]{finke2021csmc}. This model is classically used as a benchmark for high dimensional SMC-related methods~\citep[see also][]{guarniero2017iterated}. It is given by homogeneous auto-regressive Gaussian latent dynamics $p_t(x_t \mid x_{t-1}) = \mathcal{N}(x_t; F\, x_{t-1}, Q)$ and a potential defined as a multidimensional observation model 
\begin{equation}
g(x_{0:T}) = \prod_{t=0}^T h(y_t \mid x_t),\, \text{where}\quad h(y_t \mid x_t) = \prod_{d=1}^{d_x}\mathcal{N}(y_t(d); 0, \exp(x_t(d))). 
\end{equation} 

As per \citet{finke2021csmc}, we take $F = \phi I_d$, $Q_{ij} = \tau (\delta(i=j) + \delta(i \neq j) \rho)$ for $\phi = 90\%$, $\tau = 2$, and $\rho = 25\%$. Similarly, the initial distribution $p_0(x_0)$ is also taken to be the stationary distribution of the latent Gaussian dynamics and we take $d_x = 30$. However, we increase the number of time steps to $T=250$, rather than $50$ and we take the number of particles for all the auxiliary cSMC algorithms to be $N=25$. %

The different methods we compare here are the following: (i) auxiliary Kalman sampler with first order linearisation~\eqref{eq:gaussian-FK-proposal} (both on CPU and GPU), (ii) with second order linearisation~\eqref{eq:second-order} (both on CPU and GPU), (iii) auxiliary cSMC sampler with backward sampling for the proposals $\mathcal{N}(\cdot; u_t, \frac{\delta_t}{2}I)$ corresponding to \citet{finke2021csmc} (on CPU), (iv) auxiliary cSMC sampler with parallel-in-time~\citep{corenflos2022sequentialized} sampling for the proposals $\mathcal{N}(\cdot; u_t, \frac{\delta_t}{2}I)$ (on GPU), (v) auxiliary cSMC sampler with backward sampling for the gradient-informed proposals~\eqref{eq:differentiable-model} (on CPU), (vi) auxiliary cSMC sampler with parallel-in-time sampling for the gradient-informed proposals~\eqref{eq:differentiable-model} (on GPU), and (vii) the guided auxiliary cSMC sampler with backward sampling for both the proposals~\eqref{eq:guided-prop} and~\eqref{eq:guided-prop-grad} (on CPU).

\begin{figure}[htb!]
\centering
\begin{subfigure}[t]{.45\textwidth}
  \centering
  \resizebox{\columnwidth}{!}{\begin{tikzpicture}[scale=1.5]
\begin{axis}[
    grid=both,
    grid style=dotted,
    xmin=0,
    xmax=250,
    xlabel={$t$ time step},
    ymin=0,
    ymax=18,
    scale only axis=true,
    width=\textwidth,
    height=0.5\textwidth,
    smooth,
    line join=round,
    mark options={scale=1},
    mark repeat = 15,
    mark phase = 2,
    ]
\addplot[black, mark=o] table [x=t, y=Kalman_1, col sep=comma]{experiments/figures/stoch_vol_ESJD_False.csv};\label{line:stochvol-ESJD-Kalman-1_cpu}
\addplot[black, mark=x] table [x=t, y=Kalman_2, col sep=comma]{experiments/figures/stoch_vol_ESJD_False.csv};\label{line:stochvol-ESJD-Kalman-2_cpu}
\addplot[black, mark=star] table [x=t, y=cSMC_grad, col sep=comma]{experiments/figures/stoch_vol_ESJD_False.csv};\label{line:stochvol-ESJD-cSMC_grad_cpu}
\addplot[black, mark=square] table [x=t, y=cSMC, col sep=comma]{experiments/figures/stoch_vol_ESJD_False.csv};\label{line:stochvol-ESJD-cSMC_cpu}
\addplot[black, mark=otimes] table [x=t, y=cSMC_guided, col sep=comma]{experiments/figures/stoch_vol_ESJD_False.csv};\label{line:stochvol-ESJD-cSMC-guided_cpu}
\addplot[black, mark=triangle] table [x=t, y=cSMC_guided_grad, col sep=comma]{experiments/figures/stoch_vol_ESJD_False.csv};\label{line:stochvol-ESJD-cSMC-guided-grad_cpu}

\end{axis}
\end{tikzpicture}}
    \caption{CPU: auxiliary first order Kalman sampler~\ref{line:stochvol-ESJD-Kalman-1_cpu}, second order Kalman sampler~\ref{line:stochvol-ESJD-Kalman-2_cpu}, cSMC sampler~\ref{line:stochvol-ESJD-cSMC_cpu}, gradient-informed cSMC~\ref{line:stochvol-ESJD-cSMC_grad_cpu} sampler, the guided cSMC~\ref{line:stochvol-ESJD-cSMC-guided_cpu} sampler, and the gradient-informed guided cSMC~\ref{line:stochvol-ESJD-cSMC-guided-grad_cpu} sampler.}
    \label{fig:esjd-stoch-vol-cpu}
\end{subfigure}%
\hfill
\begin{subfigure}[t]{.45\textwidth}
  \centering
  \resizebox{\columnwidth}{!}{\begin{tikzpicture}[scale=1.]
\begin{axis}[
    grid=both,
    grid style=dashed,
    xmin=0, xmax=250,
    xlabel={$t$ time step},
    ymin=0,
    ymax=18,
    scale only axis=true,
    width=\textwidth,
    height=0.5\textwidth,
    smooth,
    line join=round,
    mark options={scale=1},
    mark repeat = 15,
    mark phase = 2,
    ]
\addplot[black, mark=o] table [x=t, y=Kalman_1, col sep=comma]{experiments/figures/stoch_vol_ESJD_True.csv};\label{line:stochvol-ESJD-Kalman-1_gpu}
\addplot[black, mark=x] table [x=t, y=Kalman_2, col sep=comma]{experiments/figures/stoch_vol_ESJD_True.csv};\label{line:stochvol-ESJD-Kalman-2_gpu}
\addplot[black, mark=star] table [x=t, y=cSMC_grad, col sep=comma]{experiments/figures/stoch_vol_ESJD_True.csv};\label{line:stochvol-ESJD-cSMC_grad_gpu}
\addplot[black, mark=square] table [x=t, y=cSMC, col sep=comma]{experiments/figures/stoch_vol_ESJD_True.csv};\label{line:stochvol-ESJD-cSMC_gpu}
\end{axis}
\end{tikzpicture}}
    \caption{GPU: auxiliary first order Kalman sampler~\ref{line:stochvol-ESJD-Kalman-1_gpu}, second order Kalman sampler~\ref{line:stochvol-ESJD-Kalman-2_gpu}, cSMC sampler~\ref{line:stochvol-ESJD-cSMC_gpu}, and gradient-informed cSMC~\ref{line:stochvol-ESJD-cSMC_grad_cpu} sampler. The latter two are hard to distinguish, but the gradient-informed cSMC is generally above the non-informed.}
    \label{fig:esjd-stoch-vol-gpu}
\end{subfigure}
\caption{Average (across 10 different experiments) expected squared jump distance per iteration for all the samplers considered on the stochastic volatility model of Section~\ref{subsec:stoch-vol}.}
\end{figure}
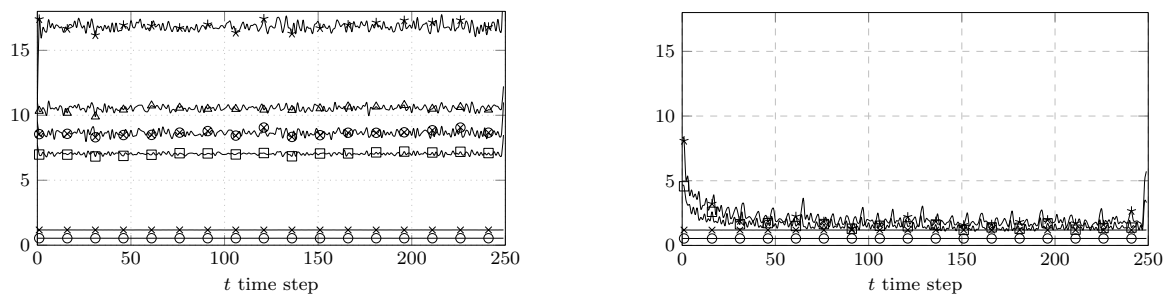

In order to compare the samplers in this example, we generate $10$ different datasets. For every dataset, we run each sampler for $2\,500$ adaptation steps. After this, we run $10\,000$ more iterations to compute the empirical expected squared jump distance~\citep[ESJD,][]{pasarica2010adaptively}\footnote{The ESJD is, in first approximation, proportional to the effective sample size~\citep[ESS, see, e.g.,][]{geyer1992practical} which measures the ``equivalent'' number of independent samples that would have resulted in an estimator with the same variance. 
The reason why we use the ESJD and not the ESS directly here is because the latter requires storing long sequences of sample trajectories, which is memory intensive and artificially decreased the performance of the GPU based methods, as GPUs have less memory available.} for each sampler, defined as, for each time step $t$, the empirical value of 
\begin{equation}
    \frac{1}{L} \sum_{l=1}^{L-1}\sum_{i,j=1}^d\left[X^{A + l + 1}_{t+1}(i,j) - X^{A + l}_{t+1}(i,j)\right]^2.
\end{equation}
All samplers, in both the sequential and parallel case, were targeting 50\% acceptance rate across all time steps and the effective acceptance rate ranged between 47\% and 52\% for all samplers and time steps. The averaged (across the 10 experiments) ESJD is reported in Figure~\ref{fig:esjd-stoch-vol-cpu} for the sequential versions of the algorithm, and in Figure~\ref{fig:esjd-stoch-vol-gpu} for the parallel counterparts (noting that there is, as expected, no statistical difference between the sequential and parallel implementations of the Kalman samplers). 
As highlighted by Figure~\ref{fig:esjd-stoch-vol-cpu}, the gradient-informed auxiliary cSMC statistically dominates the alternatives for all time steps on both CPU and GPU (although this is less obvious on the GPU). 

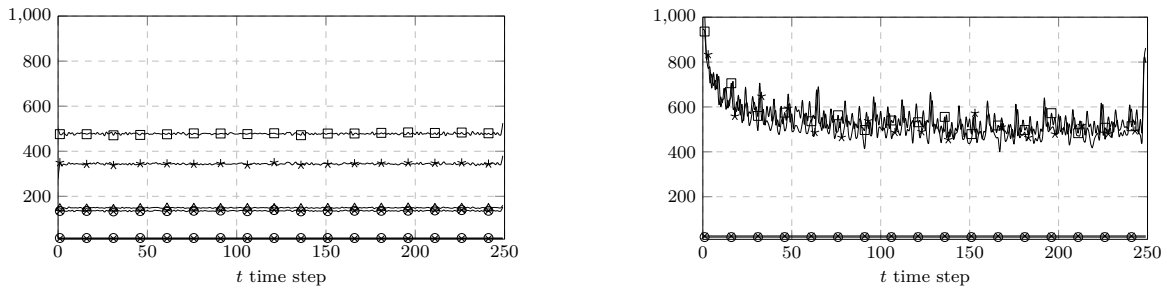
\begin{figure}[htb!]
\centering
\begin{subfigure}[t]{.45\textwidth}
  \centering
  \resizebox{\columnwidth}{!}{\begin{tikzpicture}[scale=1.5]
\begin{axis}[
    grid=both,
    grid style=dashed,
    xmin=0,
    xmax=250,
    xlabel={$t$ time step},
    ymin=10,
    ymax=1000,
    scale only axis=true,
    width=\textwidth,
    height=0.5\textwidth,
    smooth,
    line join=round,
    mark options={scale=1},
    mark repeat = 15,
    mark phase = 2,
    ]
\addplot[black, mark=o] table [x=t, y=Kalman_1, col sep=comma]{experiments/figures/stoch_vol_ESJD_time_False.csv};\label{line:stochvol-ESJD-Kalman-1-time_cpu}
\addplot[black, mark=x] table [x=t, y=Kalman_2, col sep=comma]{experiments/figures/stoch_vol_ESJD_time_False.csv};\label{line:stochvol-ESJD-Kalman-2-time_cpu}
\addplot[black, mark=star] table [x=t, y=cSMC_grad, col sep=comma]{experiments/figures/stoch_vol_ESJD_time_False.csv};\label{line:stochvol-ESJD-cSMC_grad-time_cpu}
\addplot[black, mark=square] table [x=t, y=cSMC, col sep=comma]{experiments/figures/stoch_vol_ESJD_time_False.csv};\label{line:stochvol-ESJD-cSMC-time_cpu}
\addplot[black, mark=otimes] table [x=t, y=cSMC_guided, col sep=comma]{experiments/figures/stoch_vol_ESJD_time_False.csv};\label{line:stochvol-ESJD-cSMC-guided-time_cpu}
\addplot[black, mark=triangle] table [x=t, y=cSMC_guided_grad, col sep=comma]{experiments/figures/stoch_vol_ESJD_time_False.csv};\label{line:stochvol-ESJD-cSMC-guided-grad-time_cpu}

\end{axis}
\end{tikzpicture}}
    \caption{CPU: auxiliary first order Kalman sampler~\ref{line:stochvol-ESJD-Kalman-1_cpu}, second order Kalman sampler~\ref{line:stochvol-ESJD-Kalman-2_cpu}, cSMC sampler~\ref{line:stochvol-ESJD-cSMC_cpu}, gradient-informed cSMC~\ref{line:stochvol-ESJD-cSMC_grad_cpu} sampler, the guided cSMC~\ref{line:stochvol-ESJD-cSMC-guided_cpu} sampler, and the gradient-informed guided cSMC~\ref{line:stochvol-ESJD-cSMC-guided-grad_cpu} sampler.}
    \label{fig:esjd-time-stoch-vol-cpu}
\end{subfigure}%
\hfill
\begin{subfigure}[t]{.45\textwidth}
  \centering
  \resizebox{\columnwidth}{!}{\begin{tikzpicture}[scale=1.]
\begin{axis}[
    grid=both,
    grid style=dashed,
    xmin=0, xmax=250,
    xlabel={$t$ time step},
    ymin=10,
    ymax=1000,
    scale only axis=true,
    width=\textwidth,
    height=0.5\textwidth,
    smooth,
    line join=round,
    mark options={scale=1},
    mark repeat = 15,
    mark phase = 2,
    ]
\addplot[black, mark=o] table [x=t, y=Kalman_1, col sep=comma]{experiments/figures/stoch_vol_ESJD_time_True.csv};\label{line:stochvol-ESJD-Kalman-1-time_gpu}
\addplot[black, mark=x] table [x=t, y=Kalman_2, col sep=comma]{experiments/figures/stoch_vol_ESJD_time_True.csv};\label{line:stochvol-ESJD-Kalman-2-time_gpu}
\addplot[black, mark=star] table [x=t, y=cSMC_grad, col sep=comma]{experiments/figures/stoch_vol_ESJD_time_True.csv};\label{line:stochvol-ESJD-cSMC_grad-time_gpu}
\addplot[black, mark=square] table [x=t, y=cSMC, col sep=comma]{experiments/figures/stoch_vol_ESJD_time_True.csv};\label{line:stochvol-ESJD-cSMC-time_gpu}
\end{axis}
\end{tikzpicture}}
    \caption{GPU: auxiliary first order Kalman sampler~\ref{line:stochvol-ESJD-Kalman-1_gpu}, second order Kalman sampler~\ref{line:stochvol-ESJD-Kalman-2_gpu}, cSMC sampler~\ref{line:stochvol-ESJD-cSMC_gpu}, and gradient-informed cSMC~\ref{line:stochvol-ESJD-cSMC_grad_cpu} sampler. The latter two are hard to distinguish, with no clear difference in terms of performance.}
    \label{fig:esjd-time-stoch-vol-gpu}
\end{subfigure}
\caption{Average (across 10 different experiments) expected squared jump distance per second for all the samplers considered on the stochastic volatility model.}
\end{figure}

This picture, however, is modified when looking at the ESJD per second,
rather than per iteration in Figures~\ref{fig:esjd-time-stoch-vol-cpu} and~\ref{fig:esjd-time-stoch-vol-gpu}. In this case, on the CPU, the method of \citet{finke2021csmc} dominates the other ones. This is because it offers reasonable statistical efficiency ($\sim70\%$ the ESJD of the most efficient sampler tested here) with a rather small time-complexity overall (no gradient calculation and no matrix inversion like in the Kalman samplers is needed here). On the other hand, the Kalman samplers are here completely dominated by all Monte Carlo alternatives. The GPU picture is more mixed, and both the gradient-informed and uninformed proposals seem to provide the same overall efficiency in this case but still completely dominate the Kalman alternatives here too. 

This underwhelming performance of the Kalman sampler was in fact to be expected given the need to solve $250$ matrix systems of dimension $30$ per iteration (albeit some are done in parallel on GPU). 
In fact, this had another deleterious effect: the parallel versions of the auxiliary Kalman sampler suffered from numerical divergence in this experiment when using single precision floats (32-bits representation). This problem, due to the numerical instability of the covariance matrices calculations is well known in the literature~\citep[see, e.g.][]{Bierman1977sqrt} and prompted the development of a square-root version of the parallel Kalman filtering and smoothing algorithms in \citet{Yaghoobi2022sqrt}. Here, we simply used double precision floats instead of the square-root method as this sufficed to fix the numerical instability. This numerical instability is an important drawback of Kalman methods in general and is particularly salient on parallel hardware which is often optimised to run on lower precision arithmetic~\citep{jouppi2017datacenter}. The issue did not arise for the sequential version of the algorithms, and we, therefore, stuck to single float precision arithmetic for these. In the next section, we show how latent structure can be leveraged to bypass the dimensionality problem.

\subsection{Spatio-temporal model}
    \label{subsec:spatio-temporal}
    We now consider the spatio-temporal model of \citet[Section 4.2]{cruscino2022highdim} which was recently introduced as a benchmark for high-dimensional state inference in non-linear systems. It consists of independent latent dynamics for a state $X_t(i, j)$ located on a two-dimensional lattice $\{1, \ldots, d\} \times \{1, \ldots, d\} \ni (i,j)$, for $d=8$, with an observation model that does not factorise over the nodes of the lattice, thereby creating non-trivial posterior structure between the states. We are given a $8^2 = 64$ dimensional model
\begin{equation}\label{eq:spatio-temporal}
    \begin{split}
        X_t(i, j) &= X_{t-1}(i, j) + U_t(i, j), \quad i, j = 1, \ldots, d, \\
        Y_t(i, j) &= X_{t}(i, j) + V_t(i, j), \quad i, j = 1, \ldots, d,
    \end{split}
\end{equation}
where, for all $t, i, j$, the $U_t(i, j)$ are i.i.d. according to $\mathcal{N}(0, \sigma_X^2)$, and for all $t$, the $V_t$'s are i.i.d. according to a multivariate t-distribution with $\nu$ degrees of freedom centred on $0$. The precision matrix of the $V_t$'s is  given by $\Sigma^{-1} = \tau^{D[(i, j), (i', j')]}$ if $D[(i, j), (i', j')] \leq r_y$ and $0$ otherwise, where $D[(i, j), (i', j')]$ is a graph distance, and $\tau < 0$ a given parameter.

In \citet{cruscino2022highdim}, the parameters are chosen to be $\sigma_X = 1$, $\nu = 10$, $\tau = -1/4$, $r_y = 1$, and $D[(i, j), i', j')] = \abs{i - i'} + \abs{j - j'}$, so that an observation is mostly corrupted by its direct neighbours. We keep all the parameters unchanged, with the exception that, in order to make the problem more difficult, we take $\nu=1$ so that the observation model does not have first or second moments, and to showcase the parallelisation in time, we also consider a substantially higher number of time steps $T=1\,024$~\citep[vs. $T=10$ in][]{cruscino2022highdim}. Overall, the total dimension of the target model is therefore of the order of $65\,000$.
The first-order auxiliary Kalman sampler is particularly suited to this type of model, even if the underlying state dimension is large. This is due to the fact that the prior factorises across all dimensions, so that the auxiliary LGSSM proposal~\eqref{eq:auxiliary-LGSSM} factorises too, even if the target $\pi(x_{0:T})$ does not. As a consequence, we are left with sampling from $d \times d$ independent one-dimensional LGSSMs rather than a $d \times d$ dimensional one. 
In other terms, instead of needing to compute conditional Gaussian distributions of dimension $d \times d$, and therefore needing to solve systems of size $d \times d$ (ocurring a cost $\bigO({(d^2)}^3)$ on CPU), we only need to solve one-dimensional systems, that is, divide and multiply by scalars. 
This property extends to some extent to auxiliary cSMC samplers where the proposal is chosen to factorise across dimensions too. 
This means that the cost will be dominated by the computation of the log-likelihood of the multivariate t-distribution at each time step and (for specialised implementations) the complexity of the auxiliary cSMC will then be a direct multiple of the complexity of the auxiliary Kalman sampler (under no parallelisation, we can expect is to be roughly $N+1$ times more expensive, where $N+1$ is the total number of particles).

The experiment design is as follows: we simulate $20$ datasets from~\eqref{eq:spatio-temporal}. For each of these, we set the initial trajectory of the MCMC chain to be the result of a single trajectory formed from the backward sampling~\citep{Godsill2004FFBS} of a bootstrap filter algorithm with $1\,000$ particles (this gives bad smoothing statistics but is a good starting point for an MCMC chain) and run $A=5\,000$ adaptation steps, after which the statistics of the chain are collected over $L = 20\,000$ iterations. 
For this experiment, all the sequential versions of the auxiliary Kalman and cSMC samplers were dramatically slower than the parallel-in-time alternatives: they took in the order of a second per iteration, both on CPU and GPU, compared to the PIT versions that took in the order of a millisecond per iteration on GPU. 
As a consequence, we do not report their results here but do so in Appendix~\ref{app:spatial}.
Instead, we focus on (i) the parallel-in-time version of \citet{finke2021csmc} given by using \citet{corenflos2022sequentialized} on step~\ref{step:pmcmc-2} of Algorithm~\ref{alg:aux-pgibbs}, (ii) the parallel-in-time version of the gradient auxiliary proposal~\eqref{eq:differentiable-model} of Section~\ref{subsec:guided-smc-sampler}, which we refer to as gradient-informed, and finally (iii) the auxiliary Kalman sampler~\eqref{eq:gaussian-FK-proposal} of Section~\ref{subsec:auxiliary-lgssm}, with first-order linearisation only, noting that the second order would remove the computational benefits of having a separable prior.

As per \citet{titsias2018}, we target a $50\%$ acceptance rate for the auxiliary Kalman sampler.
The final average acceptance rates were a little lower, with the auxiliary Kalman sampler accepting $34\%$ of the trajectories. This is most likely due to our calibration algorithm being too optimistic, but did not seem to impact the final results beyond reason and therefore did not, in our opinion, warrant further investigation.

\begin{figure}[htb!]
\centering
\begin{subfigure}[t]{.45\textwidth}
  \centering
  \resizebox{\columnwidth}{!}{\begin{tikzpicture}[scale=1.]
\begin{axis}[
    grid=both,
    grid style=dashed,
    xmin=0, xmax=1023,
    xlabel={$t$ time step},
    ymax=4,
    ymin=0,
    scale only axis=true,
    width=\textwidth,
    height=0.5\textwidth,
    ]
\addplot[black, line width=1pt] table [x=t, y=Kalman, col sep=comma]{experiments/figures/spatial_ESJD_gpu.csv};\label{line:ESJD-Kalman-gpu}
\addplot[lightgray, line width=1pt] table [x=t, y=cSMC_grad, col sep=comma]{experiments/figures/spatial_ESJD_gpu.csv};\label{line:ESJD-cSMC-grad-gpu}
\addplot[gray, line width=1pt] table [x=t, y=cSMC, col sep=comma]{experiments/figures/spatial_ESJD_gpu.csv};\label{line:ESJD-cSMC-gpu}
\end{axis}
\end{tikzpicture}}
    \caption{Expected squared jump distance for the auxiliary Kalman sampler~\ref{line:ESJD-Kalman-gpu}, the auxiliary cSMC sampler~\ref{line:ESJD-cSMC-gpu}, and the auxiliary cSMC sampler with gradient-informed proposals~\ref{line:ESJD-cSMC-grad-gpu}. Kalman~\ref{line:ESJD-Kalman-gpu} shows as a roughly horizontal line at the bottom.}
    \label{fig:esjd-spatio-temporal}
\end{subfigure}%
\hfill
\begin{subfigure}[t]{.45\textwidth}
  \centering
  \resizebox{\columnwidth}{!}{\begin{tikzpicture}[scale=1.]
\begin{axis}[
    grid=both,
    grid style=dashed,
    xmin=0, xmax=1023,
    xlabel={$t$ time step},
    ymax=1000,
    ymin=0,
    scale only axis=true,
    width=\textwidth,
    height=0.5\textwidth,
    ]
\addplot[black, line width=1pt] table [x=t, y=Kalman, col sep=comma]{experiments/figures/spatial_ESJD_time_gpu.csv};\label{line:ESJD-Kalman-time-gpu}
\addplot[lightgray, line width=1pt] table [x=t, y=cSMC_grad, col sep=comma]{experiments/figures/spatial_ESJD_time_gpu.csv};\label{line:ESJD-cSMC-grad-time-gpu}
\addplot[gray, line width=1pt] table [x=t, y=cSMC, col sep=comma]{experiments/figures/spatial_ESJD_time_gpu.csv};\label{line:ESJD-cSMC-time-gpu}

\end{axis}
\end{tikzpicture}}
    \caption{Expected squared jump distance per second for the auxiliary Kalman sampler~\ref{line:ESJD-Kalman-gpu}, the auxiliary cSMC sampler~\ref{line:ESJD-cSMC-gpu}, and the auxiliary cSMC sampler with gradient-informed proposals~\ref{line:ESJD-cSMC-grad-gpu}.}
    \label{fig:esjd-spatio-temporal-time}
\end{subfigure}
\caption{Average (across 20 different experiments) expected squared jump distance per iteration and second for all the samplers considered on the spatio-temporal model~\eqref{eq:spatio-temporal}.}
\end{figure}
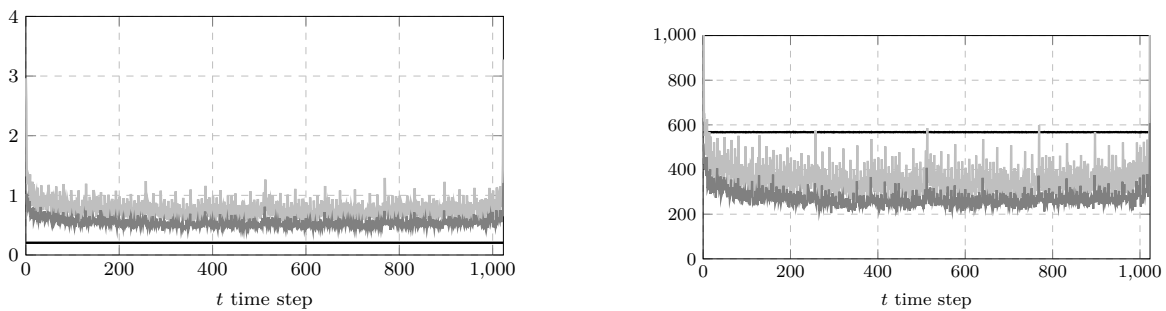

The ESJD is shown, averaged over all experiments, in Figure~\ref{fig:esjd-spatio-temporal}, while the time-scaled ESJD, namely ESJD divided by the number of seconds taken to run one step of the sampler is shown, averaged over all experiments, in Figure~\ref{fig:esjd-spatio-temporal-time}. 
The gradient-enhanced PIT auxiliary cSMC has a better ESJD than the basic PIT auxiliary cSMC which in turn has a better ESJD than the auxiliary Kalman sampler. The ordering of these methods however changes if one takes into account the additional complexity incurred by SMC, and after rescaling by the time taken by iteration, the auxiliary Kalman sampler dominates the gradient-enhanced PIT auxiliary cSMC which still dominates its basic counterpart.

In practice, the auxiliary Kalman, conditional SMC, and gradient-enhanced conditional SMC samplers took respectively on average $0.52$, $2.1$, and $2.2$ milliseconds per iteration. 
While some idiosyncrasies may be present, we believe that this performance gap could be further improved by careful consideration of the structure of the model in the Kalman sampler --- we have not undertaken this here in order to preserve the general applicability of our implementation.

\subsection{Parameter estimation in a continuous-discrete diffusion smoothing problem}
    \label{subsec:lorenz}
    In this section, we consider the same experiment as in \citet[Section 6.1]{mider2021continuous}, which consists of a joint sampling of the state of a discretely observed chaotic Lorenz stochastic differential equation, and of the parameter defining its drift.
The SDE is given, conditionally on a parameter $\theta = (\theta_1, \theta_2, \theta_3)$ as a three-dimensional SDE $\dd{x} = \beta_{\theta}(x) \dd{t} + \sigma \dd{W_t}$,
where $W$ is a three-dimensional standard Wiener process and
\begin{equation}
    \beta_{\theta}(x) = \begin{pmatrix}\theta_1 (x_2 - x_1) \\ \theta_2 x_1 - x_2 - x_1 x_3 \\ x_1 x_2 - \theta_3 x_3 \end{pmatrix}.
\end{equation} 
The state is then observed at regular intervals (every $t_0=0.01, t_1=0.02, \ldots, t_K=2$) through its second and third component only, giving an observation model $Y_k \sim \mathcal{N}(H x(t_k), 5 I_2)$, for $H=\begin{pmatrix}0 & 1 & 0 \\ 0 & 0 & 1\end{pmatrix}$ and where $I_2$ denotes the identity matrix in $\mathbb{R}^2$. In order to provide comparable results to \citet{mider2021continuous}, we use the code they provided to generate the same dataset and pick the same parametrisation of the model, including the same prior for the parameters. The Markov chain is then initialised according to the prior dynamics conditionally on the same initial parameter values as in \citet{mider2021continuous}. As per their experiment, we sample from the joint distribution 
\begin{equation}
    \pi(x(t'_0), \ldots, x(t'_L), \theta \mid y_{0:K}),
\end{equation}
where $t'_0=0, t'_1=2e-4, \ldots, t'_L=2$ is a finer grid, making for a total sampling space dimension of $3 + 30\,000$. To do so, we use the conjugacy relationship of $\theta$ given the full path for $x$, implementing a Hastings-within-Gibbs routine which samples $\theta$ conditionally on $x(t'_0), \ldots, x(t'_L)$ using its closed-form Gaussian posterior~\citet[Proposition 4.5]{mider2021continuous}, and then the auxiliary Kalman sampler to sample $x(t'_0), \ldots, x(t'_L)$ conditionally on $\theta$. 

In our case, because the observation model is linear, we use the following proposal in the Kalman sampler: first, given the current trajectory $(x^k(t'_l))_{l=0}^L$ and parameter $\theta^k$ state of the MCMC chain we linearise 
\begin{equation}
    \mathbb{E}[x(t'_l) \mid x(t'_{l-1})] = \beta_{\theta^k}(x(t'_{l-1})) (t'_l - t'_{l-1})
\end{equation}
around $x^k(t'_{l-1})$ using the method of Section~\ref{subsec:auxiliary-lgssm} with extended linearisation, obtaining approximations 
\begin{equation}
    p(x(t'_l) \mid x(t'_{l-1})) \approx \mathcal{N}(x(t'_l); F_{l-1} x(t'_{l-1}) + b_{l-1}, Q_{l-1}).
\end{equation}
For $l=0, \ldots, L$, we then sample $u_l \sim \mathcal{N}(x(t'_l), \frac{\delta}{2} I_3)$, where $I_3$ denotes the identity matrix in $\mathbb{R}^3$, and then form the proposal
\begin{equation}\label{eq:proposal-lorenz}
    \begin{split}
        q\left((z(t'_l))_{l=0}^L \mid u_{0:L}, (x^k(t'_l))_{l=0}^L, y_{0:L}\right) 
           &\propto \mathcal{N}\left(z(t'_0); m_0, P_0\right) \left\{\prod_{l=1}^L \mathcal{N}\left(z(t'_l) ; F_{l-1}z(t'_{l-1}) + b_{l-1}, Q_{l-1}\right)\right\} \\
                &\left\{\prod_{k=0}^K \mathcal{N}\left(y_k ; H z(t_k), 5 I_2\right)\right\} \left\{\prod_{l=0}^L\mathcal{N}\left(u_l ; z(t'_l), \frac{\delta}{2} I_3\right)\right\}
    \end{split}
\end{equation}
targeting the augmented model
\begin{equation}
    \begin{split}
        \pi\left((z(t'_l))_{l=0}^L \mid u_{0:L}, y_{0:K}\right)
            &\propto \mathcal{N}\left(z(t'_0); m_0, P_0\right) \left\{\prod_{l=1}^L p\left(z(t'_l) \mid z(t'_{l-1})\right)\right\} \\
                &\left\{\prod_{k=0}^K \mathcal{N}\left(y_k ; H z(t_k), 5 I_2\right)\right\}
                \left\{\prod_{l=0}^L\mathcal{N}\left(u_l ; z(t'_l), \frac{\delta}{2} I_3\right)\right\}.
    \end{split}
\end{equation}
We run $2\,500$ adaptation steps, during which we modify $\delta$ to target an average acceptance rate of $23.4\%$ (as per \citet{mider2021continuous}). Interestingly, our actual acceptance rate after adaptation was closer to $70\%$, and the resulting $\delta$ was virtually infinite. This means that the proposal distribution is almost reversible with respect to the target distribution. This high acceptance rate did not negatively impact the convergence of our algorithm. In fact, our resulting effective sampling size was larger than the best one reported by \citet{mider2021continuous} for both the parameters and the smoothing marginals (while the posterior distributions were similar). We report this in Table~\ref{tab:ess_lorenz}.
\begin{table}
\centering
\caption{Effective sample size (ESS) for the auxiliary sampler, computed using chains of length $10^5$. The results for \citet{mider2021continuous} are reported for ease of comparison.}
\label{tab:ess_lorenz}
\begin{tabular}{ lrrrrrr }
    \toprule
    & $X_{1,1.5}$ & $X_{2,1.5}$ & $X_{3,1.5}$ & $\theta_1$ & $\theta_2$ & $\theta_3$ \\
    \midrule
    This paper                  & 31254.0     & 35469.9     & 36584.7     & 11850.0    & 22960.5    & 12240.5    \\
    \citet{mider2021continuous} & 10480.3     & 22890.5     & 24070.2     & 4592.4     & 15379.5    & 10917.7    \\
    \bottomrule
\end{tabular}
\end{table}

In practice, our sampler took $3\,149$ seconds (52 minutes) to run on the GPU, and $9\,424$ seconds (2h30mn) on the CPU. \citet{mider2021continuous}, on the other hand, resulted in much faster run times (approximately 3--4 minutes). While this difference may seem massive, it can be imputed in totality to the difference in software for this experiment. Indeed, because they too rely on Gaussian filtering, the theoretical serial complexity of the two methods (when run on CPU) are exactly the same. While they use the programming language Julia~\citep{Julia-2017}, we use the JAX library~\citep{jax2018github} written in the Python language. Our choice comes with the benefit of direct GPU support but also presents the inconvenience of not supporting varying-size arrays. Consequently, rather than running Kalman filtering on the proposal LGSSM~\eqref{eq:proposal-lorenz} optimally by alternatively considering independent observations of size $3$ ($u_l$) and $2$ ($y_k$), we have to consider stacked observations $(u_l, y'_l)$ of dimension 5 and treat the $y'_l$ as being missing when $t'_l$ is not part of the $t_k'$. This technical limitation would be removed by considering instead a specialised implementation in a framework allowing for such optimisations.

\subsection{Failure modes}
    \label{subsec:failure-modes}
    In this section, we highlight the different failure modes of both the local (cSMC-based) and global (auxiliary Kalman-based) methods.
To do so, we consider a much simpler model than the ones presented in the previous sections, which is aimed at interpolating between the different regimes in which the methods outperform (or not) each other.

The model is given as a two time-step one-dimensional linear Gaussian state-space model, with a single (unlikely) observation at the second time step. 
The latent dynamics are given by an $\mathcal{N}(0, 1)$ stationary autoregressive process, that is $x_t \sim \mathcal{N}(\rho x_{t-1}, 1 - \rho^2)$, $x_0 \sim \mathcal{N}(0, 1)$, and the observation model $y \sim \mathcal{N}(x_1, r^2)$.
In other terms, the higher the value of $\rho$, the more `sticky' the dynamics are and the more the model is likely to be in a regime where the global Kalman samplers are expected to outperform the local cSMC ones.
On the other hand, the smaller the value of $\rho$, the more separable the dynamics are and the more the model is likely to be in a regime where the local cSMC samplers are expected to outperform the global Kalman ones.
Lowering the observation noise $r$ corresponds to modeling a case where an observation is very unlikely (or equivalently highly informative) relative to the rest of the observations. 
This is a case where the global Kalman samplers are expected to underperform as their scale parameter $\delta$ will shrink to mostly account for this single time step, while the rest of the time series may have required a much higher $\delta$ to achieve good mixing.

\begin{figure}[t]
    \centering
    \includegraphics[width=\columnwidth]{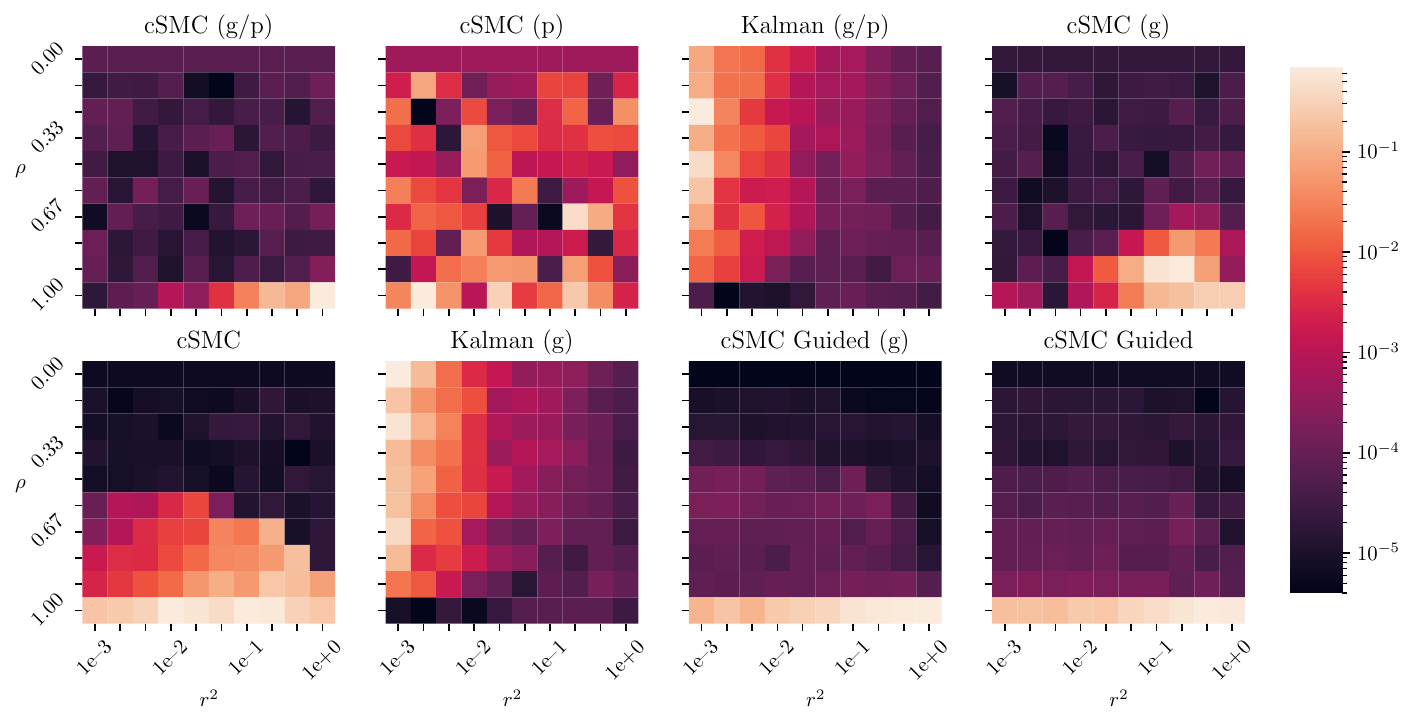}
    \caption{Heatmap of average square error of the estimated mean of the first time step $x_0$, scaled by the true standard deviation for the different samplers as a function of the observation noise $r^2$ and the autocorrelation of the latent dynamics $\rho$.
    Here ``cSMC'' stands for~\citet{finke2021csmc}, with (g) indicating gradient information (Section~\ref{subsubsec:diff}), and (p) indicating parallelisation in time of~\citet{corenflos2022sequentialized}.
    ``Kalman'' stands for the first-order auxiliary Kalman samplers of Section~\ref{sec:auxiliary_samplers}, and
    ``Guided cSMC'' stands for the methods of Sections~\ref{subsubsec:guided} and~\ref{subsubsec:guided-diff} (with (g) indicating gradient information for the latter).
    }
    \label{fig:failure-modes}
\end{figure}

We set $y = 5$ to be a rare observation, and make $\rho$ vary between $0$ and $0.999$ and $r^2$ between $0.001$ and $1$.
For each combination of $\rho$ and $r^2$, we run all the samplers started at stationarity (note that we can do so because the true model is linear Gaussian), run $5\,000$ adaptation steps, and then $8$ times $20\,000$ iterations to compute the empirical mean of the first time step $x_0$. 
The experiment is then repeated $10$ times for each $\rho$ and $r^2$ to account for the randomness in the initialisation of the samplers, after which we compute the mean squared error of the mean approximation computed as
\begin{equation}
    \text{MSE} = \frac{1}{10}\sum_{i=1}^{10}\left(\frac{\hat{x}_{0, i} - m_0}{s_0}\right)^2,
\end{equation}
where $\hat{x}_{0, i}$ is the mean estimator of the $0$-th time step of the latent state for the $i$-th out of ten experiments, and $m_0$, $s_0$ are the true posterior mean and standard deviation of the $0$-th time step.
The results are reported in Figure~\ref{fig:failure-modes}.

As expected, the Kalman samplers collapse when the observation noise $r^2$ is very low and the autocorrelation of the latent dynamics $\rho$ is low too. 
This is because the Kalman samplers adapt their step-size $\delta$ to account for the most informative time step (here the second one, $x_1$), and not the rest of the time series. 
This results in a very slow mixing of the first time step $x_0$.

On the other hand, the cSMC samplers are much more robust to the low correlation setting, as they can adapt their step-sizes $\delta_t$ per time step, and thus mix ``locally as well'' for all time steps.
However, they collapse when the correlation is very high, as proposals at time $0$ do not account for the information at time $1$ aside from the information given by the auxiliary variable $u_0$. 
This is less of an issue for the guided cSMC samplers, as they do account for the correlation between the latent states, but they still collapse when the correlation is extremely high.

Finally, the method of~\citet{finke2021csmc}, i.e. Algorithm~\ref{alg:local-csmc}, and its parallel implementation via~\citet{corenflos2022sequentialized} discussed in Section~\ref{sec:pgibbs_samplers} are not robust to varying degrees of correlation or informativeness. 
Gradient information, however, mitigates this issue, and the gradient-informed cSMC of~\citet{finke2021csmc} and its parallel counterpart are competitive with the guided cSMC and the Kalman samplers in nearly all regimes, despite the fact that they do not \emph{explicitly} account for the correlation between the latent states.

\section{Discussion}\label{sec:conclusion}
In this article, we have presented a principled approach to doing MCMC-based inference in general tractable Feynman--Kac models. At the core, the method corresponds to augmenting the model by introducing an artificial observation model, and then proceeding to sample from the augmented model using a two-step approach: first sample the observations conditionally on a trajectory, and, second, sample from a MCMC kernel keeping the distribution of the trajectory (conditionally on the artificial observations) invariant.

To summarise, we have described two versions of this class of samplers. 
The first one, which we coined auxiliary Kalman sampler can be seen as an extension/specialisation of \citet{titsias2018} to models with latent dynamics, and is particularly useful when the latent model is quasi-Gaussian and of relatively small dimension. 
We believe that this class of samplers opens the door to using the Gaussian approximations developed in the signal processing community for exact inference in state-space models. 
The second class, which considers using conditional SMC to sample the trajectory conditional to the auxiliary observations, can be seen as a generalisation of \citet{finke2021csmc} which allows for more flexibility (and therefore performance) in the design of proposal distributions. 
Importantly, we have shown that both methods introduced could be parallelised across time steps on hardware such as GPUs, while retaining good statistical properties. 
Formally, the sequential and parallel versions of the auxiliary Kalman sampler are fully statistically equivalent, while the particle Gibbs ones are not, but the parallel-in-time auxiliary particle Gibbs does not suffer from severely worse mixing properties, in particular when run time is taken into account.

At least two classes of latent Markovian models elude our auxiliary Kalman samplers:
\begin{enumerate}
    \item Models with multi-modal posteriors, which are hard for MCMC methods in general due to the ``local'' perspective they take. This can, however, be handled by combining the method with meta-algorithms, such as parallel tempering~\citep{geyer1991markov}.
    \item Models with very non-Gaussian latent dynamics or observations, such as those exhibiting multiplicative noise or presenting boundary constraints akin to discontinuities.
\end{enumerate}
Other, softer, issues comprise the following: (i) because Kalman filtering and backward sampling relies on recursive Gaussian conditioning, it requires computing matrices inverses of size $d_X \times d_X$ (or more precisely, solving systems of the same size), and, in models where no specific structure alleviates these computations, they can quickly become computationally overwhelming as the dimension of the latent space increases; (ii) the method is based on a global acceptance step, which means that its performance will naturally degrade as the number of time steps increases, and will be sensitive to a single bad time step, making it somewhat brittle to heterogeneously informative observations.

Replacing the LGSSM proposal of Section~\ref{subsec:auxiliary-lgssm} by a local conditional SMC update as per Section~\ref{sec:pgibbs_samplers} allowed us to trade the single expensive accept-reject step for a series of cheaper local ones.
This solved the brittleness issue, because time steps are considered more independently, and the calibration of the method can happen more locally.
Additionally, the conditional SMC instance of the method naturally inherits the scalability in time of the underlying cSMC algorithm, and, contrary to the auxiliary Kalman sampler, does not require specific treatment to handle increasing numbers of time steps~\citep{finke2021csmc,lee2020coupled,karjalainen2024mixingtime}.
However, the usual issues with cSMC remain: several trajectories need to be simulated, and the fully adapted auxiliary cSMCs of Section~\ref{subsec:guided-smc-sampler} cannot be parallelised-in-time, which we showed to be a significant bottleneck in the case of the spatio-temporal model of Section~\ref{subsec:spatio-temporal}.
They also do not solve the problem of intractable densities, or multimodality.

The reformulation of \citet{finke2021csmc} as a conditional SMC within a Gibbs sampler is a particularly promising avenue as it invites the direct application of the many cSMC practical and theoretical technologies developed over the past decade. 
Our experiments showed that leveraging this representation to design better auxiliary proposal distributions already largely improved the statistical properties of the algorithm at a very low additional computational cost. 
We believe that this can still be improved upon many-fold in a number of settings and a natural first step would be to combine these with methods developed to tackle degeneracy in particle Gibbs~\citep[e.g.][]{lindsten2015degenerate} or very long time series~\citep{Karppinen2022bridge}.

In addition to these, we mention that, since the first version of this article, a follow-up work, \citet{corenflos2024particlemala}, has built upon the guided cSMC perspective to unify conditional SMC and Metropolis adjusted Langevin algorithms~\citep[MALA,][]{besag1994comments} as well as the prior-informed samplers of~\citet{titsias2018} and other related methods.
While the methods of~\citet{corenflos2024particlemala} are not parallelisable, contrary to most of the methods proposed here, they overcome some limitations highlighted in Section~\ref{subsec:failure-modes}, in particular the collapse of cSMC in the highly-informative prior regime.

A final remark is concerned with the implementation of the prefix-sum algorithm~\citet{blelloch1989scans} in the JAX library~\citep{jax2018github}. 
At the time of writing this article, the JAX implementation can be considered high-level, by which we mean that the algorithm is implemented in Python~\citep{Rossum2009Python} rather than natively using the CUDA~\citep{cuda} GPU backend. 
This is in contrast to other control flow primitives such as ``for loops'' and ``if-else'' branching, and a native implementation of the algorithm, fully GPU-focused would improve the time-performance of the Kalman samplers.

\section*{Individual contributions}\label{contribs}
The original idea, methodology, implementation, and redaction of the first version of this article are due to Adrien Corenflos. Simo S\"arkk\"a contributed the divide-and-conquer sampling method and reviewed the final version of the manuscript.

\begin{acks}[Acknowledgments]
  The first author would like to warmly thank Nicolas Chopin for pointing out the link between the first method presented in this article and~\citet{titsias2018}. Some cSMC ideas presented in this article also stemmed from discussions and presentations that took place at the ``Computational methods for unifying multiple statistical analyses'' (Fusion) workshop organised by R\'emi Bardenet, Kerrie Mengersen, Pierre Pudlo, and Christian Robert in Centre International de Rencontres Mathématiques (CIRM) in October 2022.
  Finally, the authors would like to thank two anonymous reviewers for their constructive feedback. 
\end{acks}

\begin{funding}
  Both authors gratefully acknowledge funding from the
  Academy of Finland, project 321891 (ADAFUME), and project 321900 (PARADIST).
  Adrien Corenflos also acknowledges the financial support provided by UKRI for OCEAN (a 2023-2029 ERC Synergy grant co-sponsored by UKRI).
\end{funding}

\bibliographystyle{imsart-number} %
\bibliography{main}       %

\appendix
\section{Sampling and evaluating LGSSM pathwise smoothing distributions}
\label{app:kalman}
In this Section, we describe how the smoothing distribution forming the proposal of the auxiliary Kalman samplers in Section~\ref{subsec:auxiliary-lgssm} can be sampled and evaluated efficiently. 
We first review the ``classical'' sequential Kalman filter and backward sampling algorithms, which are used to sample from the smoothing distribution of a linear Gaussian state-space model (LGSSM) in $\bigO(T)$ steps on sequential hardware.
We then show how these algorithms can be parallelised to run in $\bigO(\log T)$ steps on parallel hardware, either by using prefix-sum algorithms or by divide-and-conquer strategies. 

In this section only, and in contrast with the notations of the main text, we consider that we are given a LGSSM of the form
\begin{equation}\label{eq:lgssm-app}
    \begin{split}
        p_t(x_t \mid x_{t-1}) &= \mathcal{N}(x_t; F_{t-1} x_{t-1} + b_{t-1}, Q_{t-1}), \quad p_0(x_0) = \mathcal{N}(x_0; m_0, P_0),\\
        p_t(y_t \mid x_t) &= \mathcal{N}(y_t; H_t x_t + c_t, R_t),
    \end{split}
\end{equation}
and we want to sample from the smoothing distribution as well as evaluate its likelihood 
\begin{equation}\label{eq:smoothing-likelihood-app}
    p(x_{0:T} \mid y_{0:T}) = \frac{p(x_{0:T}, y_{0:T})}{p(y_{0:T})} = \frac{p(x_{0:T}, y_{0:T})}{\int p(x_{0:T}, y_{0:T}) \dd x_{0:T}}.
\end{equation}
Noting that, without loss of generality, we can assume that $c_t = 0$, we will omit it from the notation in the remainder of this section.

\subsection{Sequential implementations}\label{subsec:kalman-sequential}
The Kalman filter~\citep{kalman1960} and Rauch--Tung--Striebel smoother~\citep{rauch1965maximum} are well-known algorithms to compute the \emph{marginal} filtering and smoothing distributions of a LGSSM in $\bigO(T)$ steps~\citep[see, e.g.,][for a review]{sarkka2023bayesian}.
The Kalman filter computes the filtering distribution $p(x_t \mid y_{0:t}) = \mathcal{N}(x_t; m^f_t, P^f_t)$ recursively for $t=1, \ldots, T$ as
\begin{equation}\label{eq:kalman-filter-app}
    \begin{split}
        m^f_t &= m^p_t + K_t (y_t - H_t m^p_t), \\
        P^f_t &= P^p_{t} - K_t H_t P^p_{t},
    \end{split}
\end{equation}
where $K_t = P^p_{t} H_t^{\top} (H_t P^p_{t} H_t^{\top} + R_t)^{-1}$ is the Kalman gain and 
\begin{equation}
    m^p_t = F_{t-1} m^f_{t-1} + b_{t-1}, \quad P^p_t = F_{t-1} P^f_{t-1} F_{t-1}^{\top} + Q_{t-1}
\end{equation}
are the predicted mean and covariance of the filtering distribution.
The initialisation is done with $m^p_0 = m_0$, $P^p_0 = P_0$ and
\begin{equation}
    m^f_0 = m_0 + K_0 (y_0 - H_0 m_0), \quad P^f_0 = P_0 - K_0 H_0 P_0,
\end{equation}
for $K_0 = P_0 H_0^{\top} (H_0 P_0 H_0^{\top} + R_0)^{-1}$.
Importantly, the marginal likelihood of the observations $y_{0:T}$ can be computed recursively as
\begin{equation}\label{eq:kalman-likelihood-app}
    p(y_{0:T}) = \prod_{t=0}^T \mathcal{N}(y_t; H_t m^f_t, H_t P^f_t H_t^{\top} + R_t).
\end{equation}
Put together, this results in Algorithm~\ref{alg:kalman-filter} for the Kalman filter.

\begin{algorithm}[!htb]
    \DontPrintSemicolon
    \caption{Kalman filter}\label{alg:kalman-filter}
    \KwResult{The filtering means and covariances $m^f_{0:T}$, $P^f_{0:T}$ and the likelihood $p(y_{0:T})$}
    \KwData{The observations $y_{0:T}$ and the LGSSM parameters}
    \Fn{\textsc{KalmanFilter}$\big(y_{0:T}$, $m_0$, $P_0$, $F_{0:T-1}$, $b_{0:T-1}$, $Q_{0:T-1}$, $H_{0:T}$, $R_{0:T}\big)$}{
        Initialise $m^p_0 = m_0$, $P^p_0 = P_0$\;
        Set $K_0 = P_0 H_0^{\top} (H_0 P_0 H_0^{\top} + R_0)^{-1}$\;
        Initialise $m^f_0 = K_0 y_0 + (I - K_0 H_0) m_0$, $P^f_0 = P_0 - K_0 H_0 P_0$\;
        Initialise $p(y_{0:T}) = \mathcal{N}(y_0; H_0 m^f_0, H_0 P^f_0 H_0^{\top} + R_0)$\;
        \For{$t=1, \ldots, T$}{
            Compute $m^p_t = F_{t-1} m^f_{t-1} + b_{t-1}$, $P^p_t = F_{t-1} P^f_{t-1} F_{t-1}^{\top} + Q_{t-1}$\;
            Compute $K_t = P^p_{t} H_t^{\top} (H_t P^p_{t} H_t^{\top} + R_t)^{-1}$\;
            Compute $m^f_t = m^p_t + K_t (y_t - H_t m^p_t)$, $P^f_t = P^p_{t} - K_t H_t P^p_{t}$\;
            Update $p(y_{0:T}) \leftarrow p(y_{0:T}) \times \mathcal{N}(y_t; H_t m^p_t, H_t P^p_t H_t^{\top} + R_t)$\;
        }
        \Return{$m^f_{0:T}$, $P^f_{0:T}$, $p(y_{0:T})$}
    }
\end{algorithm}

Once the marginal likelihood $p(y_{0:T})$ has been computed, it is then easy to evaluate the smoothing distribution $p(x_{0:T} \mid y_{0:T})$ using the identity
\begin{equation}\label{eq:smoothing-app}
    \begin{split}
        p(x_{0:T} \mid y_{0:T}) = \frac{p(x_{0:T}, y_{0:T})}{p(y_{0:T})}
    \end{split}
\end{equation}
noting that the numerator can be computed as the product
\begin{equation}\label{eq:smoothing-terms-app}
    \begin{split}
        p(x_{0:T}, y_{0:T}) 
            &= p_0(x_0) \left\{\prod_{t=1}^T p(x_t \mid x_{t-1})\right\} \prod_{t=0}^T p(y_t \mid x_t)\\
            &= \mathcal{N}(x_0; m_0, P_0) \left\{\prod_{t=1}^T \mathcal{N}(x_t; F_{t-1} x_{t-1} + b_{t-1}, Q_{t-1})\right\} \prod_{t=0}^T \mathcal{N}(y_t; H_t x_t, R_t),
    \end{split}
\end{equation}
or more numerically stably with the sum of the logarithms of the terms in~\eqref{eq:smoothing-terms-app} and similarly for the denominator $p(y_{0:T})$.

The backward sampler then proceeds recursively to obtain a sample from the smoothing distribution $p(x_{0:T} \mid y_{0:T})$ in $\bigO(T)$ steps via the recursive identity
\begin{equation}\label{eq:general-smoother-app}
    \begin{split}
        p(x_{t-1} \mid y_{0:T}, x_{t+1:T}) = p(x_{t-1} \mid y_{0:t}, x_{t+1}),
    \end{split}
\end{equation}
which, for LGSSMs is given as a conditional Gaussian distribution
\begin{equation}\label{eq:rauch-tung-striebel-app}
    \begin{split}
        p(x_{t-1} \mid y_{0:t}, x_{t+1}) = \mathcal{N}\left(x_t; m^f_t + G_{t} (x_{t+1} - m^p_{t+1}), P^f_t - G_t P^p_{t+1} G_t^{\top}\right), \quad t < T,
    \end{split}
\end{equation}
and with initialisation $p(x_T \mid y_{0:T}) = \mathcal{N}(x_T; m^f_T, P^f_T)$.
The backward sampler is summarised in Algorithm~\ref{alg:rauch-tung-striebel} which we refer to as the Rauch--Tung--Striebel \emph{sampler} due to its similarity with the Rauch--Tung--Striebel smoother~\citep{rauch1965maximum}.
Other approaches to sampling from the smoothing distribution of a LGSSM exist~\citep[see, e.g.,][]{Doucet:2010}, but do not necessarily improve the computational complexity and are not detailed here.

\begin{algorithm}[!htb]
    \DontPrintSemicolon
    \caption{Rauch--Tung--Striebel sampler}\label{alg:rauch-tung-striebel}
    \KwResult{A sample from the smoothing distribution $p(x_{0:T} \mid y_{0:T})$}
    \Fn{\textsc{SequentialSampler}$\big(m^f_{0:T}$, $P^f_{0:T}$, $F_{0:T-1}$, $b_{0:T-1}$, $Q_{0:T-1}$, $H_{0:T}$, $R_{0:T}$, $y_{0:T}\big)$}{
        Initialise $x_T \sim \mathcal{N}(m^f_T, P^f_T)$\;
        \For{$t=T-1, \ldots, 0$}{
            Compute $G_t = P^f_t F_t^{\top}\left(F_t P^f_t F_t^{\top} + Q_t\right)^{-1}$\;
            Sample $x_t \sim \mathcal{N}\left(m^f_t + G_{t} (x_{t+1} - m^p_{t+1}), P^f_t - G_t P^p_{t+1} G_t^{\top}\right)$\;
        }
        \Return{$x_{0:T}$}
    }
\end{algorithm}

\subsection{Parallel implementations}\label{subsec:kalman-parallel}
We now turn to two different strategies to parallelise the Kalman filter and Rauch--Tung--Striebel sampler algorithms on parallel hardware.
The first strategy is to use prefix-sum algorithms~\citep{blelloch1989scans} to parallelise the backward sampler, while the second strategy is to use divide-and-conquer strategies.
Because both these rely on having pre-computed the filtering means and covariances $m^f_{0:T}$, $P^f_{0:T}$, we first describe how to parallelise the Kalman filter algorithm in $\bigO(\log T)$ steps following the methods of~\citet{Sarkka2021temporal}.

\subsubsection{Prefix-sums and the parallel Kalman filter}\label{subsubsec:prefix-sums}
Prefix-sum algorithms~\citep{blelloch1989scans} are a class of parallel algorithms that can be used to compute the \emph{cumulative} composition $e_1 \circ \ldots \circ e_t$, $t=1, \ldots, T$ of a sequence of $T$ elements in $\bigO(\log T)$ steps on parallel hardware.
It relies on the associative property of the operator $\circ$, whereby we have
\begin{equation}\label{eq:prefix-sum}
    \begin{split}
        (e_1 \circ e_2) \circ e_3 = e_1 \circ (e_2 \circ e_3).
    \end{split}
\end{equation}
A typical example is when the $e_t$'s are scalars and the operator $\circ$ is the addition, in which case the prefix-sum of the sequence $e_t$ is the cumulative sum
$s_t = e_1 + \cdots + e_t$, $t=1, \ldots, T$ of the sequence.
Several parallel implementations of prefix-sums are available, with different memory/parallelisation properties.
In Algorithm~\ref{alg:prefix-sums} we illustrate the simplest such algorithm, known as the Hillis--Steele scan~\citep{hillis1986data}.
A visual representation of the algorithm is also given in Figure~\ref{fig:prefix-sums}.
As can be seen, the algorithm performs $\lfloor{\log_2 T}\rfloor$ iterations, each of which requires (at most) $T$ operations but which are embarrassingly parallel, and thus the full algorithm runs in $\bigO(\log T)$ steps on parallel hardware provided enough parallel resources are available.
In practice, more efficient implementations exist, such as the work-efficient scan~\citep{blelloch1989scans} but we do not detail them here.

\begin{algorithm}
    \caption{Hillis-Steele algorithm.}\label{alg:prefix-sums}
    \DontPrintSemicolon
    \KwResult{Prefix-sums $e_1 \circ \ldots \circ e_t$, for $t=1, \dots, T$.}
    \Fn{\textsc{PrefixSum}$\big(e_1, \ldots, e_T\big)$}{
    \For{$d \gets 0$ \KwTo $\lfloor{\log_2 T}\rfloor$}
    {
        \For(\tcp*[f]{in parallel}){$t \gets T-1$ \KwTo $0$}
        {
            \If{$t - 2^d \geq 0$}
            {
                $e_t \gets e_{t - 2^d} \circ e_t$
            }
        }
    }
    \Return{$e_1, \ldots, e_T$}
    }
\end{algorithm}
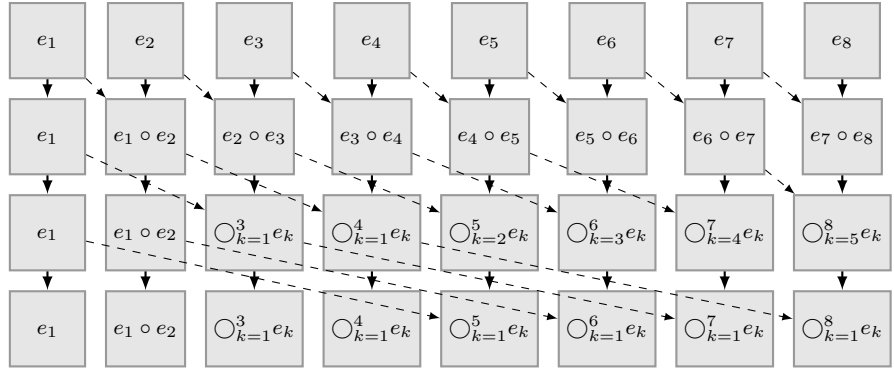
\begin{figure}
    \centering
    \tikzstyle{matrx}=[rectangle,
thick,
minimum size=1cm,
draw=gray!80,
fill=gray!20]

\begin{tikzpicture}[>=latex,text height=1.3ex,text depth=0.25ex]
    \matrix[row sep=0.25cm,column sep=0.25cm] {
        \node (e_1) [matrx]{$e_1$}; &
        \node (e_2) [matrx]{$e_2$}; &
        \node (e_3) [matrx]{$e_3$}; &
        \node (e_4) [matrx]{$e_4$}; &
        \node (e_5) [matrx]{$e_5$}; &
        \node (e_6) [matrx]{$e_6$}; &
        \node (e_7) [matrx]{$e_7$}; &
        \node (e_8) [matrx]{$e_8$};
        \\
        \node (e_11) [matrx]{$e_1$}; &
        \node (e_12) [matrx]{$e_1 \circ e_2$}; &
        \node (e_23) [matrx]{$e_2 \circ e_3$}; &
        \node (e_34) [matrx]{$e_3 \circ e_4$}; &
        \node (e_45) [matrx]{$e_4 \circ e_5$}; &
        \node (e_56) [matrx]{$e_5 \circ e_6$}; &
        \node (e_67) [matrx]{$e_6 \circ e_7$}; &
        \node (e_78) [matrx]{$e_7 \circ e_8$};
        \\
        \node (e_111) [matrx]{$e_1$}; &
        \node (e_112) [matrx]{$e_1 \circ e_2$}; &
        \node (e_123) [matrx]{$\bigcirc_{k=1}^3 e_k$}; &
        \node (e_1234) [matrx]{$\bigcirc_{k=1}^4 e_k$}; &
        \node (e_2345) [matrx]{$\bigcirc_{k=2}^5 e_k$}; &
        \node (e_3456) [matrx]{$\bigcirc_{k=3}^6 e_k$}; &
        \node (e_4567) [matrx]{$\bigcirc_{k=4}^7 e_k$}; &
        \node (e_5678) [matrx]{$\bigcirc_{k=5}^8 e_k$};
        \\
        \node (e_1111) [matrx]{$e_1$}; &
        \node (e_1112) [matrx]{$e_1 \circ e_2$}; &
        \node (e_1233) [matrx]{$\bigcirc_{k=1}^3 e_k$}; &
        \node (e_12344) [matrx]{$\bigcirc_{k=1}^4 e_k$}; &
        \node (e_12345) [matrx]{$\bigcirc_{k=1}^5 e_k$}; &
        \node (e_123456) [matrx]{$\bigcirc_{k=1}^6 e_k$}; &
        \node (e_1234567) [matrx]{$\bigcirc_{k=1}^7 e_k$}; &
        \node (e_12345678) [matrx]{$\bigcirc_{k=1}^8 e_k$};
        \\
    };
    \path[->]
    (e_1) edge[thick] (e_11)
    (e_1) edge[dashed] (e_12)
    (e_2) edge[thick] (e_12)
    (e_2) edge[dashed] (e_23)
    (e_3) edge[thick] (e_23)
    (e_3) edge[dashed] (e_34)
    (e_4) edge[thick] (e_34)
    (e_4) edge[dashed] (e_45)
    (e_5) edge[thick] (e_45)
    (e_5) edge[dashed] (e_56)
    (e_6) edge[thick] (e_56)
    (e_6) edge[dashed] (e_67)
    (e_7) edge[thick] (e_67)
    (e_7) edge[dashed] (e_78)
    (e_8) edge[thick] (e_78)
    (e_11) edge[thick] (e_111)
    (e_11) edge[dashed] (e_123)
    (e_12) edge[thick] (e_112)
    (e_12) edge[dashed] (e_1234)
    (e_23) edge[thick] (e_123)
    (e_23) edge[dashed] (e_2345)
    (e_34) edge[thick] (e_1234)
    (e_34) edge[dashed] (e_3456)
    (e_45) edge[thick] (e_2345)
    (e_45) edge[dashed] (e_4567)
    (e_56) edge[thick] (e_3456)
    (e_67) edge[thick] (e_4567)
    (e_67) edge[dashed] (e_5678)
    (e_78) edge[thick] (e_5678)
    (e_111) edge[thick] (e_1111)
    (e_112) edge[thick] (e_1112)
    (e_123) edge[thick] (e_1233)
    (e_1234) edge[thick] (e_12344)
    (e_111) edge[dashed] (e_12345)
    (e_2345) edge[thick] (e_12345)
    (e_112) edge[dashed] (e_123456)
    (e_3456) edge[thick] (e_123456)
    (e_123) edge[dashed] (e_1234567)
    (e_4567) edge[thick] (e_1234567)
    (e_1234) edge[dashed] (e_12345678)
    (e_5678) edge[thick] (e_12345678)
    ;
\end{tikzpicture}
    \caption{Illustration of the Hillis--Steele prefix-sum algorithm. The algorithm performs $\lfloor{\log_2 T}\rfloor$ iterations, each of which using operations which are embarrassingly parallel.}
    \label{fig:prefix-sums}
\end{figure}

In order to apply the prefix-sum algorithm to the Kalman filter, we now need to express the filtering means and covariances $m^f_{0:T}$, $P^f_{0:T}$ as the result of a prefix-sum operation for elements $e_t$ and operator $\circ$ to be defined.
This is done in~\citet{Sarkka2021temporal} by remarking that the Bayesian filtering recursion (for both the state and the marginal likelihood) can be written as
\begin{equation}\label{eq:kalman-filter-recursion}
    \begin{split}
        p(x_t \mid y_{0:t}) &= \frac{\int p(y_t \mid x_{t-1}) p(x_t \mid y_t, x_{t-1}) p(x_{t-1} \mid y_{0:t-1}) \dd x_{t-1}}{\int p(y_t \mid x_{t-1}) p(x_{t-1} \mid y_{0:t-1}) \dd x_{t-1}},\\
        p(y_{0:t}) &= p(y_{0:t-1}) \int p(y_t \mid x_{t-1}) p(x_{t-1} \mid y_{0:t-1}) \dd x_{t-1},
    \end{split}
\end{equation}
noting that $p(y_t \mid y_{0:t-1}) = \int p(y_t \mid x_{t-1}) p(x_{t-1} \mid y_{0:t-1}) \dd x_{t-1}$.
As a consequence, the elements $e_t$ can be identified as the pairs of (conditional) distributions $[p(x_t \mid y_t, x_{t-1}),  p(y_t \mid x_{t-1})]^\top$ appearing in~\eqref{eq:kalman-filter-recursion}, and the operator $\circ$ as the integration corresponding to
\begin{equation}\label{eq:kalman-filter-op}
    \begin{split}
        \begin{bmatrix}
            p(x_t \mid y_t, x_{t-1})\\
            p(y_t \mid x_{t-1})
        \end{bmatrix} \circ \begin{bmatrix}
            p(x_{t-1} \mid y_{0:t-1})\\
            p(y_{0:t-1})
        \end{bmatrix} = \begin{bmatrix}
            p(x_t \mid y_{0:t})\\
            p(y_{0:t})
        \end{bmatrix}.
    \end{split}
\end{equation}
Thankfully, both the element pairs and the operator can be computed\footnote{In practice, the marginal likelihood $p(y_{0:t})$ is obtained up to a multiplicative constant that may depend on the parameters of the LGSSM, and one therefore needs to perform a second step to compute it using~\eqref{eq:kalman-likelihood-app}.} for Gaussian LGSSMs~\citep[for details on their expressions, see][]{Sarkka2021temporal}, and the prefix-sum algorithm can be applied to compute the filtering means and covariances $m^f_{0:T}$, $P^f_{0:T}$ and $p(y_{0:t})$ in $\bigO(\log T)$ steps on parallel hardware.

\subsubsection{Parallel Rauch--Tung--Striebel sampler}\label{subsubsec:prefix-sum-sampling}
Now that we have the filtering means and covariances $m^f_{0:T}$, $P^f_{0:T}$, we can modify Algorithm~\ref{alg:rauch-tung-striebel} to use the prefix-sum algorithm to sample from the smoothing distribution $p(x_{0:T} \mid y_{0:T})$ in $\bigO(\log T)$ parallel steps.
Indeed, we know~\citep[][Proposition 1]{Fruhwirth1994data} that
\begin{equation}\label{eq:backward}
    \begin{split}
        p(x_T \mid y_{0:T}) &= \mathcal{N}(x_T; m^f_T, P^f_T), \\
        p(x_t \mid x_{t+1}, y_{0:t}) &= \mathcal{N}\left(x_t; m^f_t + G_{t} [x_{t+1} - F_t m^f_t - b_t], \Sigma_t\right), \quad t < T,
    \end{split}
\end{equation}
where $G_t = P^f_t F_t^{\top}\left(F_t P^f_t F_t^{\top} + Q_t\right)^{-1}$ and $\Sigma_t = P^f_t - G_t (F_t P^f_t F_t^{\top} + Q_t) G_t^{\top}$ for all $t < T$.
    
We can furthermore rearrange the terms to express $\hat{X}_T \sim \mathcal{N}(m^f_T, P^f_T)$ and $\hat{X}_t \sim p(x_t \mid \hat{X}_{t+1}, y_{0:t})$ recursively as $\hat{X}_t = G_{t} \hat{X}_{t+1} + U_t$,
where the $U_t$'s are independently distributed as Gaussians $\mathcal{N}(m^f_t - G_t (F_t m^f_t + b_t), \Sigma_t)$ for all $t < T$. We also let $G_T = 0$, so that we can then define $U_T \sim \mathcal{N}(m^f_T, P^f_T)$ to be a sample of the final marginal smoothing distribution. 
Because the means and covariances of the $U_t$'s only depend on the LGSSM coefficients and the filtering means and covariances at time $t$, they can be sampled fully in parallel. To sample from $p(x_{0:T} \mid y_{0:T})$ we then need to apply the recursion to the pre-sampled sequence $U_t$, $t=0, \ldots, T$. However, the recursive dependency in~\eqref{eq:backward} is not directly parallelisable, and we instead need to rephrase it in terms of an associative operator, which will allow us to use prefix-sum primitives~\citep{blelloch1989scans}. 
Thankfully, this is readily done by considering the elements $e_t = \begin{bmatrix}
    G_t &U_t
\end{bmatrix}^{\top}$ and the operator $\circ$ defined as follows
\begin{equation}\label{eq:sampling-op-app}
    \begin{split}
        (G_{ij}, U_{ij}) &= (G_i, U_i) \circ (G_j, U_j), \quad \text{where } G_{ij} = G_i G_j, \, \text{and } U_{ij} = G_i U_j + U_i.
    \end{split}
\end{equation}
\begin{proposition}\label{prop:prefix-sum-sampling-app}
    The backward prefix-sum of operator $\circ$ applied to the sequence $(G_t, U_t)$, $t=0, \ldots, T$, recovers the pathwise smoothing distribution $p(x_{0:T} \mid y_{0:T})$, that is, if $(\tilde{G}_t, \tilde{U}_t) = (G_t, U_t) \circ \ldots \circ (G_T, U_T)$, then $(\tilde{U}_0, \ldots, \tilde{U}_T)$ is distributed according to $p(x_{0:T} \mid y_{0:T})$.
\end{proposition}
\begin{proof}
    The operator $\circ$ defined in~\eqref{eq:sampling-op} is clearly associative. We prove that its result corresponds to sampling from the pathwise smoothing distribution by reversed induction: suppose that $(\tilde{U}_t, \ldots, \tilde{U}_T)$ is distributed according to $p(x_{t:T} \mid y_{0:T})$, then $\tilde{U}_{t-1} = G_{t-1} \tilde{U}_t + U_{t-1}$, which is distributed according to $p(x_{t-1} \mid \tilde{U}_{t}, y_{0:t-1})$ as discussed before, so that $(\tilde{U}_{t-1}, \ldots, \tilde{U}_T)$ is distributed according to $p(x_{t-1:T} \mid y_{0:T})$. The initial case follows from the definition of $U_T$.
\end{proof}

To summarise, in order to perform prefix-sum sampling of LGSSMs, it suffices to use the parallel-in-time Kalman filtering method of \citet{Sarkka2021temporal} to compute the filtering means and covariances $m^f_t$, $P^f_t$, $t=0, \ldots, T$, then form all the elements $G_t$ and sample $U_t$ fully in parallel, and finally, apply the prefix-sum primitive~\citep{blelloch1989scans} to $(G_t, U_t)_{t=0}^T$ with the associative operator $\circ$.
The parallel implementation of the Rauch--Tung--Striebel sampler is then given in Algorithm~\ref{alg:rauch-tung-striebel-parallel}.
\begin{algorithm}[!htb]
    \DontPrintSemicolon
    \caption{Parallel Rauch--Tung--Striebel sampler}\label{alg:rauch-tung-striebel-parallel}
    \KwResult{A sample from the smoothing distribution $p(x_{0:T} \mid y_{0:T})$}
    \Fn{\textsc{ParallelSampler}$\big(m^f_{0:T}$, $P^f_{0:T}$, $F_{0:T-1}$, $b_{0:T-1}$, $Q_{0:T-1}$, $H_{0:T}$, $R_{0:T}$, $y_{0:T}\big)$}{
        Initialise $U_T \sim \mathcal{N}(m^f_T, P^f_T)$\;
        \For(\tcp*[f]{in parallel}){$t=T-1, \ldots, 0$}{
            Compute $G_t = P^f_t F_t^{\top}\left(F_t P^f_t F_t^{\top} + Q_t\right)^{-1}$\;
            Sample $U_t \sim \mathcal{N}\left(m^f_t - G_t (F_t m^f_t + b_t), P^f_t - G_t (F_t P^f_t F_t^{\top} + Q_t) G_t^{\top}\right)$\;
        }
        Apply the prefix-sum algorithm to $(G_t, U_t)_{t=T}^0$ with the operator $\circ$\;
        \Return{$U_{0:T}$}
    }
\end{algorithm}

\subsubsection{Divide-and-conquer strategies}\label{subsubsec:divide-and-conquer}
An alternative strategy to parallelise the Kalman filter and Rauch--Tung--Striebel sampler algorithms is to use divide-and-conquer strategies.
Again, we assume that the filtering means and covariances $m^f_{0:T}$, $P^f_{0:T}$ have been computed using the parallel-in-time Kalman filter of~\citet{Sarkka2021temporal} or similar methods.

We now present a divide-and-conquer alternative to Section~\ref{subsubsec:prefix-sum-sampling} for PIT sampling from the pathwise smoothing distribution of LGSSMs. The method is based on recursively finding tractable Gaussian expressions for the ``bridging'' $p(x_l \mid y_{0:T}, x_k, x_m)$, $0 \leq k < l < m \leq T$ of the smoothing distribution. This will allow us to derive a tree-based divide-and-conquer sampling mechanism for the pathwise smoothing distribution $p(x_{0:T} \mid y_{0:T})$.

Suppose we are given the LGSSM~\eqref{eq:lgssm-app}, then given three indices $0 \leq k < l < m \leq T$. We have
\begin{equation}
\begin{split}
  p(x_l \mid y_{0:T}, x_k, x_m)
  = \frac{p(x_k, x_l \mid y_{0:T}, x_m)}
    {p(x_k \mid y_{0:T}, x_m)}
\end{split}
\end{equation}
with, furthermore, 
\begin{equation}
    p(x_k, x_l \mid y_{0:T}, x_m) = p(x_k \mid y_{0:T}, x_l) p(x_l \mid y_{0:T}, x_m)
\end{equation}
thanks the to Markovian structure of the model.
Now let $p(x_k \mid y_{0:T}, x_l)$ and $p(x_l \mid y_{0:T}, x_m)$ be given by 
\begin{equation}
    \begin{split}
        p(x_k \mid y_{0:T}, x_l) &= \mathcal{N}(x_k; E_{k:l} x_l + g_{k:l}, L_{k:l})\\
        p(x_l \mid y_{0:T}, x_m) &= \mathcal{N}(x_k; E_{l:m} x_m + g_{l:m}, L_{l:m})
    \end{split}
\end{equation}
for some parameters $E_{k:l}$, $g_{k:l}$, $L_{k:l}$, $E_{l:m}$, $g_{l:m}$, and $L_{l:m}$ that we will define below. Then we can write 
\begin{equation}
\begin{split}
  &p(x_k, x_l \mid y_{0:T}, x_m)\\
  &= \mathcal{N}\left( \begin{pmatrix} x_l \\ x_k \end{pmatrix} ; 
      \begin{pmatrix}
        E_{l:m} x_{m} + g_{l:m} \\
       E_{k:l} E_{l:m} x_{m} + E_{k:l} g_{l:m} + g_{k:l}
       \end{pmatrix}, 
      \begin{pmatrix}
        L_{l:m}     &      L_{l:m} E_{k:l}^\top \\
       E_{k:l} L_{l:m}  & E_{k:l} L_{l:m} E_{k:l}^\top + L_{k:l}
       \end{pmatrix} \right)
\end{split}
\end{equation}
giving both the marginal distribution of $x_k$
\begin{equation}
\begin{split}
  p(x_k \mid y_{0:T}, x_m)
  &= \mathcal{N}(x_k; 
      E_{k:l} E_{l:m} x_{m} + E_{k:l} g_{l:m} + g_{k:l}, 
      E_{k:l} L_{l:m} E_{k:l}^\top + L_{k:l} ) \\
  &= \mathcal{N}( x_k;  E_{k:m} x_{m} + g_{k:m}, L_{k:m} ),
\end{split}
\end{equation}
where
\begin{equation}
\begin{split}
    E_{k:m} = E_{k:l} E_{l:m}, \quad g_{k:m} = E_{k:l} g_{l:m} + g_{k:l}, \quad
    L_{k:m} = E_{k:l} L_{l:m} E_{k:l}^\top + L_{k:l},
\end{split}
\label{eq:EgL_comb}
\end{equation}
and (after simplification for~\eqref{eq:EgL_comb}) the conditional distribution of $x_l$
\begin{equation}
\begin{split}
  p(x_l \mid y_{0:T}, x_k, x_m) = \mathcal{N}(x_l; G_{k:l:m} x_k + \Gamma_{k:l:m} x_m + w_{k:l:m}, V_{k:l:m} ),
\end{split}
\label{eq:gauss_l_km}
\end{equation}
for 
\begin{equation}\label{eq:recursive_params}
\begin{split}
    G_{k:l:m} &= L_{l:m} E_{k:l}^\top L_{k:m}^{-1}, \\
    \Gamma_{k:l:m} &= E_{l:m} - G_{k:l:m} E_{k:m}, 
\end{split} \qquad
\begin{split}
    w_{k:l:m} &= g_{l:m} - G_{k:l:m} g_{k:m}, \\
    V_{k:l:m} &= L_{l:m} - G_{k:l:m} L_{k:m} G_{k:l:m}^\top.
\end{split}
\end{equation}

This construction provides a recursive tree structure for sampling from $p(x_{0:T} \mid y_{0:T})$ which can be initialised by 
\begin{equation}
\begin{split}
  p(x_t \mid y_{0:T}, x_{t+1})
  &= \mathcal{N}(x_t; 
      E_{t:t+1} x_{t+1} + g_{t:t+1}, L_{t:t+1} ),
\end{split}
\end{equation}
with
\begin{equation}\label{eq:init-dnc}
\begin{split}
  E_{t:t+1} = P^f_t F_t^\top (F_t P^f_t F_t^\top + Q_t)^{-1}, \quad
  g_{t:t+1} = m^f_t - E_{t:t+1} (F_t m^f_t + b_t), \quad
  L_{t:t+1} = P^f_t - E_{t:t+1} F_t P^f_t,
\end{split}
\end{equation}
and $p(x_T \mid y_{0:T}) = \mathcal{N}(x_T; m^f_T, P^f_T)$. Finally, noting that
\begin{equation}
    p(x_0 \mid y_{0:T}, x_T) = \mathcal{N}(x_0; E_{0:T} m^f_T + g_{0:T}, L_{0:T}),
\end{equation} 
we can combine these identities to form a divide-and-conquer algorithm.

To summarise, in order to perform divide-and-conquer sampling of LGSSMs, it suffices, as in Section~\ref{subsubsec:prefix-sum-sampling}, to use the parallel-in-time Kalman filtering method of \citet{Sarkka2021temporal} to compute the filtering means and covariances $m^f_t$, $P^f_t$, $t=0, \ldots, T$. After this, we can recursively compute the tree of elements $E_{k:m}, g_{k:m}, L_{k:m}$, together with the auxiliary variables $G_{k:l:m}, w_{k:l:m}, \Gamma_{k:l:m}, V_{k:l:m}$, starting from $E_{t:t+1}, g_{t:t+1}, L_{t:t+1}$, for $t=0, 1, \ldots, T-1$, then $E_{t-1:t+1}, g_{t-1:t+1}, L_{t-1:t+1}$, for $t=1, 3, 5, \ldots, 2\lfloor (T - 1)/2\rfloor + 1$, etc. Once this has been done, we can then sample from $p(x_T \mid y_{0:T})$, then from $p(x_0 \mid y_{0:T}, x_T)$, then $x_{\lfloor T/2 \rfloor}$ conditionally on $x_0$ and $x_T$, then, in parallel $x_{\lfloor T/4 \rfloor}$ and $x_{\lfloor 3 T/4 \rfloor}$, conditionally on the rest, and continue until all have been sampled.

\newpage
\section{Generalised statistical linear regression}
\label{app:gslr}
We now describe how to linearise state-space models arising in Section~\ref{sec:auxiliary_samplers} using the generalised statistical linear regression (GSLR) framework of \citet{Garcia:2017,Tronarp2018iterative}, which requires the existence of the first two conditional moments $\mathbb{E}[X_t \mid X_{t-1}]$ and $\mathbb{V}[X_t \mid X_{t-1}]$ of the transition model $p_t$.
This approach comprises, as a special case, the extended and unscented linearisation methods of~\citet{Jazwinski:1970,julier2004unscented}.
For the sake of completeness, we also describe how to handle the potential $g(x_{0:T})$, when it is given as a product of observation models $h_t(y_t \mid x_t)$, in the same framework.

Following \citet{Tronarp2018iterative}, we suppose that the first two conditional moments 
\begin{align}\label{eq:conditional-moments}
    m^X(x_{t-1}) &\coloneqq \int x_t p_t(x_t \mid x_{t-1}) \dd x_t,\\
    V^X(x_{t-1}) &\coloneqq \int (x_t - m^X(x_{t-1})) (x_t - m^X(x_{t-1}))^\top p_t(x_t \mid x_{t-1}) \dd x_t,
\end{align}
and 
\begin{align}
    m^Y(x_{t}) &\coloneqq \int y_t h_t(y_t \mid x_t) \dd y_t,\\
    V^Y(x_{t}) &\coloneqq \int (y_t - m^Y(x_{t})) (y_t - m^Y(x_{t}))^\top h_t(y_t \mid x_t) \dd y_t,
\end{align}
of, respectively, the transitions and observation models appearing in~\eqref{eq:ssm} can easily be either computed in closed form, or approximated well enough. Similarly, we suppose that the two first moments $m_0$ and $P_0$ of $p_0$ are known at least approximately.
As described in Section~\ref{subsec:auxiliary-general}, in order to form a proposal distribution $q(x_{0:T} \mid u_{0:T}, y_{0:T})$ for $p(\ft{x}{0}{T} \mid \ft{y}{0}{T}, \ft{u}{0}{T})$, we linearise the state-space model~\eqref{eq:ssm} around the trajectory at hand.
Let $x_{0:T} \in \bbR^{T \times d_x}$ be the current states of the auxiliary Markov chain, and let $\Gamma_{0:T}$ be a set of reference covariance matrices in $\bbR^{T \times d_x \times d_x}$, by which we mean that $\Gamma_t \in \bbR^{d_x \times d_x}$ needs to be positive definite for all $t$. We can apply the generalised statistical linear regression (GSLR) framework of \citet{Tronarp2018iterative} for the reference random variables $\zeta_t \sim \mathcal{N}(x_t, \Gamma_t)$, $t=0, \ldots, T$ to derive Gaussian approximations of the transition and observation models as follows:
\begin{equation}\label{eq:general_approx}
    \begin{split}
        p_{t}(z_t \mid z_{t-1}) 
            &\approx \mathcal{N}(z_t ; F_{t-1} z_{t-1} + b_{t-1}, Q_{t-1}), \\
        h_{t}(y_t \mid z_{t})
            &\approx \mathcal{N}(y_t ; H_{t} z_{t} + c_{t}, R_{t}),
    \end{split}
\end{equation}
with,
\begin{align}\label{eq:coeffs}
    \begin{split}
        F_{t-1} &= C^X_{t-1} \Gamma_{t-1}^{-1},\\
        b_{t-1} &= \mu^X_{t-1} - F_{t-1} x_{t-1},\\
        Q_{t-1} &= S^X_{t-1} - F_{t-1} \Gamma_{t-1} F_{t-1}^{\top},
    \end{split}
    &
    \begin{split}
        H_{t} &= C^Y_{t} \Gamma_{t}^{-1},\\
        c_{t} &= \mu^Y_{t} - H_{t} x_{t},\\
        R_{t} &= S^Y_{t} - H_{t} \Gamma_{t} H_{t}^{\top},
    \end{split}
\end{align}
and where, for the sake of readability, we do not notationally emphasise the dependency on $x$ and $\Gamma$. These Gaussian approximations are known to minimise a forward KL divergence with respect to the transition and observation models for the Gaussian variational family. The coefficients appearing in~\eqref{eq:coeffs} are in turn given by the general formulae
\begin{align}\label{eq:general-GSLR-coeffs}
    \begin{split}
        C^X_{t-1} &= \mathbb{C}\left[m^X(\zeta_{t-1}), \zeta_{t-1}\right], \\
        \mu^X_{t-1} &= \mathbb{E}\left[m^X(\zeta_{t-1})\right], \\
        S^X_{t-1} &= \mathbb{E}\left[V^X(\zeta_{t-1})\right] + \mathbb{V}\left[m^X(\zeta_{t-1})\right],\\
    \end{split}
    &
    \begin{split}
        C^Y_t &= \mathbb{C}\left[m^Y(\zeta_{t}), \zeta_{t}\right], \\
        \mu^Y_t &= \mathbb{E}\left[m^Y(\zeta_{t})\right], \\
        S^Y_t &= \mathbb{E}\left[V^Y(\zeta_{t})\right] + \mathbb{V}\left[m^Y(\zeta_{t})\right].
    \end{split}
\end{align}
Clearly, the quantities in~\eqref{eq:general-GSLR-coeffs} are not typically available in closed-form, and we instead need to resort to further approximations. Such approximations are given by, for example, Taylor series expansions or sigma-point methods, such as Gauss--Hermite or unscented methods~\citep[see, e.g.,][Ch. 5]{sarkka2023bayesian}.

\newpage
\section{Backward sampling and parallel-in-time particle Gibbs}
\label{app:pGibbs}
For the sake of completeness, in this Section, we present details of the backward sampling method of~\citet{whiteley2010discussion}, to be used instead of the genealogy selection step in Algorithm~\ref{alg:csmc} for improved mixing.
Additionally, we present the parallel-in-time particle Gibbs algorithm~\citep[Section 3]{corenflos2022sequentialized}, specialised to the method of~\citet[][see also Algorithm~\ref{alg:local-csmc}]{finke2021csmc}, in its auxiliary form presented in Section~\ref{subsec:smc-samplers}. 
The description extends to the gradient-informed proposals of Section~\ref{subsubsec:diff} almost \emph{verbatim} by corresponding a different auxiliary proposal mechanism.

\subsection{Backward sampling}\label{subsec:backward-sampling}
As discussed in Section~\ref{subsec:smc-samplers}, the genealogy selection step of Algorithm~\ref{alg:csmc}, lines~\ref{line:genealogy} and beyond, can be replaced by a backward sampling step~\citep{whiteley2010discussion} to improve mixing.
This modified version of Algorithm~\ref{alg:csmc} is given in Algorithm~\ref{alg:local-csmc-backward} and is only implementable provided the quantity $p_t(x_t \mid x_{t-1}) g_t(x_t, x_{t-1})$ can be evaluated pointwise.
While other techniques exist when this is not the case~\citep{Dau2022complexity}, we focus on this method given all our examples verify this assumption.
The algorithm is given in Algorithm~\ref{alg:local-csmc-backward}.
\begin{algorithm}[!htb]
    \SetAlgoLined
    \DontPrintSemicolon
    \caption{Conditional SMC}\label{alg:local-csmc-backward}
    \KwResult{An updated trajectory $z_{0:T}$}
    \Fn{\textsc{cSMC}$\big(x_{0:T}$, $N\big)$}{
        \tcp{Forward propagation: same as Algorithm~\ref{alg:csmc}}
        \tcp{Genealogy selection}\label{line:backward-sampling}
        Sample $B_T$ with $\mathbb{P}(B^n_T = k) \propto w^k_{T}$ and set $z_T = X^{B_T}_T$\;
        \For{$t=T-1,\ldots, 0$}{
            \For{$n=1, \ldots, N$}{
                Compute $\bar{w}^n = w_{t}^n g_{t+1}(z_{t+1}, X^n_t) p(z_{t+1} \mid X^n_t)$\;
            }
            Sample $B_t$ with $\mathbb{P}(B^n_t = k) \propto \bar{w}^k$ and set $z_t = X^{B_t}_t$\;        
        }
        \Return{$z_{0:T}$}
    }
    \end{algorithm}
    The algorithm can further be augmented to use different acceptance probabilities for the backward sampling step, as in~\citet{chopin2015particlegibbssampling}, but we do not consider this here.
    Contrary to simple genealogy tracing, as implemented in Algorithm~\ref{alg:csmc}, backward sampling obtains mixing properties that do not degrade with the number of time steps $T$, even for a fixed number of particles $N \geq 2$~\citep{andrieu2018uniform,lee2020coupled,karjalainen2024mixingtime}.

\subsection{Parallel-in-time particle Gibbs}\label{subsec:parallel-in-time-pgibbs}
Consider the auxiliary model~\eqref{eq:auxiliary-feynman-kac}
\begin{equation}\label{eq:auxiliary-feynman-kac-again}
    \begin{split}
        \pi(\ft{x}{0}{T}, \ft{u}{0}{T})
            &\propto g_0(x_0) \, p_0(x_0) \left\{\prod_{t=1}^T g_t(x_t, x_{t-1}) \, p_t(x_t \mid x_{t-1}) \right\}\left\{\prod_{t=0}^T \mathcal{N}\left(u_t; x_t, \frac{\delta_t}{2} \Sigma_t\right)\right\},\\
        &\coloneqq \Gamma_0(x_0) \left\{\prod_{t=1}^T \Gamma_t(x_t, x_{t-1})\right\}\left\{\prod_{t=0}^T \mathcal{N}\left(x_t; u_t, \frac{\delta_t}{2} \Sigma_t\right)\right\},
    \end{split}
\end{equation}
for $\Gamma_t(x_t, x_{t-1}) = g_t(x_t, x_{t-1}) \, p_t(x_t \mid x_{t-1})$ $t>1$ and $\Gamma_0(x_0) = g_0(x_0) \, p_0(x_0)$.
\citet{corenflos2022sequentialized} then proceeds from the enabling recursion on ``partial'' smoothing distributions
\begin{equation}\label{eq:auxiliary-recursion}
    \begin{split}
    \pi_{a:b}(x_{a:b} \mid u_{a:b})
    &\coloneqq
    \frac{1}
    {L_{a:b}}
    \mathcal{N}(x_a; u_a, \frac{\delta_a}{2} \Sigma_a) \prod_{t=a+1}^b \Gamma_t(x_t, x_{t-1}) \mathcal{N}\left(x_t; u_t, \frac{\delta_t}{2} \Sigma_t\right)
    \\
    &= 
    \frac{L_{a:c-1} L_{c:b}}
    {L_{a:b}}
    \Gamma_c(x_c, x_{c-1})
    \pi_{a:c-1}(x_{a:c-1} \mid u_{a:c-1})
    \pi_{c:b}(x_{c:b} \mid u_{c:b}),
    \end{split}
\end{equation}
and we have $\pi_{0:T}$ as the target distribution as well as $\pi_{a:a}(x_a \mid u_a) = \mathcal{N}(x_a; u_a, \frac{\delta_a}{2} \Sigma_a)$ for all $a=1, \ldots, T$, and $\pi_{0:0}(x_0 \mid u_0) = \mathcal{N}(x_0; u_0, \frac{\delta_0}{2} \Sigma_0) \Gamma_0(x_0)$.\footnote{Or equivalently $\pi_{0:0}(x_0 \mid u_0) = \mathcal{N}(x_0; u_0, \frac{\delta_0}{2} \Sigma_0)$ and the weight $\Gamma_0$ is added to $\Gamma_1$: i.e., $\Gamma_1 \leftarrow \Gamma_0 \times \Gamma_1$.}

The recursion~\eqref{eq:auxiliary-recursion} is then used to implement the parallel-in-time particle Gibbs algorithm.
Indeed, if $\frac{1}{N}\sum_{n=1}^N \delta_{X^n_{a:c-1}}$ and $\frac{1}{N}\sum_{n=1}^N \delta_{X^n_{c:b}}$ are two independent Monte Carlo approximations of $\pi_{a:c-1}$ and $\pi_{c:b}$, respectively, then the `stitched' empirical distribution
\begin{equation}\label{eq:product-form}
    \sum_{m, n=1}^N W^{mn} \delta_{[X^m_{a:c-1}, X^n_{c:b}]},
\end{equation}
where 
\begin{equation}
    W^{mn} = \frac{\Gamma_c(X^n_c, X^m_{c-1})}{\sum_{i,j=1}^N \Gamma_c(X^j_c, X^i_{c-1})},
\end{equation}
is an approximation of $\pi_{a:b}$.
We can then resample $N$ trajectories out of the $N^2$ following the weights $W^{mn}$ to then obtain an $N$-sized sample from $\pi_{a:b}$.
The conditional version of this approach is then implemented similarly as for standard conditional SMC (Algorithm~\ref{alg:csmc}), by ensuring that one of the trajectories remains the current state of the Markov chain at each time step, until the last `stitching' step where the genealogy is selected.
For more details on the implementation of this algorithm, we refer the reader to~\citet[Section 3]{corenflos2022sequentialized}.

\newpage
\section{Sequential results for the spatio-temporal experiment of Section~\ref{subsec:spatio-temporal}}
\label{app:spatial}
\begin{figure}[htb!]
    \centering
    \begin{subfigure}[t]{.45\textwidth}
      \centering
      \resizebox{\columnwidth}{!}{\begin{tikzpicture}[scale=1.]
\begin{axis}[
    grid=both,
    grid style=dashed,
    xmin=0, xmax=1023,
    xlabel={$t$ time step},
    ymin=0,
    ymax=4,
    scale only axis=true,
    width=\textwidth,
    height=0.5\textwidth,
    ]
\addplot[black, line width=1pt] table [x=t, y=Kalman, col sep=comma]{experiments/figures/spatial_ESJD_cpu.csv};\label{line:ESJD-Kalman-cpu}
\addplot[lightgray, line width=1pt] table [x=t, y=cSMC_grad, col sep=comma]{experiments/figures/spatial_ESJD_cpu.csv};\label{line:ESJD-cSMC-grad-cpu}
\addplot[gray, line width=1pt] table [x=t, y=cSMC, col sep=comma]{experiments/figures/spatial_ESJD_cpu.csv};\label{line:ESJD-cSMC-cpu}
\end{axis}
\end{tikzpicture}}
        \caption{Expected squared jump distance for the sequential versions of the auxiliary Kalman sampler~\ref{line:ESJD-Kalman-gpu}, the auxiliary cSMC sampler~\ref{line:ESJD-cSMC-cpu}, and the auxiliary cSMC sampler with gradient-informed proposals~\ref{line:ESJD-cSMC-grad-cpu}. Kalman~\ref{line:ESJD-Kalman-cpu} shows as a roughly horizontal line at the bottom.}
        \label{fig:esjd-spatio-temporal-cpu}
\end{subfigure}%
\hfill
\begin{subfigure}[t]{.45\textwidth}
    \centering
    \resizebox{\columnwidth}{!}{\begin{tikzpicture}[scale=1.]
\begin{axis}[
    grid=both,
    grid style=dashed,
    xmin=0, xmax=1023,
    xlabel={$t$ time step},
    ymax=1000,
    ymin=0,
    scale only axis=true,
    width=\textwidth,
    height=0.5\textwidth,
    ]
\addplot[black, line width=1pt] table [x=t, y=Kalman, col sep=comma]{experiments/figures/spatial_ESJD_time_cpu.csv};\label{line:ESJD-Kalman-time-cpu}
\addplot[lightgray, line width=1pt] table [x=t, y=cSMC_grad, col sep=comma]{experiments/figures/spatial_ESJD_time_cpu.csv};\label{line:ESJD-cSMC-grad-time-cpu}
\addplot[gray, line width=1pt] table [x=t, y=cSMC, col sep=comma]{experiments/figures/spatial_ESJD_time_cpu.csv};\label{line:ESJD-cSMC-time-cpu}

\end{axis}
\end{tikzpicture}}
    \caption{Expected squared jump distance per second for the auxiliary Kalman sampler~\ref{line:ESJD-Kalman-cpu}, the auxiliary cSMC sampler~\ref{line:ESJD-cSMC-cpu}, and the auxiliary cSMC sampler with gradient-informed proposals~\ref{line:ESJD-cSMC-grad-cpu}.}
    \label{fig:esjd-spatio-temporal-time-cpu}
\end{subfigure}
\caption{Average (across 20 different experiments) expected squared jump distance per iteration and second for all the sequential samplers considered on the spatio-temporal model~\eqref{eq:spatio-temporal}.}
\end{figure}
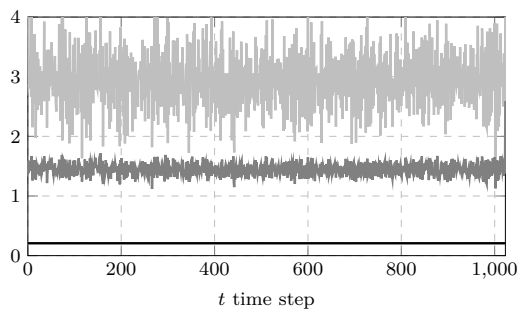
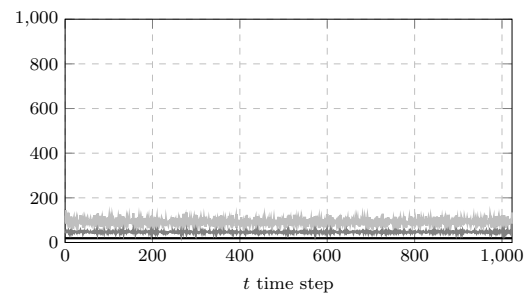

We now report the sequential counterpart of the experiment run in Section~\ref{subsec:spatio-temporal}.
It is worth noting that the sequential and parallel implementations of the two Kalman samplers are fully equivalent and only differ in their actual implementation.
Consequently, the expected squared jump distance for both should be (and is indeed) the same up to some variance coming from differences in generating the random variables for the sampling procedure.
This is not the case for the cSMC implementations, and while their properties should be similar (from using both the same proposal mechanism), they are not expected to behave exactly similarly.
The ESJD and ESJD per second are reported in Figure~\ref{fig:esjd-spatio-temporal-cpu} and Figure~\ref{fig:esjd-spatio-temporal-time-cpu}, respectively, where we have kept the same y-axis scale as in the parallel case for ease of comparison.
As discussed already in Section~\ref{subsec:spatio-temporal}, the non-sequential version are comparatively so much slower (up to 5 times as slow) in this instance than the parallel ones, that their comparative statistical performances are fully erased by their computational drawbacks.

\end{document}